\documentclass[12pt,english,aps,manuscript,12pt,A4]{revtex4}
\usepackage[T1]{fontenc}
\usepackage[latin9]{inputenc}
\usepackage{color}
\usepackage{babel}
\usepackage{verbatim}
\usepackage{bm}
\usepackage{amsmath}
\usepackage{amssymb}
\usepackage{graphicx}
\usepackage{esint}
\PassOptionsToPackage{normalem}{ulem}
\usepackage{ulem}
\usepackage[unicode=true,pdfusetitle,
 bookmarks=true,bookmarksnumbered=false,bookmarksopen=false,
 breaklinks=false,pdfborder={0 0 1},backref=false,colorlinks=true]
 {hyperref}
\hypersetup{
 pdfborderstyle=,linkcolor=blue,citecolor=cyan}

\makeatletter
\@ifundefined{textcolor}{}
{%
 \definecolor{BLACK}{gray}{0}
 \definecolor{WHITE}{gray}{1}
 \definecolor{RED}{rgb}{1,0,0}
 \definecolor{GREEN}{rgb}{0,1,0}
 \definecolor{BLUE}{rgb}{0,0,1}
 \definecolor{CYAN}{cmyk}{1,0,0,0}
 \definecolor{MAGENTA}{cmyk}{0,1,0,0}
 \definecolor{YELLOW}{cmyk}{0,0,1,0}
}

\usepackage{babel}
\newcommand{\modDY}[1]{{{\color{blue}{#1}}}}

\usepackage{xcolor}

\makeatother

\begin{document}
\title{Hydrodynamic helicity polarization in relativistic heavy ion collisions}
\author{Cong Yi}
\email{congyi@mail.ustc.edu.cn}

\affiliation{Department of Modern Physics, University of Science and Technology
of China, Hefei, Anhui 230026, China}
\author{Shi Pu}
\email{shipu@ustc.edu.cn}

\affiliation{Department of Modern Physics, University of Science and Technology
of China, Hefei, Anhui 230026, China}
\author{Jian-Hua Gao}
\email{gaojh@sdu.edu.cn}

\affiliation{Shandong Provincial Key Laboratory of Optical Astronomy and Solar-Terrestrial
Environment, Institute of Space Sciences, Shandong University, Weihai,
Shandong 264209, China}
\author{Di-Lun Yang}
\email{dlyang@gate.sinica.edu.tw}

\affiliation{Institute of Physics, Academia Sinica, Taipei, 11529, Taiwan}
\begin{abstract}
We study helicity polarization through the (3+1) dimensional
relativistic viscous hydrodynamic models at $\sqrt{s_{NN}}=200$GeV
Au+Au collisions. Similar to the local spin polarization, we consider
the helicity polarization beyond global equilibrium and investigate
the contributions induced by thermal vorticity, shear viscous tensor,
and the fluid acceleration.  
We find that the local helicity polarization induced by 
thermal vorticity dominates over other contributions. It also implies 
that in the low-energy collisions, the the fluid
vorticity as part of thermal vorticity may play the crucial role to the total
helicity polarization.
Such a finding could be useful for probing the local strength of vorticity
in rotational quark gluon plasmas by measuring helicity polarization.
Our simulation confirms the strict space reversal symmetry, whereas
we also compare our numerical results with approximated relations derived
from ideal Bjorken flow.
Our studies also provide a baseline for the future investigation on
local parity violation through the correlations of helicity polarization.
\end{abstract}
\maketitle

\section{Introduction}


In non-central heavy-ion collisions, large orbital angular
momentum (OAM) of the order of $10^{5}\hbar$ is produced, part of
which transfers into quark gluon plasma (QGP) as the form of vortical
fields. 
The large OAM is deposited in the QGP with fast rotation.
Such rotation can lead to the spin polarization of the hadrons similar
to the famous Barnett effect \citep{RevModPhys.7.129}. The global
polarization of $\Lambda$ and $\bar{\Lambda}$ hyperons created in
relativistic heavy ion collisions through spin-orbital coupling
was first proposed by Liang and Wang in Refs.  \citep{ZTL_XNW_2005PLB,ZTL_XNW_2005PRL}. In 2017,
the STAR collaboration observed the global polarization
of $\Lambda$ hyperons \citep{STAR:2017ckg}. There are many theoretical
approaches to investigate the global polarization including
the pioneer works based on the statistical field theory \citep{Becattini:2007nd,Becattini:2007sr,Becattini:2013fla}
and Winger-function approach near equilibrium \citep{Fang_2016PRC_polar},
as theoretical predictions even before experimental measurements,
from which the derived modified Cooper-Frye formula paves
the way for  numerical simulations.
In light of this formula, the results from numerical simulations
\citep{Karpenko_epjc2017_kb,XieYilong_prc2017_xwc,LiHui_prc2017_lpwx,Sun:2017xhx,Shi_plb2019_sll,WeiDexian_prc2019_wdh,Shi:2019wzi,Fu:2020oxj,Ryu:2021lnx,Lei:2021mvp}
are consistent with the experimental measurements for the
global polarization. See also Refs. \citep{Aziz:2021HADES,STAR:2021beb,Ivanov_prc2019_its,Deng:2020ygd,Guo:2021udq,Ayala:2021xrn,Deng:2021miw}
for recent studies of the spin polarization in low-energy collisions.


Later, in order to study the structure of the local vorticity in the
QGP, STAR collaboration measures the local spin polarization of $\Lambda$
hyperons as a function of azimuthal angle along the global angular
momentum and the beam directions \citep{Niida:2018hfw,Adam:2019srw},
dubbed as the transverse and longitudinal polarization.
Surprisingly, the numerical simulations from the same models mentioned
above for the global polarizations, disagree with the experiment data.
See the disagreements in e.g. relativistic hydrodynamics \citep{Becattini_prl2018_bk,Fu:2020oxj}
and transport models \citep{WeiDexian_prc2019_wdh,XieYilong_prc2017_xwc,Xia:2018tes}.
For longitudinal polarization, these theoretical calculations
obtain the results with qualitatively an opposite sign compared to
experimental observations. This discrepancy is called the ``sign''
problem for spin polarization in relativistic heavy-ion collisions.
It is found in Refs.~\citep{Xia:2019fjf,Becattini_epjc2019_bcs,Li:2021jvn}
that the feed-down effect cannot explain this disagreement either.
Although some phenomenological models \citep{Liu:2019krs,Wu:2020yiz,Wu:2019eyi,Voloshin_2018epjWeb}
qualitatively describe the experimental data, the ``sign'' problem
is still an open question in the community. 


Nevertheless, most of the theoretical studies have assumed
that the spin degree of freedom is in global thermal equilibrium at
a freezeout hypersurface, which is actually not justified from
the first principle. A lot of efforts are made to investigate
dynamical spin polarization with non-equilibrium effects both from
macroscopic and microscopic approaches. One of the macroscopic
theories is relativistic spin hydrodynamics \citep{Hattori:2019lfp,Fukushima:2020qta,Fukushima:2020ucl,Li:2020eon,She:2021lhe,Montenegro:2017lvf,Montenegro:2017rbu,Florkowski:2017ruc,Florkowski:2018myy,Becattini:2018duy,Florkowski:2018fap,Yang:2018lew,Bhadury:2020puc,Shi:2020qrx,Gallegos:2021bzp,Hongo:2021ona,Florkowski:2017dyn,Florkowski:2018ahw,Florkowski:2019qdp,Florkowski:2019voj,Bhadury:2020cop,Shi:2020htn,Singh:2020rht,Wang:2021ngp,Liu:2020ymh,Peng:2021ago,Florkowski:2021wvk,Copinger:2022jgg,Wang:2021wqq},
which includes the spin degree of freedom and spin-orbit
interaction  by coupling the hydrodynamic equations
with the conservation of angular momentum. 
On the other hand, one of the microscopic descriptions that
complements the macroscopic approach is the quantum kinetic theory
(QKT) for massive fermions with collisions \citep{Gao:2019znl,Weickgenannt:2019dks,Weickgenannt:2020aaf,Hattori:2019ahi,Wang:2019moi,Yang:2020hri,Weickgenannt:2020sit,Li:2019qkf,Liu:2020flb,Weickgenannt:2021cuo,Wang:2021qnt,Sheng:2021kfc,Huang:2020wrr},
which is an extension of the chiral kinetic theory (CKT) for massless
fermions \citep{Stephanov:2012ki,Son:2012zy,Gao:2012ix,Chen:2012ca,Manuel:2013zaa,Manuel:2014dza,Chen:2014cla,Chen:2015gta,Hidaka:2016yjf,Hidaka:2017auj,Mueller:2017lzw,Hidaka:2018ekt,Hidaka:2018mel,Gao:2018wmr,Huang:2018wdl,Liu:2018xip,Lin:2019ytz,Lin:2019fqo,Yamamoto:2020zrs}.
Also see Ref. \citep{Hidaka:2022dmn} for a recent review of QKT. There is also a distinct microscopic
model incorporating the spin-orbital interaction in collisions in Ref. \citep{Zhang:2019xya} . In
addition, there have been further studies on the QKT for polarized
photons \cite{Huang:2020kik,Hattori:2020gqh,Lin:2021mvw} with possible
generalization to weakly coupled gluons and the inclusion of background
chromo-electromagnetic fields for the QKT of massless and massive
fermions \cite{Luo:2021uog,Muller:2021hpe,Yang:2021fea}.


Recently, the shear induced polarization, which was found
for massless fermions in Ref.~\citep{Hidaka:2017auj} and later obtained
for massive fermions \citep{Liu:2020dxg,Liu:2021uhn,Becattini:2021suc}
in local thermal equilibrium, has drawn lots of attentions. 
Including such an effect, the local spin polarization from
numerical simulations could qualitatively match the experimental observations
\citep{Fu:2021pok,Becattini:2021iol}, while the numerical results
depend on different approximations adopted. It is also pointed
out that the polarization of strange quarks is sensitive
to the equation of state and other parameters \citep{Yi:2021ryh, Wu:2022mkr}.
See also Ref. \citep{Sun:2021nsg} for similar studies on the parameter dependences in the $\sqrt{s_{NN}}=19.6\textrm{GeV}$ collisions and Refs.~\citep{Shi:2020htn,Liu:2021nyg,Florkowski:2021xvy,Wu:2022mkr}
for related studies. Therefore, solving the ``sign'' problem requires
more systematic studies on the off-equilibrium effects, which
may be obtained from spin hydrodynamics or QKT. 


Moreover, 
helicity polarization defined as the local spin polarization projected
to the momentum direction of polarized hadrons has been proposed
in Refs.~\citep{Becattini:2020xbh,Gao:2021rom}, which could be implemented
to probe local parity violation
characterized by an axial chemical potential in quantum chromodynamics
(QCD) matter at finite temperature (see Ref.~\cite{Muller:2021hpe}
for a different proposal) to complement the longstanding search for
the chiral magnetic effect \cite{Vilenkin:1980fu,Kharzeev:2007jp,Fukushima:2008xe,Kharzeev:2015znc,STAR:2009wot,STAR:2021mii}.
See also Refs.~\citep{Jacob:1987sj,Ambrus:2019khr,Ambrus:2020oiw,Ambrus:2019ayb}
for other studies and phenomenological applications related to particle
helicity in relativistic heavy ion collisions. In order to extract
the signal of local-parity violation from helicity-helicity (polarization)
correlations \citep{Becattini:2020xbh,Gao:2021rom}, it is essential
to study the helicity polarization without an axial chemical potential
from hydrodynamic as a baseline for the future analysis.

In this work, we study the hydrodynamic helicity polarization. 
We focus on the local-equilibrium contributions from thermal
vorticity, shear corrections, and fluid acceleration and analyze their
features analytically.  Then, we implement the (3+1)-dimensional
relativistic viscous hydrodynamic models to simulate the hydrodynamic
helicity polarization. We examine the relations derived from
the ideal Bjorken flow \citep{Gao:2021rom} and space reversal symmetry.
We also investigate the helicity polarization for both $\Lambda$-hyperon
equilibrium and strange-quark equilibrium (abbreviated as $\Lambda$
equilibrium and s equilibrium) scenarios proposed in Refs.~\cite{Fu:2021pok,Yi:2021ryh}.

The structure of this article is as follows. In Sec. \ref{sec:Helicity-polarization},
we introduce the helicity polarization with corrections in
local equilibrium \citep{Hidaka:2017auj,Yi:2021ryh} and briefly review
the analysis of the contribution from thermal vorticity based on symmetries
\citep{Becattini:2020xbh,Gao:2021rom}.  
In Sec. \ref{sec:Numerical-results-from}, we implement the (3+1)dimensional
viscous hydrodynamic simulation to study the azimuthal angle and the
momentum rapidity dependence of helicity polarization. At last, we
summarize our results and make further discussions in Sec.
\ref{sec:Conclusion-and-discussion}. Throughout this work, we adopt
the metric $g_{\mu\nu}=\textrm{diag}\{+,-,-,-\}$, $\epsilon^{0123}=1$
and the projector $\Delta^{\mu\nu}=g^{\mu\nu}-u^{\mu}u^{\nu}$ with
$u^{\mu}$ being fluid velocity. We also use the boldface
notation such as ${\mathbf{k}}$ to denote the spatial component of
a four-vector like $k^{\mu}$.

\section{Helicity polarization \label{sec:Helicity-polarization}}

In this section, we briefly review the formalism for helicity polarization
based on Ref. \citep{Becattini_prl2018_bk,Becattini:2020xbh,Gao:2021rom}.
For simplicity, we concentrate on the helicity polarization induced
by hydrodynamic variables and neglect the contribution from axial
chemical potential.

We start from the single-particle mean spin vector $\mathcal{S}^{\mu}(p)$
by the modified Cooper-Frye formula \citep{Becattini:2013fla,Fang:2016uds}.
\begin{equation}
\mathcal{S}^{\mu}(\mathrm{p})=\frac{\int d\Sigma\cdot p\mathcal{J}_{5}^{\mu}(p,X)}{2m_{\Lambda}\int d\Sigma\cdot\mathcal{N}(p,X)},\label{eq:CooperFryeFormula}
\end{equation}
where the $m_{\Lambda}$ is the mass of $\Lambda$ hyperons, $\Sigma_{\mu}$
is the normal vector of the freeze-out surface, the $\mathcal{N}^{\mu}(p,X)$
and $\mathcal{J}_{5}^{\mu}(p,X)$ are number density and axial-charge
current density in phase space, respectively. The $\mathcal{N}^{\mu}(p,X)$
and $\mathcal{J}_{5}^{\mu}(p,X)$ can be derived from the quantum
kinetic theory \citep{Hidaka:2018mel} 
\begin{eqnarray}
\mathcal{N}^{\mu}(p,X) & = & 2\int_{p\cdot n}\left[\mathcal{J}_{+}^{\mu}(p,X)+\mathcal{J}_{-}^{\mu}(p,X)\right],\nonumber \\
\mathcal{J}_{5}^{\mu}(p,X) & = & 2\int_{p\cdot n}\left[\mathcal{J}_{+}^{\mu}(p,X)-\mathcal{J}_{-}^{\mu}(p,X)\right],
\end{eqnarray}
where $\int_{p\cdot n}\equiv\int dp\cdot np\cdot n\theta(p\cdot n)/(2\pi)$
with $n^{\mu}$ being chosen as the fluid velocity $u^{\mu}$ in thermal
equilibrium, $\mathcal{J}_{+}^{\mu}(p,X)$ and $\mathcal{J}_{-}^{\mu}(p,X)$
are the Wigner functions for the right and left-handed fermions, respectively.

Inserting the expression of $\mathcal{J_{\pm}^{\mu}}(p,X)$ into Eq.
(\ref{eq:CooperFryeFormula}) and assuming the chemical potential
for left and right handed fermions are identical $\mu_{R}=\mu_{L}\equiv\mu$,
we can further decompose $\mathcal{S}^{\mu}(\mathbf{p})$ as \citep{Yi:2021ryh},
\begin{equation}
\mathcal{S}^{\mu}(\mathbf{p})=\mathcal{S}_{\textrm{thermal}}^{\mu}(\mathbf{p})+\mathcal{S}_{\textrm{shear}}^{\mu}(\mathbf{p})+\mathcal{S}_{\textrm{accT}}^{\mu}(\mathbf{p})+\mathcal{S}_{\textrm{chemical}}^{\mu}(\mathbf{p})+\mathcal{S}_{\textrm{EB}}^{\mu}(\mathbf{p}),
\end{equation}
where 
\begin{eqnarray}
\mathcal{S}_{\textrm{thermal}}^{\mu}(\mathbf{p}) & = & \int d\Sigma^{\sigma}F_{\sigma}\epsilon^{\mu\nu\alpha\beta}p_{\nu}\partial_{\alpha}\frac{u_{\beta}}{T},\nonumber \\
\mathcal{S}_{\textrm{shear}}^{\mu}(\mathbf{p}) & = & \int d\Sigma^{\sigma}F_{\sigma}\frac{\epsilon^{\mu\nu\alpha\beta}p_{\nu}}{(u\cdot p)T}\left\{ p^{\rho}(\partial_{\rho}u_{\alpha}+\partial_{\alpha}u_{\rho}-u_{\rho}Du_{\alpha})u_{\beta}\right\} \nonumber \\
\mathcal{S}_{\textrm{accT}}^{\mu}(\mathbf{p}) & = & -\int d\Sigma^{\sigma}F_{\sigma}\frac{1}{T}\epsilon^{\mu\nu\alpha\beta}p_{\nu}u_{\alpha}(Du_{\beta}-\frac{1}{T}\partial_{\beta}T),\nonumber \\
\mathcal{S}_{\textrm{chemical}}^{\mu}(\mathbf{p}) & = & 2\int d\Sigma^{\sigma}F_{\sigma}\frac{1}{(u\cdot p)}\epsilon^{\mu\nu\alpha\beta}p_{\alpha}u_{\beta}\partial_{\nu}\frac{\mu}{T},\nonumber \\
\mathcal{S}_{\textrm{EB}}^{\mu}(\mathbf{p}) & = & 2\int d\Sigma^{\sigma}F_{\sigma}\left[\frac{1}{(u\cdot p)T}\epsilon^{\mu\nu\alpha\beta}p_{\alpha}u_{\beta}E_{\nu}+\frac{B^{\mu}}{T}\right],\label{eq:S_all}
\end{eqnarray}
and, 
\begin{equation}
F^{\mu}=\frac{\hbar}{8m_{\Lambda}N}p^{\mu}f_{V}^{(0)}(1-f_{V}^{(0)}),\;  N = \int d\Sigma^{\mu}p_{\mu}f_{V}^{(0)},
\label{eq:def_N}
\end{equation}
Here, $T$ is the temperature 
and $\text{\ensuremath{f_{V}^{(0)}}}$ is the Fermic-Dirac distribution function. The subscripts, \emph{thermal, shear, accT, chemical }and\emph{ EB,}
stand for the terms related to thermal vorticity, shear viscous tensor,
the fluid acceleration minus gradient of temperature $(Du_{\beta}-\frac{1}{T}\partial_{\beta}T)$,
the gradient of $\mu/T$ , and electromagnetic fields, respectively. The $E^{\mu}$ and $B^{\mu}$ are given by,
$E^{\mu}=F^{\mu\nu}u_{\nu}$ and $B^{\mu}=\frac{1}{2}\epsilon^{\mu\nu\alpha\beta}u_{\nu}F_{\alpha\beta}$.
Note that Ref. \citep{Yi:2021ryh} has roughly extended the case for
massless fermions \cite{Hidaka:2017auj} to the one for
massive fermions. For the related decomposition as Eq.~(\ref{eq:S_all})
for massive fermions, one may refer to Refs. \citep{Liu:2019krs,Liu:2020dxg,Liu:2021uhn,Fu:2021pok,Becattini:2021iol,Becattini:2021suc}.

Helicity polarization is defined as \citep{Becattini:2020xbh,Gao:2021rom},
\begin{eqnarray}
S^{h} & = & \text{\ensuremath{\widehat{\mathbf{p}}}\ensuremath{\cdot}}\mathcal{\boldsymbol{S}}(\mathbf{p})=\widehat{p}^{x}\mathcal{S}^{x}+\widehat{p}^{y}\mathcal{S}^{y}+\widehat{p}^{z}\mathcal{S}^{z},\label{eq:def_Sh}
\end{eqnarray}
where $\mathbf{\ensuremath{\widehat{\mathbf{p}}}}\equiv\mathbf{p}/|\mathbf{p}|$
Inserting Eq. (\ref{eq:S_all}) into Eq. (\ref{eq:def_Sh}),
we obtain 
\begin{eqnarray}
S_{\textrm{thermal}}^{h}(\mathbf{p}) & = & \int d\Sigma^{\sigma}F_{\sigma}p_{0}\epsilon^{0ijk}\widehat{p}_{i}\nabla_{j}\left(\frac{u_{k}}{T}\right),\nonumber \\
S_{\textrm{shear}}^{h}(\mathbf{p}) & = & -\int d\Sigma^{\sigma}F_{\sigma}\frac{\epsilon^{0ijk}\widehat{p}^{i}p_{0}}{(u\cdot p)T}\left\{ p^{\sigma}(\partial_{\sigma}u_{j}+\partial_{j}u_{\sigma}-u_{\sigma}Du_{j})u_{k}\right\} ,\nonumber \\
S_{\textrm{accT}}^{h}(\mathbf{p}) & = & \int d\Sigma^{\sigma}F_{\sigma}\frac{1}{T}\epsilon^{0ijk}\widehat{p}^{i}p_{0}u_{j}(Du_{k}-\frac{1}{T}\partial_{k}T),\nonumber \\
S_{\textrm{chemical}}^{h}(\mathbf{p}) & = & -2\int d\Sigma^{\sigma}F_{\sigma}\frac{1}{(u\cdot p)}p_{0}\epsilon^{0ijk}\hat{p}_{i}\left[\nabla_{j}\left(\frac{\mu}{T}\right)\right]u_{k},\nonumber\\
S_{\textrm{EB}}^{h}(\mathbf{p}) & = & 2\int d\Sigma^{\sigma}F_{\sigma}\left[\frac{1}{(u\cdot p)T}\epsilon^{0ijk}\widehat{p}^{i}p_{0}E_{j}u_{k}+\frac{\widehat{p}^{i}B^{i}}{T}\right],\label{eq:helcity_decomp_01}
\end{eqnarray}
where we implicitly impose the on-shell condition $p_{0}=\sqrt{|{\bf p}|^{2}+m^{2}}$
with $m$ being the fermionic mass in the end. We emphasize that
only the spatial components of thermal vorticity contribute to the
helicity polarization. The measurement of the helicity polarization
can provide the information of the spatial thermal vorticity, which could present  a fine structure of thermal vorticity.  

Now we review the symmetric properties
for $S_{\textrm{thermal}}^{h}(\mathbf{p})$ in hydrodynamical models.
For simplicity, we consider $S_{\textrm{thermal}}^{h}(\mathbf{p})$
in an ideal fluid and eventually implement our result in a Bjorken
flow. 
More detailed analyses based on symmetries are shown in Refs.~
\citep{Becattini_prl2018_bk,Becattini:2020xbh,Gao:2021rom}.

In an ideal fluid, the temperature vorticity,

\begin{equation}
\Omega_{T}^{\mu\nu}=\partial^{\mu}(Tu^{\nu})-\partial^{\nu}(Tu^{\mu}),\label{eq:Tvoticity}
\end{equation}
is conserved along the velocity, i.e. $\Omega_{T}^{\mu\nu}u_{\nu}=0$
\citep{Becattini:2015ska,Deng_2016PRC,Wu:2019eyi} and satisfies the
relativistic Kelvin circulation theorem \citep{MahajanPRL2010,Gao:2014coa,Yang:2017asn,Wang:2020ewj}.
As a consequence, if $\Omega_{T}^{\mu\nu}$ is zero at initial time,
it will always be vanishing during the evolution of this ideal fluid.
Then, by using this condition, we can express the $(\partial_{\mu}u_{\nu}-\partial_{\nu}u_{\mu})$
as, 
\begin{equation}
(\partial_{\mu}u_{\nu}-\partial_{\nu}u_{\mu})=-\frac{1}{T}(u_{\nu}\partial_{\mu}-u_{\mu}\partial_{\mu})T.\label{eq:Tvorticity_02}
\end{equation}
Using Eq. (\ref{eq:Tvorticity_02}), we simplify $\mathcal{S}_{\textrm{thermal}}^{\mu}(\mathbf{p})$
in Eq. (\ref{eq:S_all}), 
\begin{eqnarray}
\mathcal{S}_{\textrm{thermal}}^{\mu}(\mathbf{p}) & = & \frac{2}{T^{2}}\int d\Sigma^{\sigma}F_{\sigma}\epsilon^{\mu\nu\alpha\beta}p_{\nu}(u_{\alpha}\partial_{\beta}T).
\end{eqnarray}
To compute the $\mathcal{S}_{\textrm{thermal}}^{\mu}(\mathbf{p})$
in hydrodynamics at the freeze-out hyper-surface, we can assume that
the distribution function $f_{V}^{(0)}$ is approximately at the equilibrium.
Then, we can use the relation $\frac{\partial}{\partial p^{\sigma}}f_{V}^{(0)}=-\frac{u_{\sigma}}{T}f_{V}^{(0)}(1-f_{V}^{(0)})$
to remove the fluid velocity and obtain

\begin{eqnarray}
\mathcal{S}_{\textrm{thermal}}^{\mu}(\mathbf{p}) & = & -\frac{1}{4m_{\Lambda}N}\int d\Sigma_{\alpha}p^{\alpha}\epsilon^{\mu\nu\rho\sigma}p_{\nu}\frac{1}{T}(\partial_{\sigma}T)\frac{\partial f_{V}^{(0)}}{\partial p^{\text{\ensuremath{\rho}}}},
\end{eqnarray}
Integrating by parts, we get 
\begin{eqnarray}
\mathcal{S}_{\textrm{thermal}}^{\mu}(\mathbf{p}) & = & -\frac{1}{4m_{\Lambda}N}\epsilon^{\mu\nu\rho\sigma}p_{\nu}\frac{\partial}{\partial p^{\text{\ensuremath{\rho}}}}\left[\int d\Sigma_{\alpha}p^{\alpha}\frac{1}{T}(\partial_{\sigma}T)f_{V}^{(0)}\right]\nonumber \\
 &  & +\frac{1}{4m_{\Lambda}N}\epsilon^{\mu\nu\rho\sigma}p_{\nu}\int d\Sigma_{\ensuremath{\rho}}\frac{\partial_{\sigma}T}{T}f_{V}^{(0)}.\label{eq:thermal_ideal_02}
\end{eqnarray}
As argued in Ref. \citep{Becattini_prl2018_bk,Gao:2021rom}, the
temperature may be constant at the freeze-out hyper-surface and the
direction of $\partial^{\mu}T$ is approximately parallel to normal
vector of hyper-surface $\Sigma^{\mu}$. The second term in above
equation may therefore vanish at the freeze out hyper surface. Here,
we emphasize that in the later hydrodynamic simulations, we did not
follow this approximation.

In an ideal fluid with longitudinal boost invariant, the temperature
only depends on the proper time $\tau=\sqrt{t^{2}-z^{2}}$ and $\epsilon^{\mu\nu\rho\sigma}(\partial_{\sigma}T)=\epsilon^{\mu\nu\rho0}\text{cosh}\ensuremath{\eta\frac{dT}{d\tau}}-\epsilon^{\mu\nu\rho3}\text{\text{sinh \ensuremath{\eta\frac{dT}{d\tau}}}},$where
$\eta$ is spatial rapidity The spin vector $\mathcal{S}_{\textrm{thermal}}^{\mu}(\mathbf{p})$
in Eq. (\ref{eq:thermal_ideal_02}) is reduced to \citep{Gao:2021rom}
\begin{eqnarray}
\mathcal{S}_{\textrm{thermal}}^{\mu}(\mathbf{p}) & = & -\frac{1}{4m_{\Lambda}N}p_{\nu}\frac{\partial}{\partial p^{\text{\ensuremath{\rho}}}}\left[\int d\Sigma_{\alpha}p^{\alpha}f_{V}^{(0)}(\epsilon^{\mu\nu\rho0}\text{cosh}\ensuremath{\eta}-\epsilon^{\mu\nu\rho3}\text{\text{sinh \ensuremath{\eta}}})\frac{1}{T}\frac{dT}{d\tau}\right].\label{eq:thermal_Bjorken_01}
\end{eqnarray}
Note that only the second term in Eq. (\ref{eq:thermal_Bjorken_01})
contributes to helicity polarization $\mathcal{S}_{\text{thermal}}^{h}$.
At small rapidity $Y$ region, we can approximate that the space-time rapidity is equal to the momentum rapidity, i.e. $\eta \simeq Y$.
After taking the Fourier transformation to the azimuthal distribution at the freezeout hypersurface,
\begin{equation}
\int d\Sigma_\lambda p^\lambda f^{(0)}_V = \frac{d N}{2 \pi E_p p_T dp_T dY}\left[1+\sum_{n=1}^\infty 2v_n(p_T,Y) \cos n\phi \right],
\end{equation}
the polarization vector $\mathcal{S}_{\textrm{thermal}}^{\mu}(\mathbf{p})$ 
and helicity polarization $S^h(\mathbf{p})$ read
\citep{Becattini_prl2018_bk,Gao:2021rom}, 
\begin{eqnarray}
\mathcal{S}_{\text{thermal}}^{z}\ & = & -\frac{1}{4m_{\Lambda}N}\frac{1}{T}\left.\frac{dT}{d\tau}\right|_{\Sigma}\partial_{\phi}\int d\Sigma_{\alpha}p^{\alpha}f_{V}^{(0)}\cosh\eta  \nonumber \\
& \approx &  \frac{1}{4m_{\Lambda}}\frac{1}{T}\left.\frac{dT}{d\tau}\right|_{\Sigma}\left[ \sum_{n=1}^{\infty}2 n v_{n}(p_{T},Y)
\sin n\phi \right], \label{eq:Sthz}\\
S^{h}_{\textrm{thermal}} & = & -\frac{1}{4m_{\Lambda}N}\frac{1}{T}\left.\frac{dT}{d\tau}\right|_{\Sigma}\partial_{\phi}\int d\Sigma_{\alpha}p^{\alpha}f_{V}^{(0)}\sinh\eta
\nonumber \\
 & \approx & \frac{Y}{4m_{\Lambda}}\frac{1}{T}\left.\frac{dT}{d\tau}\right|_{\Sigma} \left[\sum_{n=1}^{\infty}2 n v_{n}(p_{T},Y)\sin n\phi \right].
 \label{eq:YSz}
\end{eqnarray}
From Eqs. (\ref{eq:Sthz}) and (\ref{eq:YSz}), it implies that in the small rapidity region $Y\sim 0$ \citep{Becattini_prl2018_bk,Gao:2021rom}, 
\begin{equation}
S_{\text{thermal}}^{h}(Y,\phi_{p}) \approx Y \mathcal{S}^z_{\text{thermal}}(Y,\phi_{p}).
\label{eq:thermal_property_01}
\end{equation}
Meanwhile, for a given rapidity, the elliptical flow coefficient $v_2$ is larger than other coefficients $v_{n}$. Therefore, Eq.~(\ref{eq:Sthz}) can also be written as in small rapidity region $Y\sim 0$,
\begin{equation}
\mathcal{S}^z_{\textrm{thermal}} \approx \frac{1}{m_{\Lambda}}\frac{1}{T}\left.\frac{dT}{d\tau}\right|_{\Sigma} v_2(p_T,0) \sin 2\phi.
\end{equation}
In hydrodynamic simulations, the system is beyond the ideal Bjorken
flow, we therefore expect that the condition (\ref{eq:thermal_property_01})
is approximately satisfied.We will revisit Eq. (\ref{eq:thermal_property_01}) after the rapidity integration in Sec. \ref{sec:Numerical-results-from}.

Meanwhile, the system should have the space reversal symmetry \citep{Becattini_prl2018_bk,Becattini:2020xbh},
i.e. $\boldsymbol{\mathcal{\boldsymbol{S}}}(\mathbf{p})=\boldsymbol{\mathcal{\boldsymbol{S}}}(-\mathbf{p}),$
or 
\begin{equation}
\boldsymbol{\mathcal{\boldsymbol{S}}}(Y,\phi_{p})=\boldsymbol{\mathcal{\boldsymbol{S}}}(-Y,\phi_{p}+\pi),\label{eq:S_space_reversal}
\end{equation}
and from Eq. (\ref{eq:def_Sh}) we can obtain the following relation,
\begin{equation}
S^{h}(Y,\phi_{p})=-S^{h}(-Y,\phi_{p}+\pi).\label{eq:thermal_property_02}
\end{equation}
Note that, Eq. (\ref{eq:thermal_property_02}) should be satisfied
for each part in Eq. (\ref{eq:helcity_decomp_01}). We will test it
in the later hydrodynamic simulations.

At last, let us discuss the frame dependence of helicity polarization. All the quantities mentioned above are chosen in the laboratory frame. In experiments, the polarization of $\Lambda$ and $\bar{\Lambda}$ hyperons are measured in their own rest frames. Since the momentum in the rest frame of hyperons $p^{\prime \mu}=(m_\Lambda, 0)$ is related to the momentum $p^\mu=(E_\Lambda,\bold{p})$ in the laboratory frame by Lorentz transformation, i.e. $p^{\prime\mu} = \Lambda^\mu_{\;\nu}p^\nu$, we obtain the polarization vector $\mathcal{S}^{\prime\mu}$ in the rest frame of hyperons, 
\begin{equation}
\mathcal{S}^{\prime \mu} = \left( 0,  \boldsymbol{\mathcal{S}}- \frac{(\bold{p}\cdot \boldsymbol{\mathcal{S}})\bold{p}}{E_\Lambda(E_\Lambda+m_\Lambda)} \right).
\end{equation}
Similar to the definition in Eq.~(\ref{eq:def_Sh}), one can also define the helicity polarization as $\hat{\bold{p}} \cdot \boldsymbol{\mathcal{S}}^\prime$, which is connected to our helicity polarization $S^h$ by,
\begin{equation}
\hat{\bold{p}} \cdot \boldsymbol{\mathcal{S}}^\prime =\frac{m_\Lambda}{E_\Lambda} \hat{\bold{p}} \cdot \boldsymbol{\mathcal{S}} 
=\frac{m_\Lambda}{E_\Lambda} S^h. 
\end{equation}
Therefore, it is straightforward to transform our simulation results to $\hat{\bold{p}} \cdot \boldsymbol{\mathcal{S}}^\prime$  \citep{Becattini_prl2018_bk,Becattini:2020xbh}.

Before end this section, we emphasize  that in general the axial chemical potential $\mu_{A}=\mu_{R}-\mu_{L}$
can also contribute to the $\mathcal{S}^{\mu}(\mathbf{p})$ and leads
to the extra helicity polarization, which could be a possible signal
of the parity violation \citep{Becattini:2020xbh}. Since we have
already assumed that $\mu_{A}=0$ in Eqs. (\ref{eq:S_all}), there
are no such corrections in our case. One can also see Refs.
\citep{Becattini:2020xbh,Shi:2020qrx,Liu:2021nyg} for other corrections
for $\mathcal{S}^{\mu}(\mathbf{p})$.

\section{Numerical results from hydrodynamics approach \label{sec:Numerical-results-from}}

In this section, we study the the azimuthal angle $\phi_{p}$ and
the momentum rapidity $Y$ dependence of helicity polarization $S^{h}$
in the hydrodynamic model and compare our results with those in an
ideal Bjorken flow shown in Sec. \ref{sec:Helicity-polarization}.

\subsection{Setup \label{subsec:Setup}}

\begin{figure}
\includegraphics[scale=0.35]{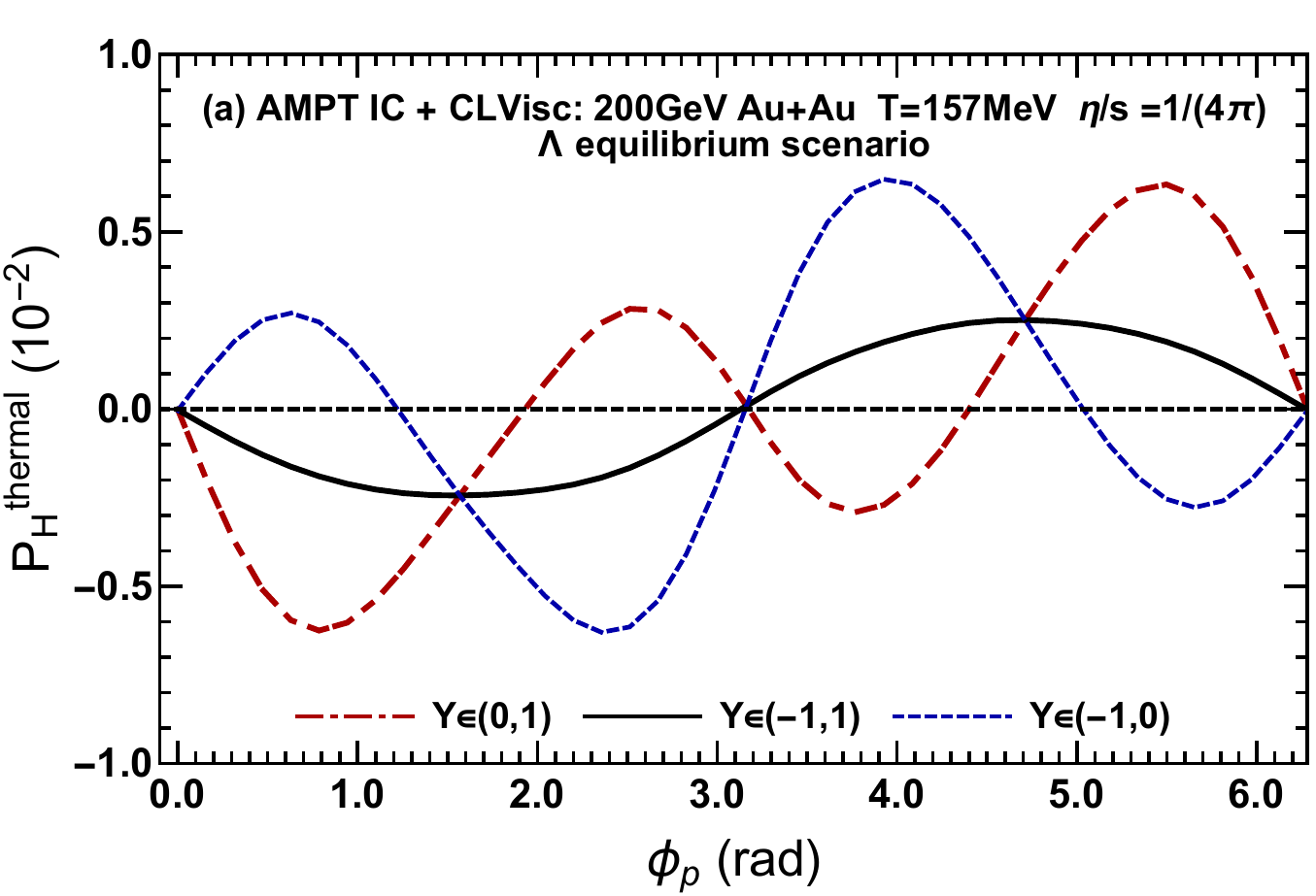}\includegraphics[scale=0.35]{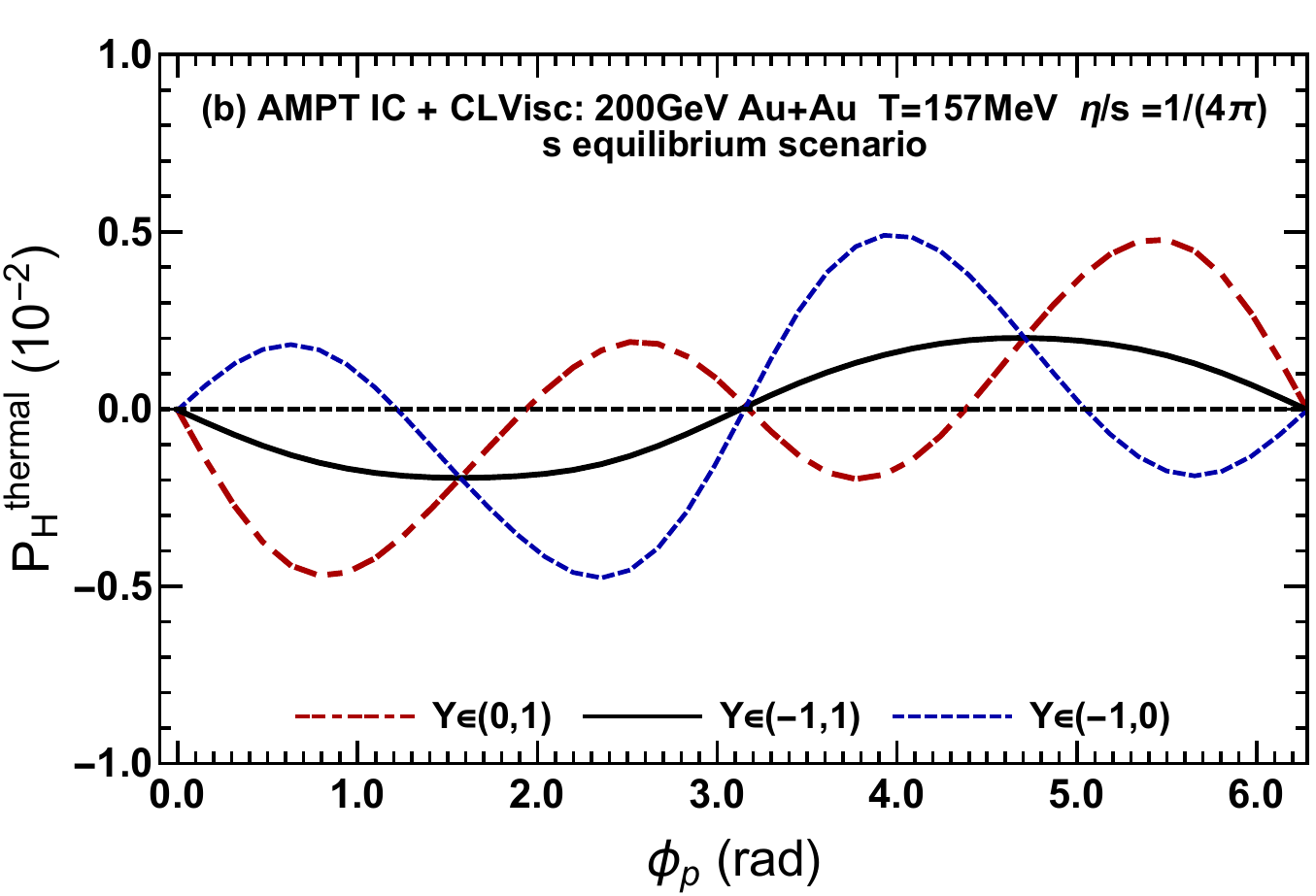}

\includegraphics[scale=0.35]{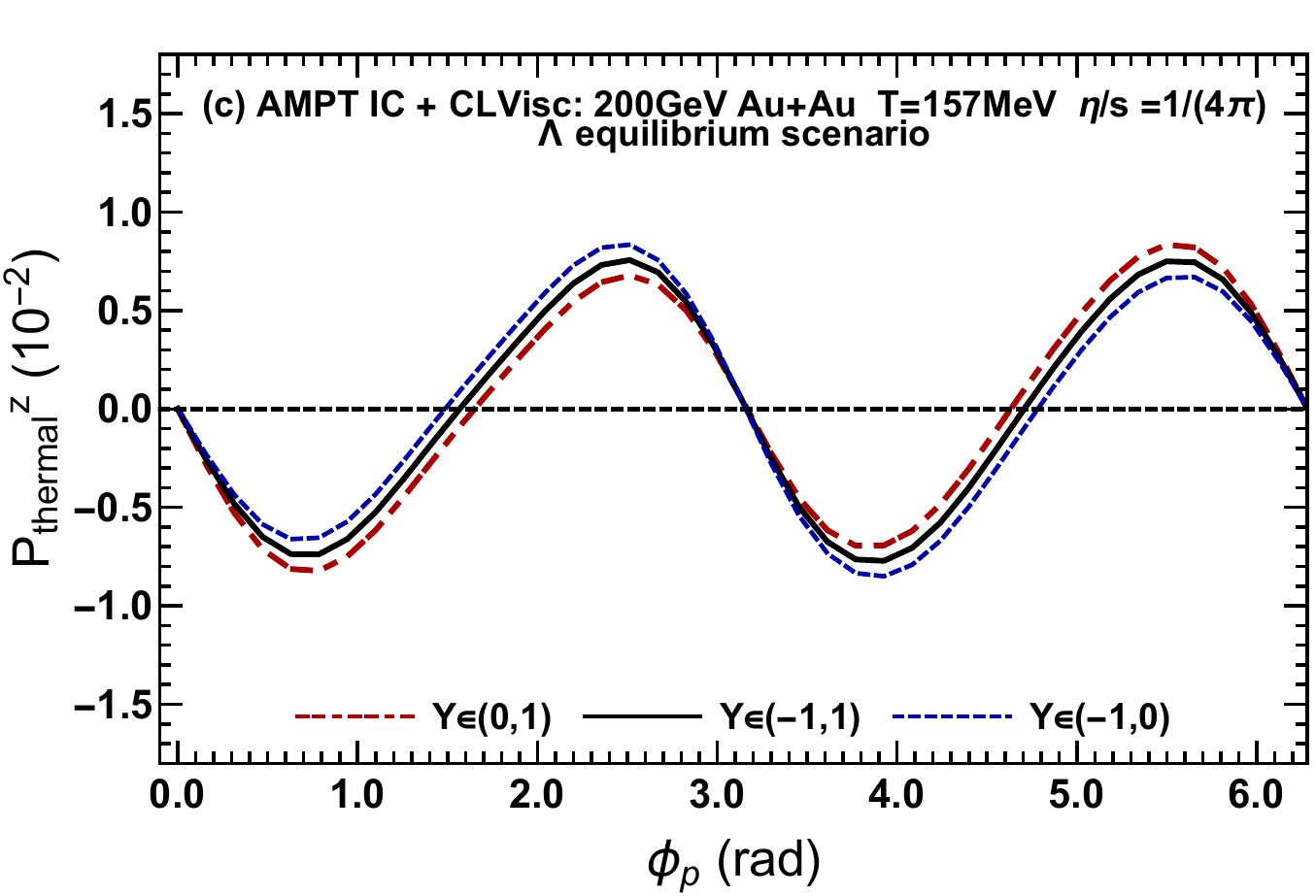}\includegraphics[scale=0.35]{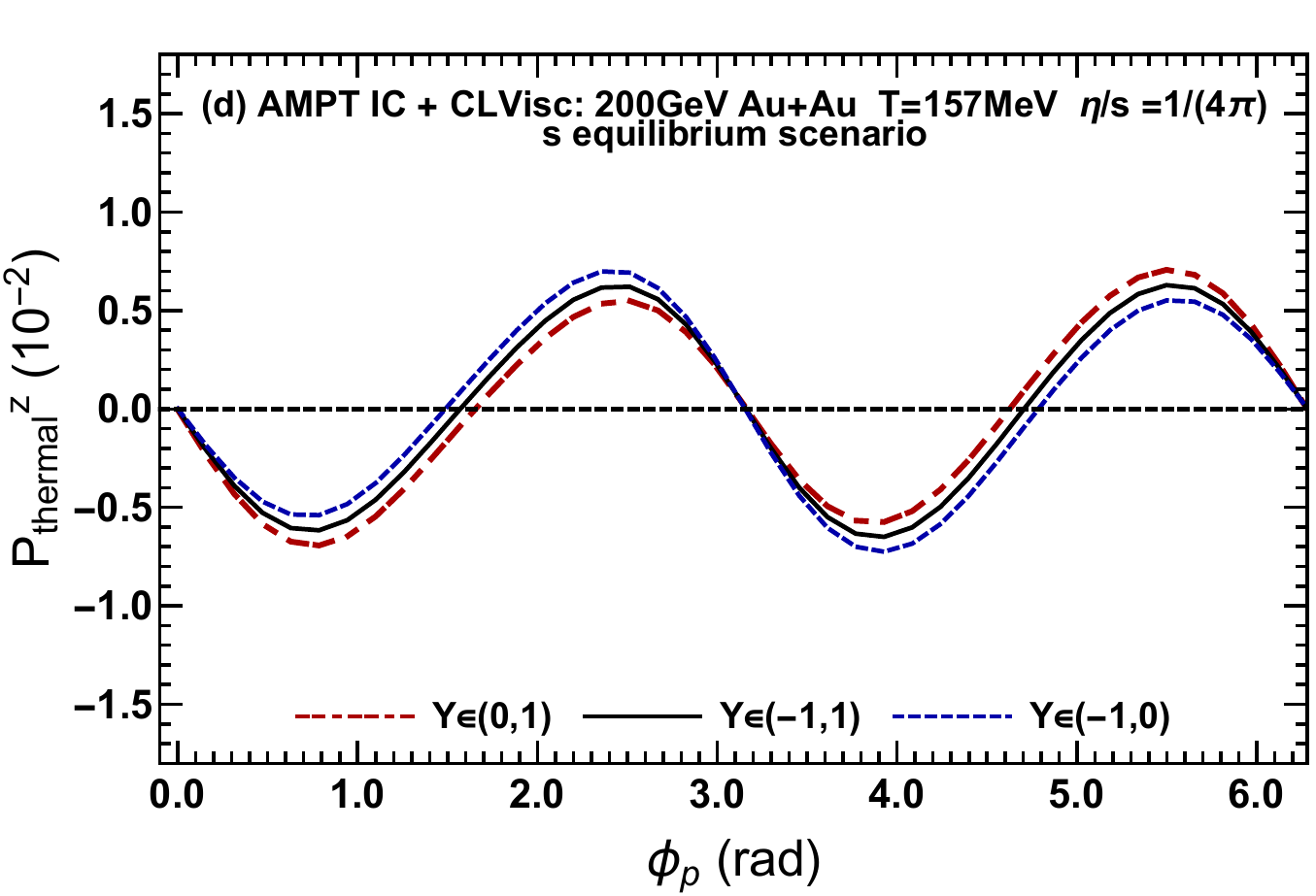}

\caption{The helicity polarization $P_{H}^{\text{thermal}}$ and spin polarization
along the beam direction $P_{\text{thermal}}^{z}$ contributed by
thermal vorticity as a function of $\phi_{p}$ in upper and lower
plane, respectively. The results for $\Lambda$ and $s$\emph{ }\textit{\emph{equilibrium
scenario}}\emph{ }are shown in the left and right handed side, respectively.
Black solid, red dash-dotted and blue dashed lines stand for helicity
polarization $P_{H}^{\text{thermal}},P_{H}^{+\text{thermal}},P_{H}^{-\text{thermal}}$
or spin polarization $P_{\text{thermal}}^{z},P_{\text{thermal}}^{+z},P_{\text{thermal}}^{-z}$
in the upper and lower plane, respectively. \label{fig:th_PH_phi}}
\end{figure}

\begin{figure}
\includegraphics[scale=0.35]{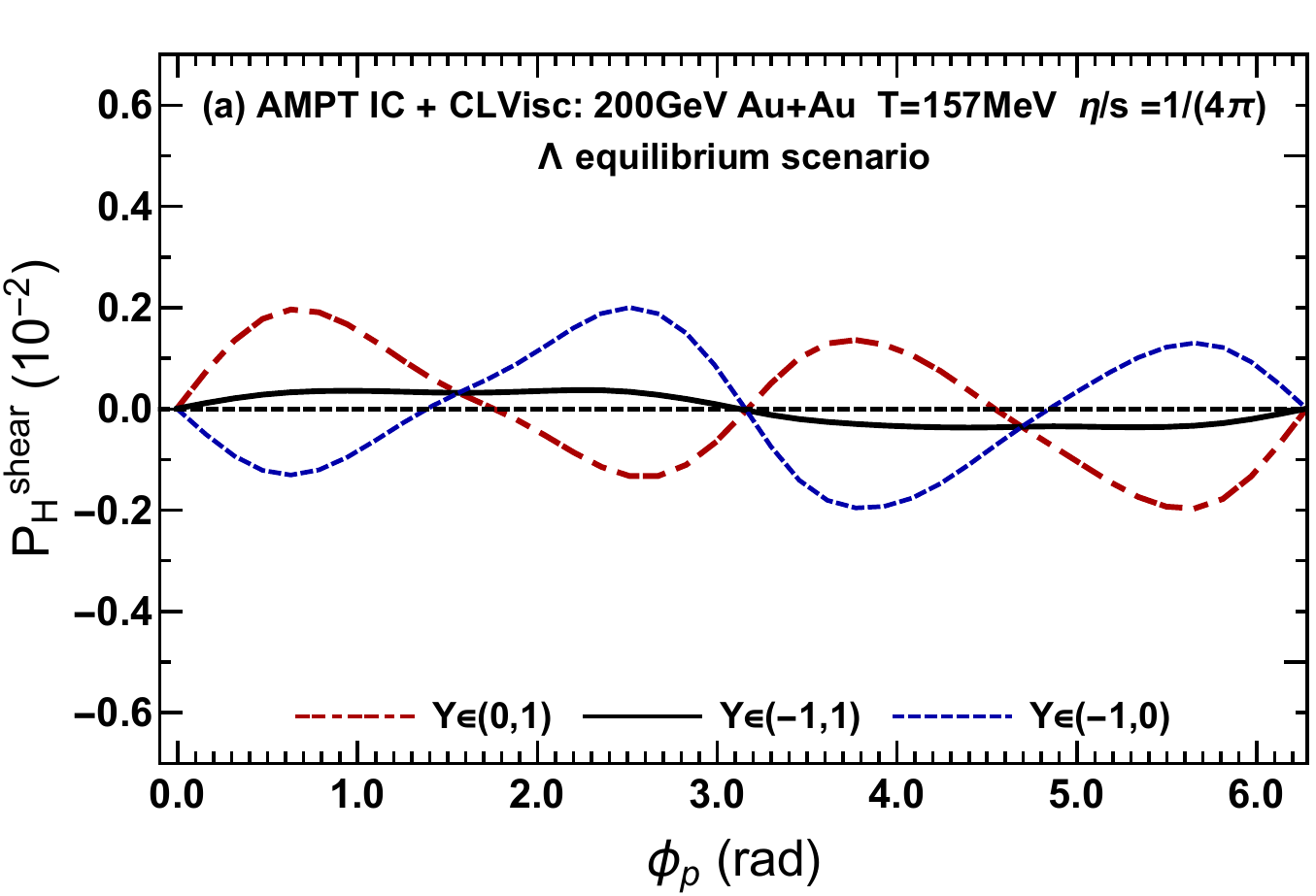}\includegraphics[scale=0.35]{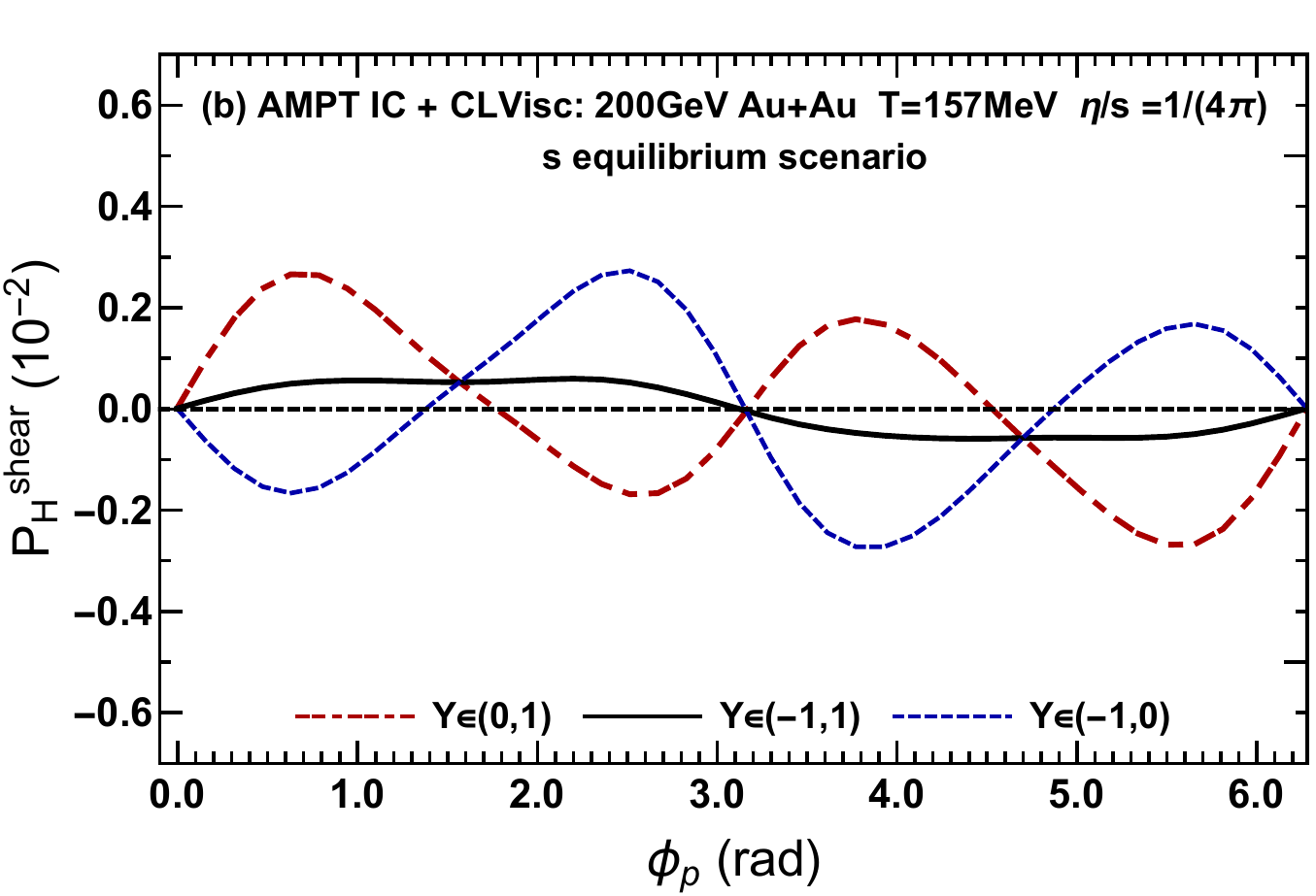}

\includegraphics[scale=0.35]{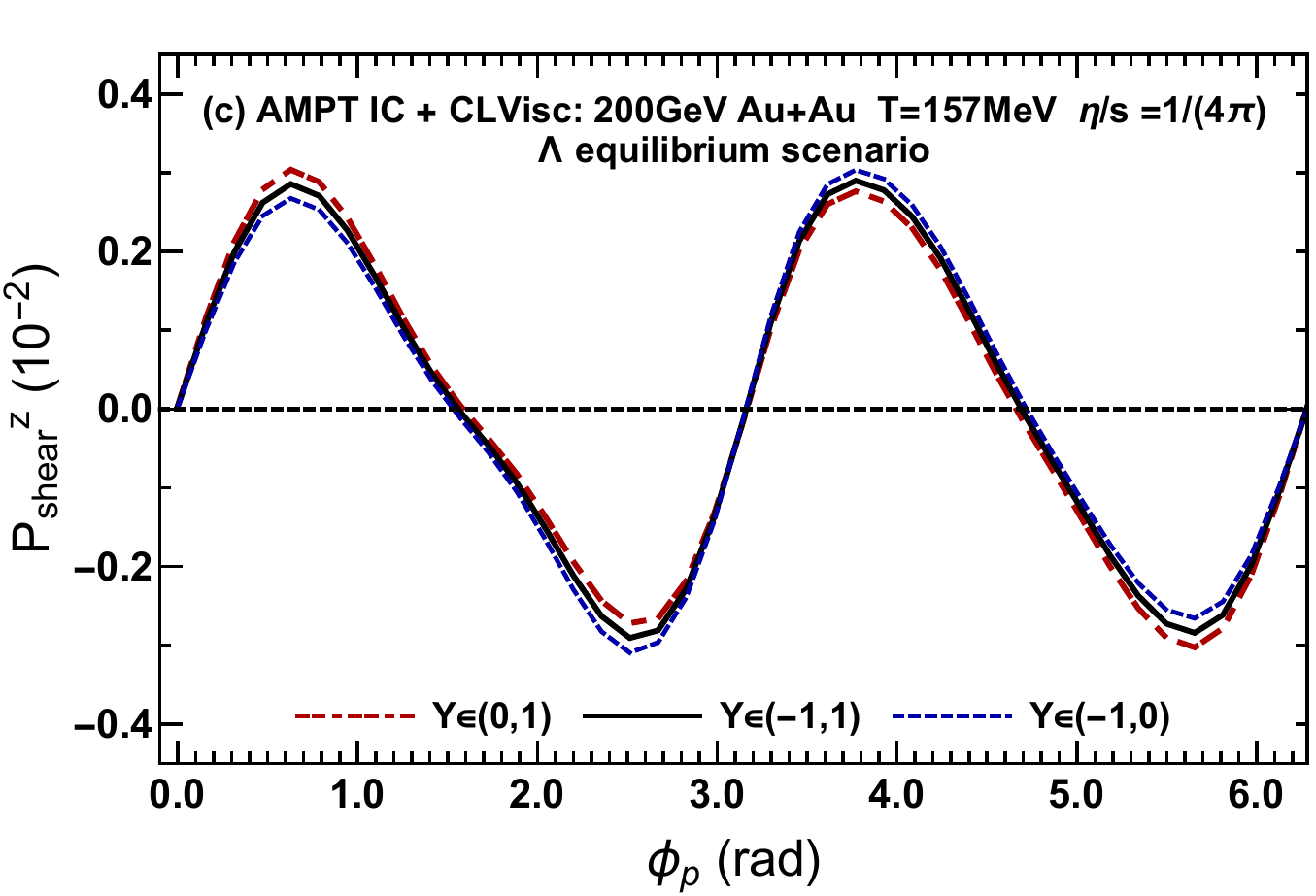}\includegraphics[scale=0.35]{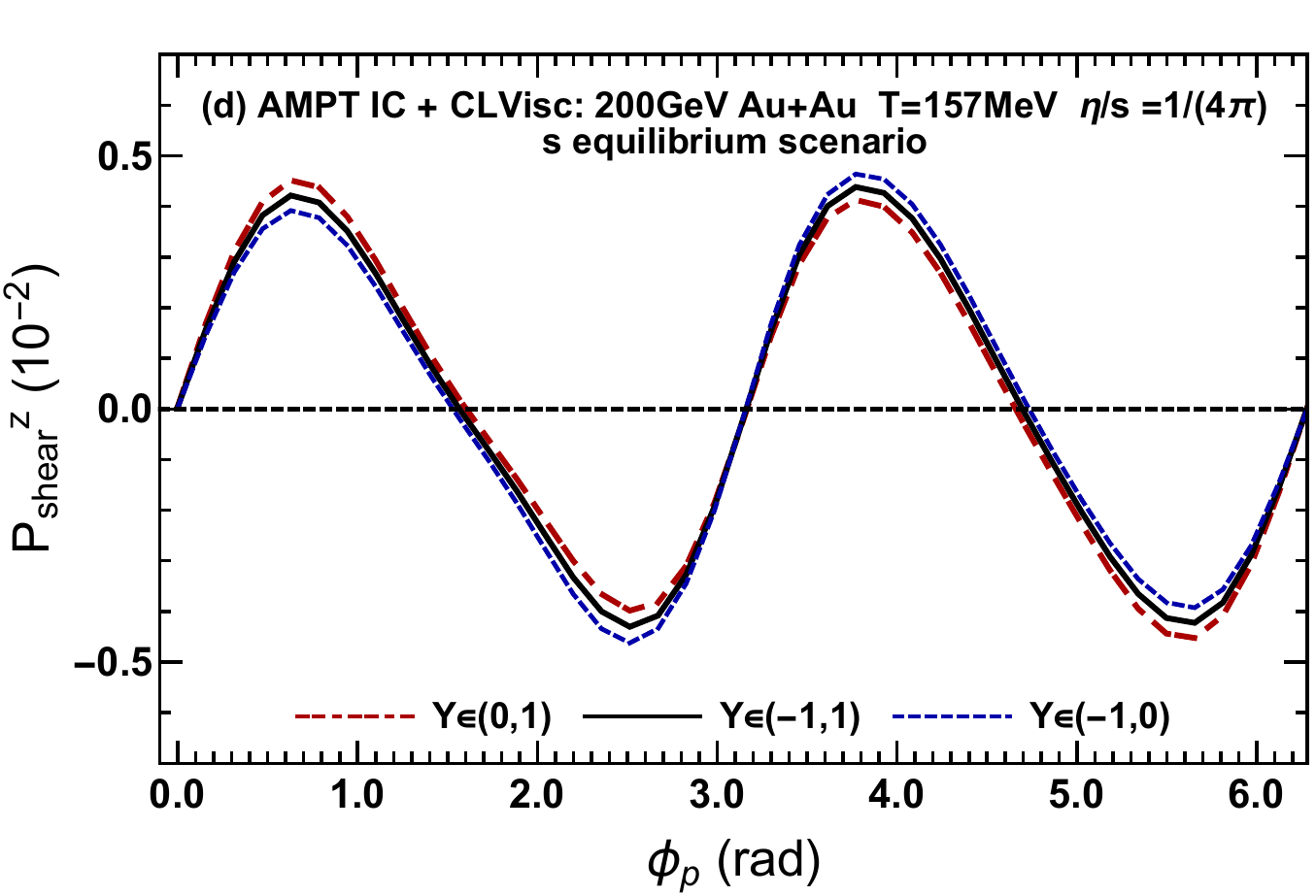}

\caption{The helicity polarization $P_{H}^{\text{shear}}$and local spin polarization
along the beam direction $P_{\text{shear}}^{z}$ induced by shear
viscous tensor as a function of $\phi_{p}$ for $\Lambda$ and $s$
\textit{\emph{equilibrium scenario}}s\emph{. }We use the same setup
and color assignments as those in Fig. \ref{fig:th_PH_phi}. \label{fig:shear_PH_phi}}
\end{figure}

\begin{figure}
\includegraphics[scale=0.35]{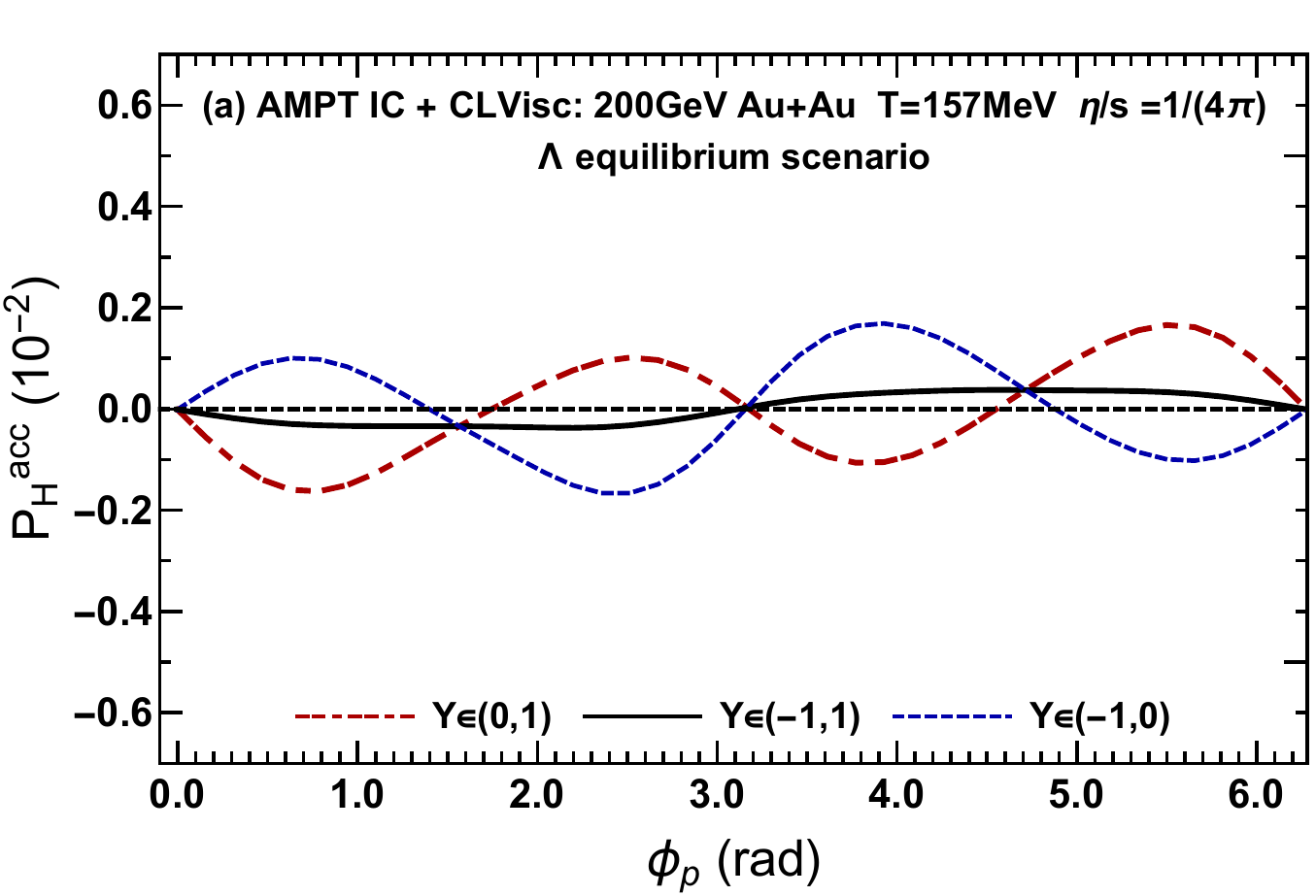}\includegraphics[scale=0.35]{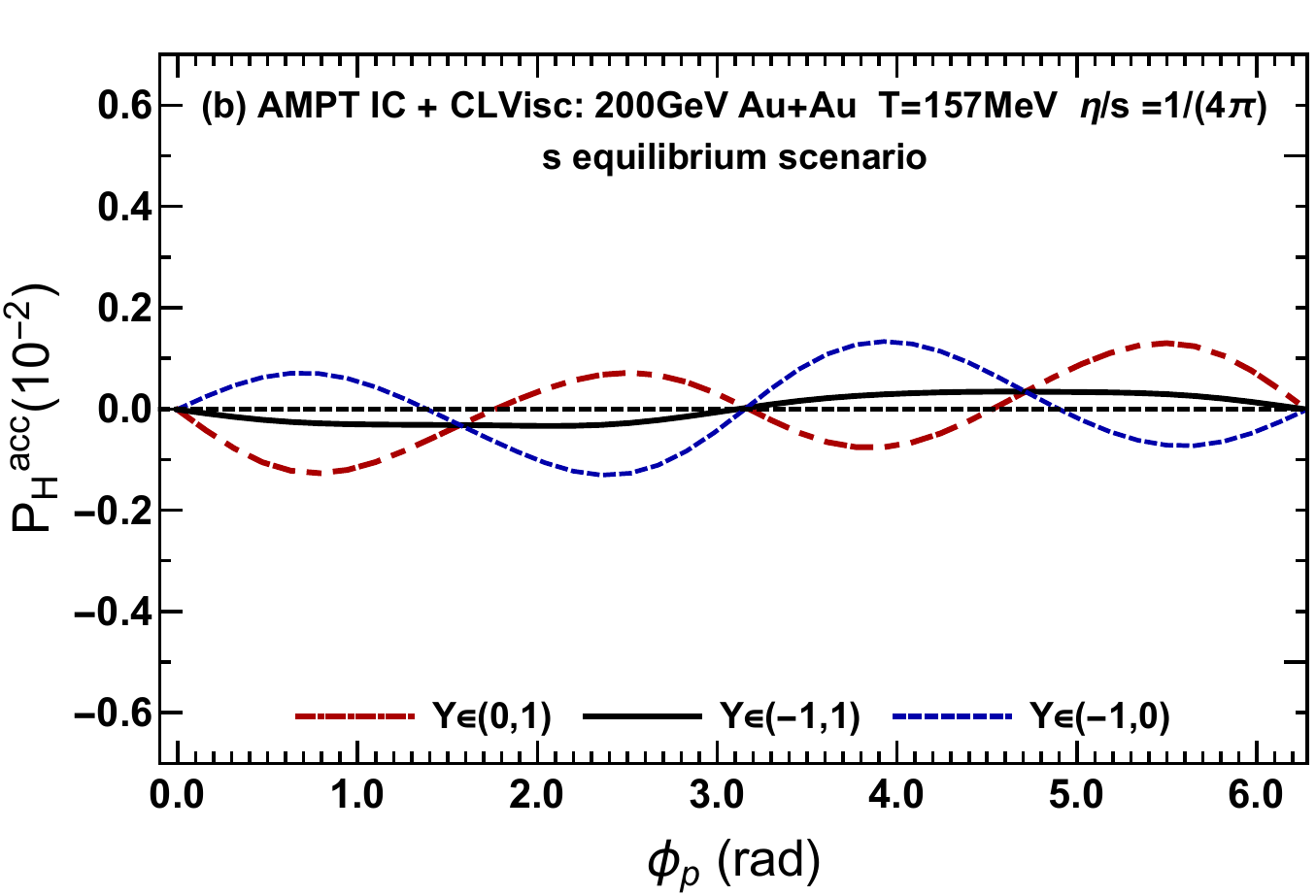}

\includegraphics[scale=0.35]{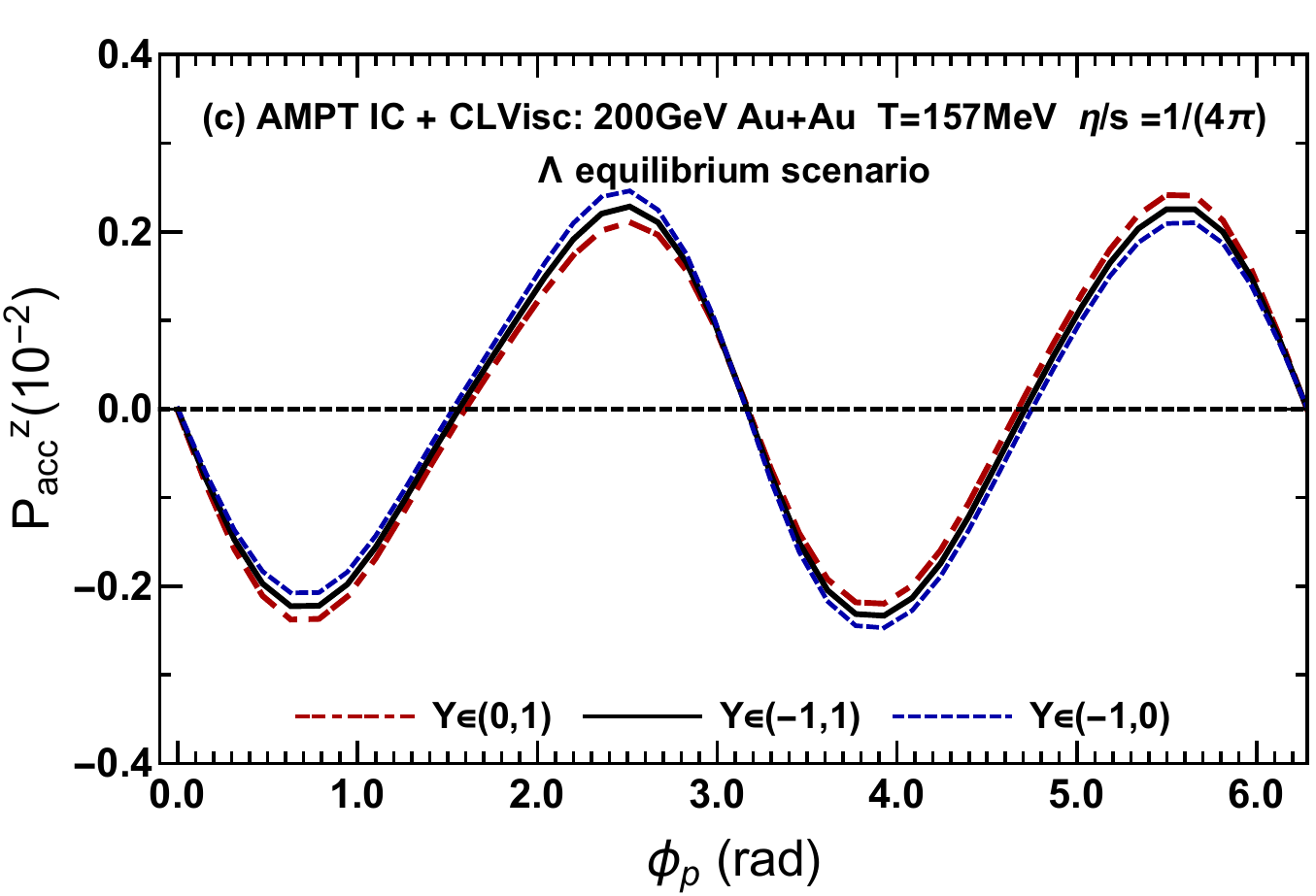}\includegraphics[scale=0.35]{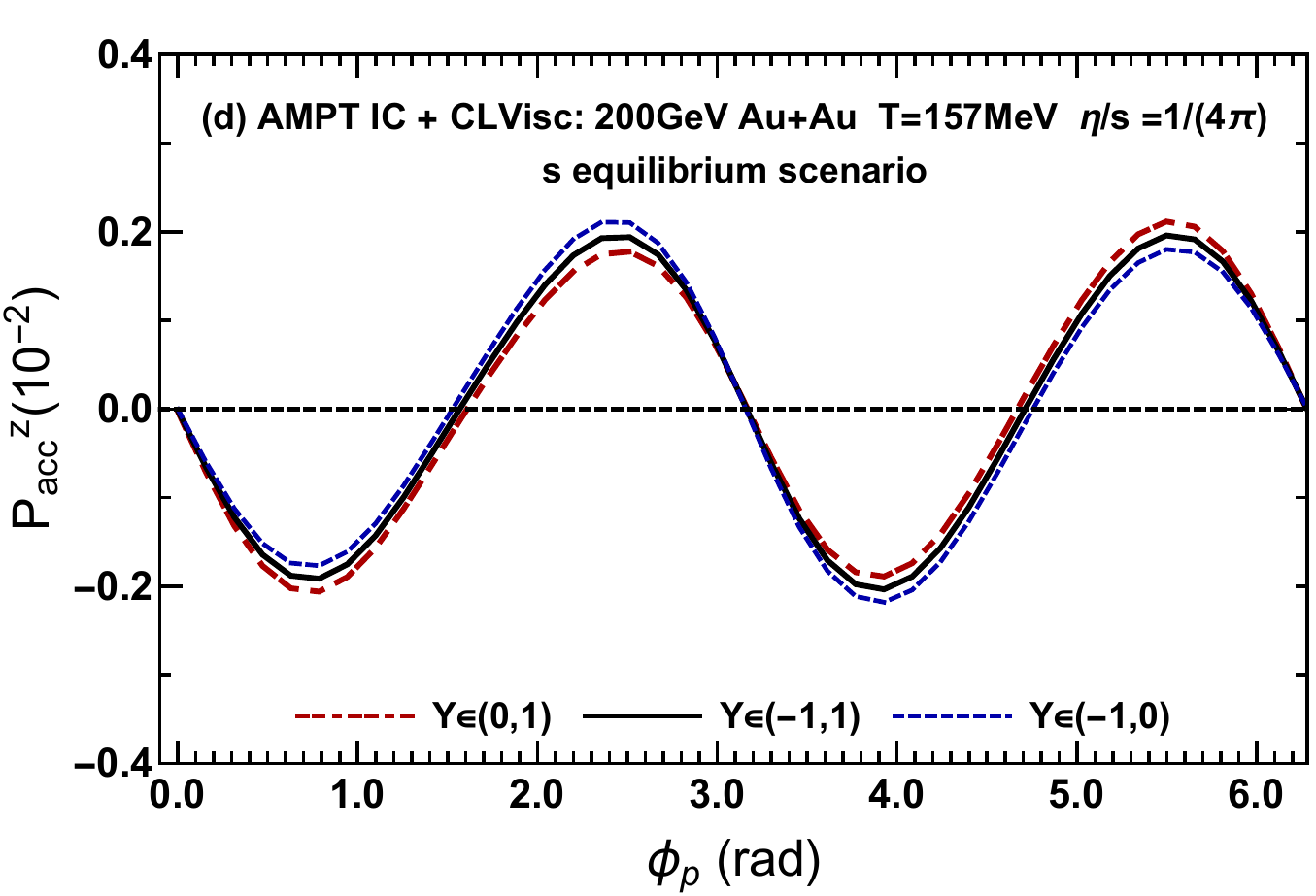}

\caption{The helicity polarization $P_{H}^{\text{acc}}$ and local spin polarization
along the beam direction $P_{\text{acc}}^{z}$ induced by fluid acceleration
as a function of $\phi_{p}$ for $\Lambda$ and $s$ \textit{\emph{equilibrium
scenario}}s\emph{. }We use the same setup and color assignments as
those in Fig. \ref{fig:th_PH_phi}. \label{fig:acc_PH_phi}}
\end{figure}

\begin{figure}
\includegraphics[scale=0.35]{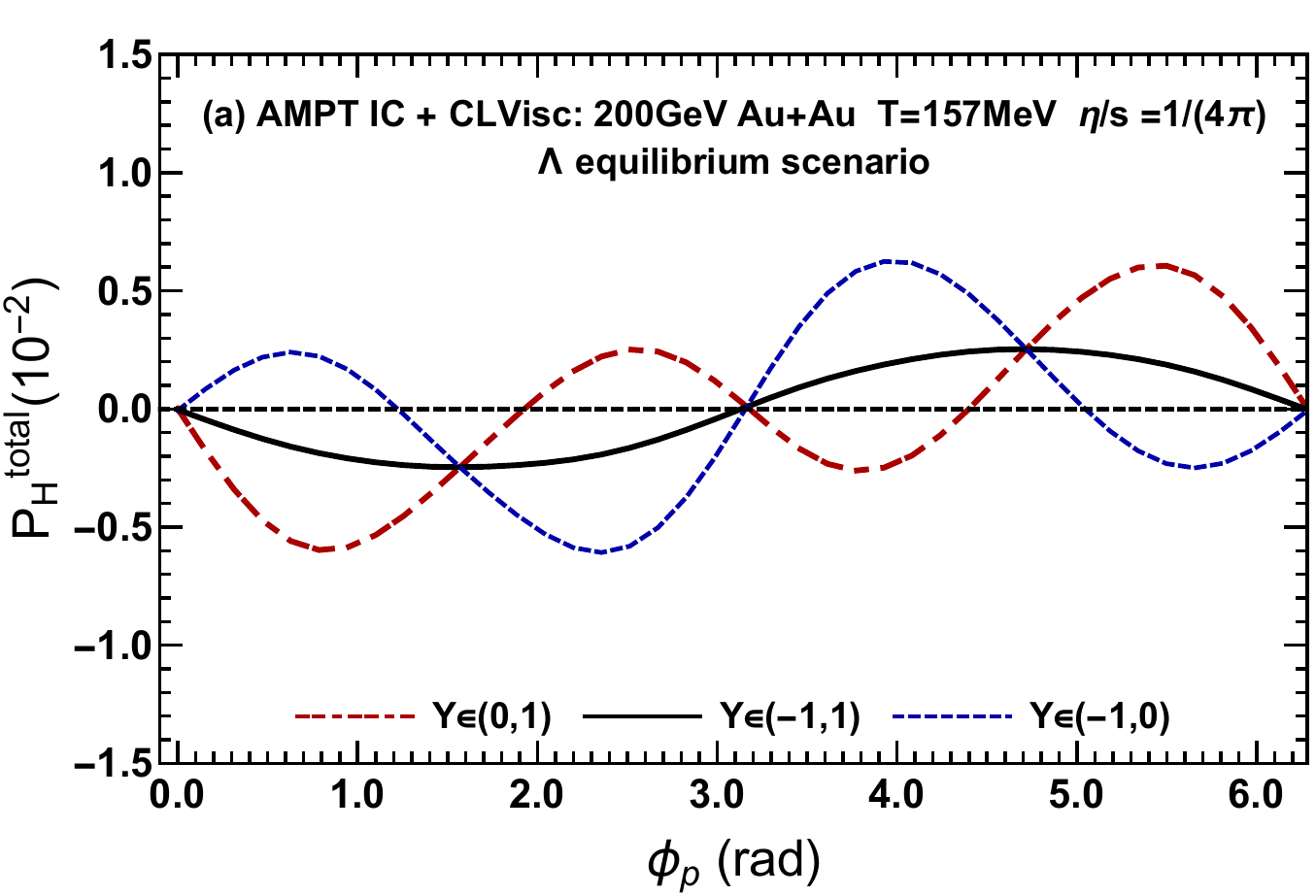}\includegraphics[scale=0.35]{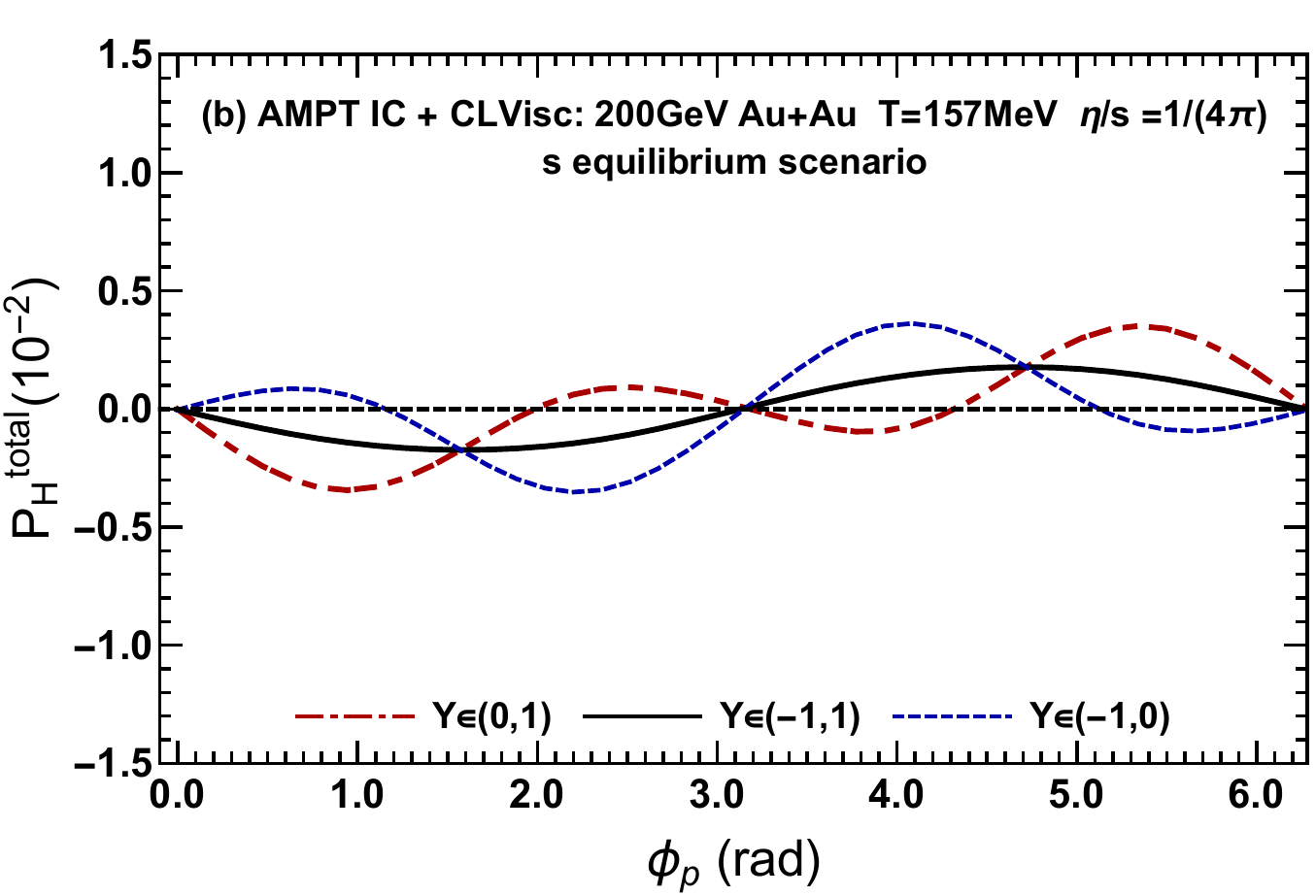}

\includegraphics[scale=0.35]{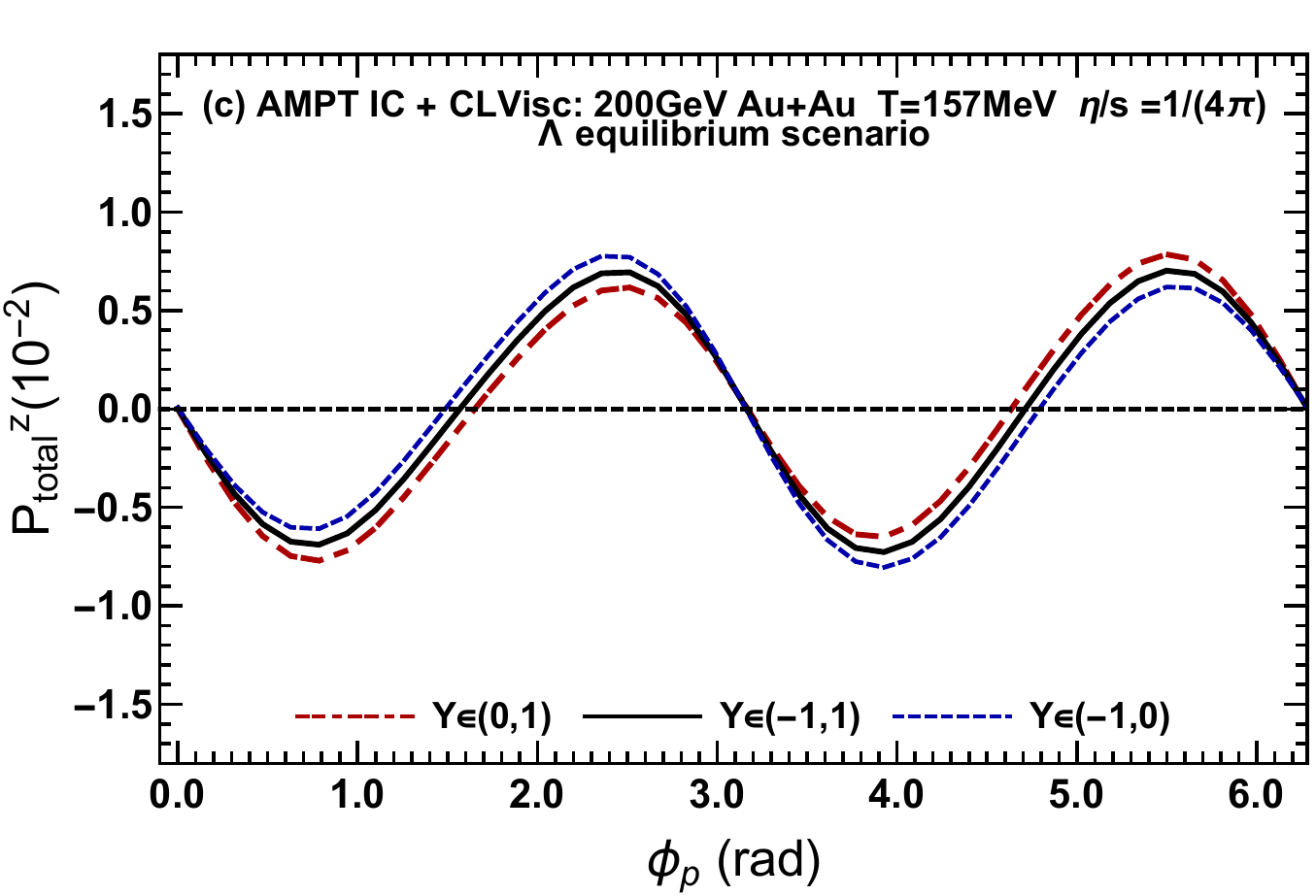}\includegraphics[scale=0.35]{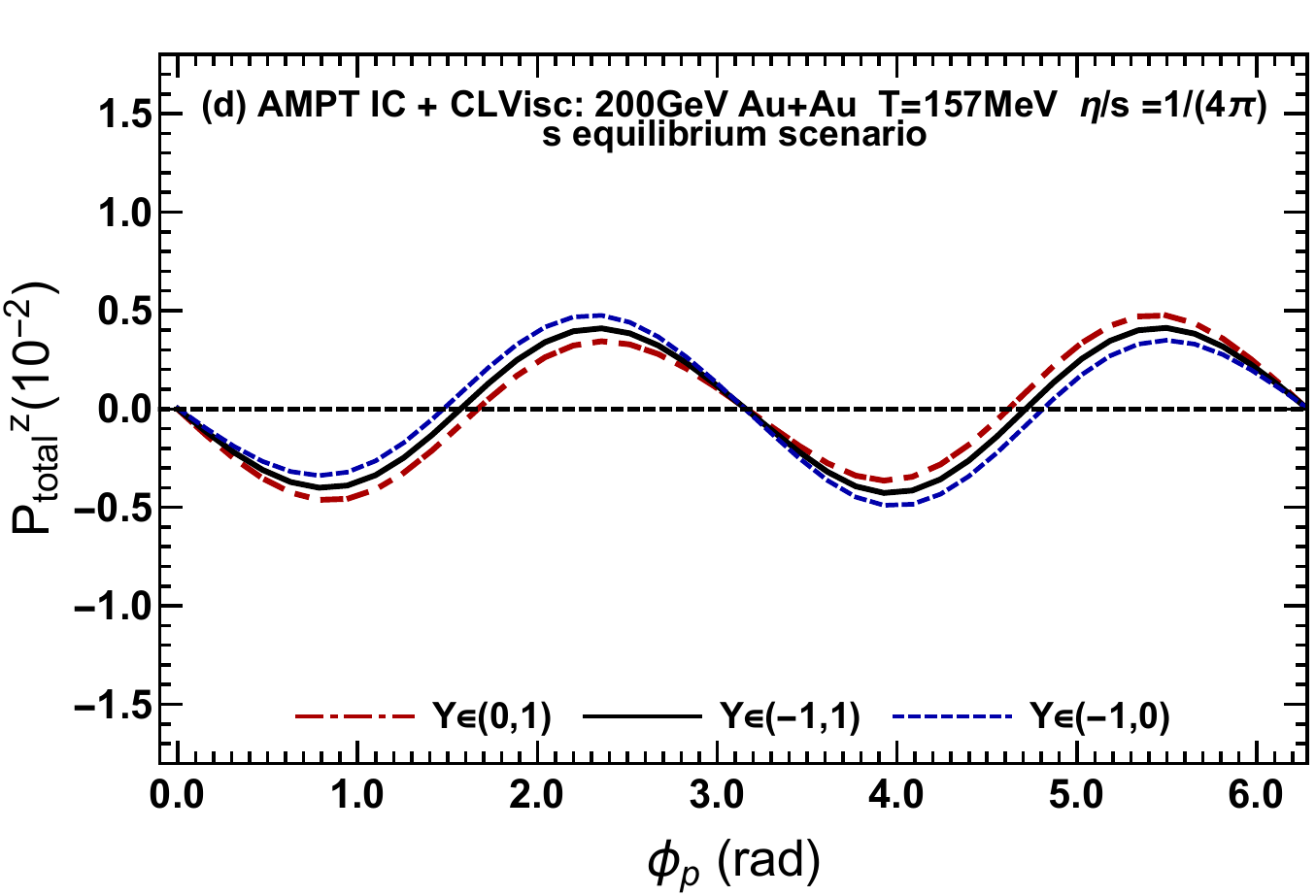}

\caption{The total helicity polarization $P_{H}^{\text{total}}$ and local
spin polarization along the beam direction $P_{\text{total}}^{z}$
as a function of $\phi_{p}$ for $\Lambda$ and $s$ \textit{\emph{equilibrium
scenario}}s\emph{. }We use the same setup and color assignments as
those in Fig. \ref{fig:th_PH_phi}. \label{fig:total_PH_phi}}
\end{figure}

\begin{figure}
\includegraphics[scale=0.35]{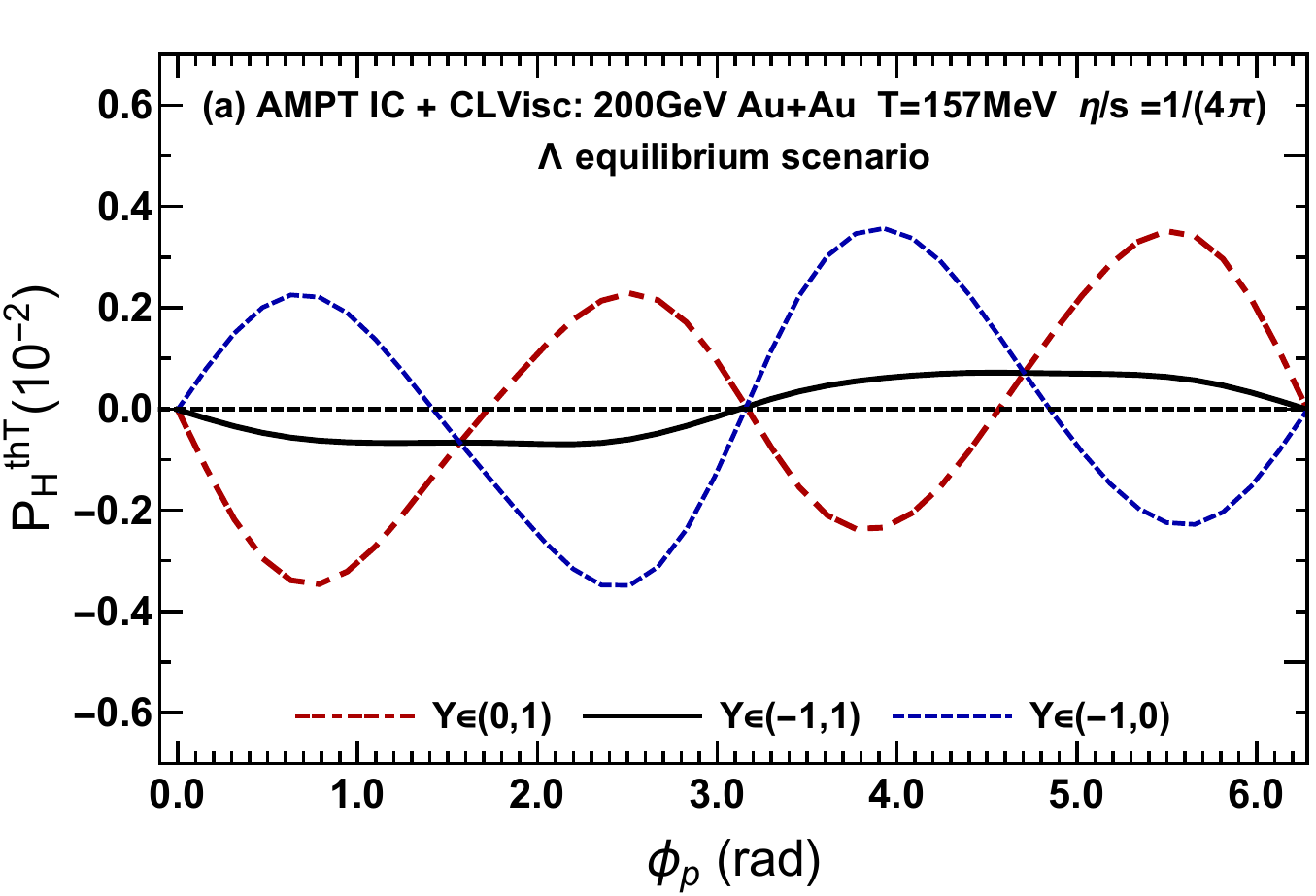}\includegraphics[scale=0.35]{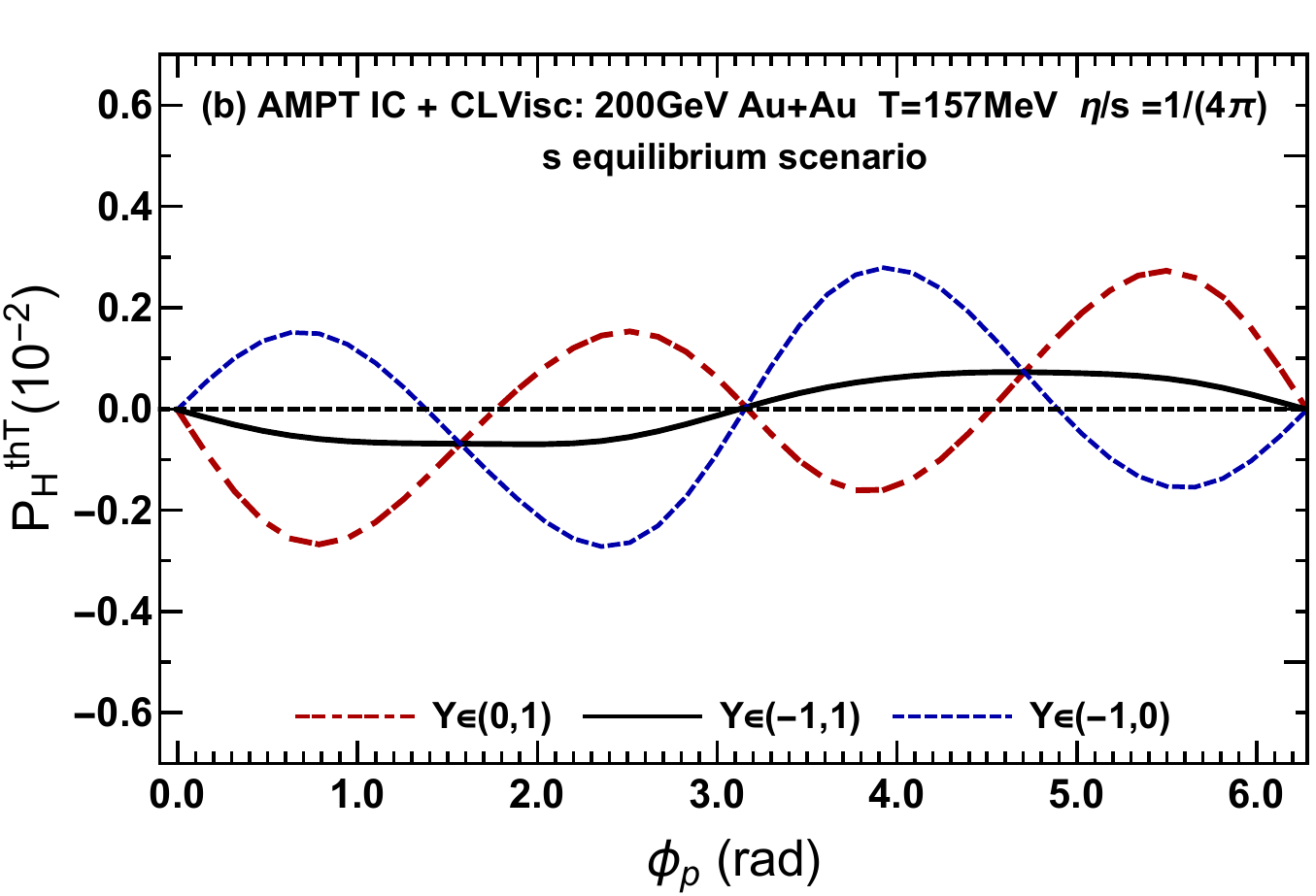}

\includegraphics[scale=0.35]{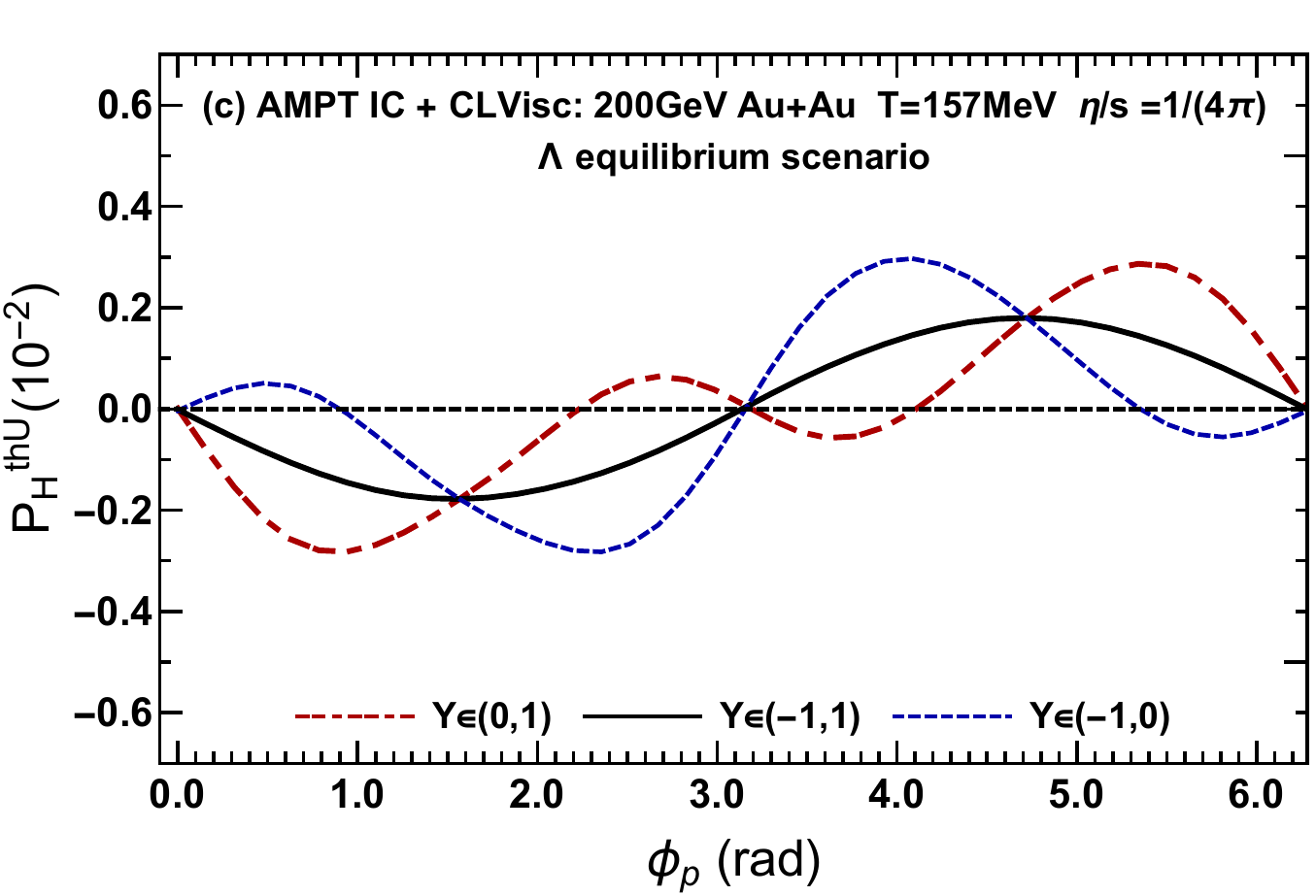}\includegraphics[scale=0.35]{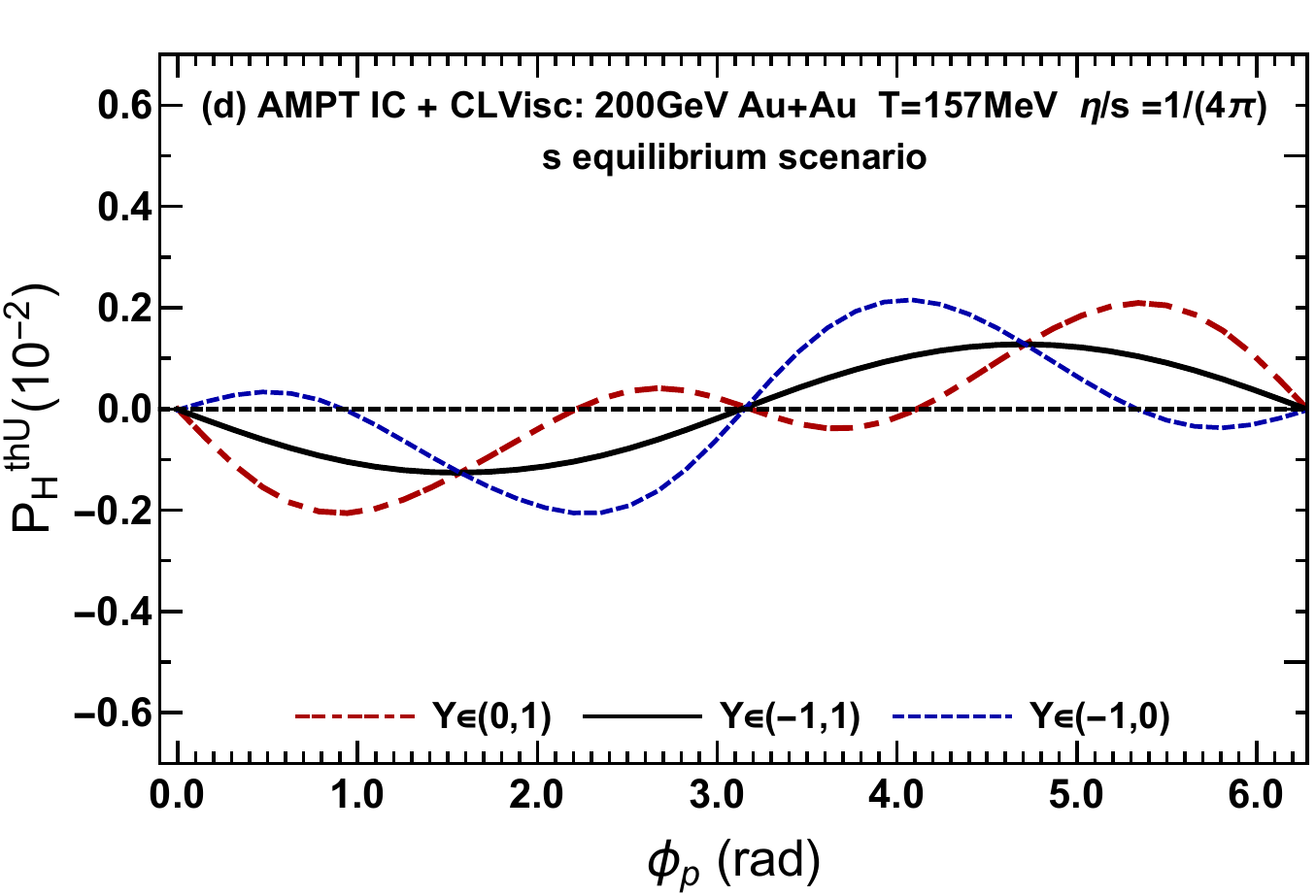}

\caption{The helicity polarization $P_{H}^{\text{thT}}$(up) and $P_{H}^{\text{thU}}$(down)
separated by $P_{H}^{\text{thermal}}$ , as a function of $\phi_{p}$
for $\Lambda$ and $s$ equilibrium scenarios. We use the same setup and color assignments as
those in Fig. \ref{fig:th_PH_phi}. \label{fig:decompose-thermal}}
\end{figure}

We implement the open source (3+1) dimensional viscous hydrodynamic
package CLVisc \citep{Pang:2012he} with AMPT initial conditions \citep{Lin:2004en,Wu:2019eyi,Wu:2020yiz}
at $20\%-50\%$ centrality to study the helicity polarization in $\sqrt{s_{NN}}=200\text{ GeV}$
Au+Au collisions. Similar to our previous work in Ref. \citep{Yi:2021ryh},
we choose $\eta/s=1/(4\pi)$, the freeze-out temperature $T=157\text{ MeV}$
and \textit{``s95p-pce'' }\citep{Huovinen:2009yb} as equation of
state to generate the information of the freeze-out hyper-surface
$\Sigma^{\mu}$.

By using the Eq. (\ref{eq:def_Sh}), we define the azimuthal angle
$\phi_{p}$ dependent helicity polarization as, 
\begin{equation}
P_{H}  (\phi_{p})=  \frac{2}{\mathcal{M}(+\Delta Y, - \Delta Y)} \int_{-\Delta Y}^{+\Delta Y}dY \int_{p_{T}^{min}}^{p_{T}^{max}}dp_{T} p_{T} [N S^{h} (\mathbf{p})],\label{eq:PHdef}
\end{equation}
where $N$ is defined in Eq.~(\ref{eq:def_N}) and the normalization factor is given by,
\begin{eqnarray}
\mathcal{M}(Y_{max},Y_{min})=  \int_{Y_{min}}^{Y_{max}}dY\int_{p_{T}^{min}}^{p_{T}^{max}}dp_{T} p_{T} N,
\end{eqnarray}
Note that, one can also define the averaged helicity polarization as, 
$P_{H}  (\phi_{p})=   \frac{1}{\Delta Y}\int_{-\Delta Y}^{+\Delta Y}dY\frac{1}{\Delta p_{T}}\int_{p_{T}^{min}}^{p_{T}^{max}}dp_{T}\text{ }S^{h}(\mathbf{p})$. 
The definition (\ref{eq:PHdef}) used in the current work is close to the local and global polarization measured in the experiments. 
where the $p_{T}=\sqrt{p_{x}^{2}+p_{y}^{2}}$ is the transverse momentum
and $\Delta p_{T}\equiv p_{T}^{max}-p_{T}^{min}$. We have chosen
$p_{T}^{max}=3\text{ GeV}$ and $p_{T}^{min}=0\text{ GeV}$ in our
calculation and concentrate on the mid-rapidity $[-1,+1]$, i.e. $\Delta Y=1$.
In order to test the theoretical result in Eq. (\ref{eq:YSz}), we
also introduce the $P_{H}^{+}(\phi_{p})$ and $P_{H}^{-}(\phi_{p})$,
\begin{eqnarray}
P_{H}^{+} & (\phi_{p})= & \frac{2}{\mathcal{M}(+\Delta Y, 0)} \int_{0}^{+\Delta Y}dY \int_{p_{T}^{min}}^{p_{T}^{max}}dp_{T}p_T [N S^{h}(\mathbf{p}) ],\nonumber \\
P_{H}^{-} & (\phi_{p})= & \frac{2}{\mathcal{M}(0,-\Delta Y)}\int_{-\Delta Y}^{0}dY\int_{p_{T}^{min}}^{p_{T}^{max}}dp_{T} p_T [ N S^{h}(\mathbf{p})].\label{eq:PHpl_def}
\end{eqnarray}

According to Eq. (\ref{eq:helcity_decomp_01}), we can also decompose
$P_{H}$, $P_{H}^{+}$, $P_{H}^{-}$ into 5 terms. Since the electromagnetic
field decays rapidly in the relativistic heavy ion collision \citep{Deng:2012pc,Roy:2015coa},
we ignore the helicity polarization contributed by EB term. Since
the gradient of chemical potential $\mu$ is negligible for the $\sqrt{s_{NN}}=200\text{ GeV}$
Au+Au high-energy collisions and the information of chemical potential
lacks in EoS\textit{ \textquotedbl s95pce\textquotedbl}, we also
neglect the helicity polarization related to chemical potential. Therefore,
we only consider the following parts,

\begin{eqnarray}
P_{H}^{\textrm{total}} & = & P_{H}^{\text{thermal}}+P_{H}^{\text{shear}}+P_{H}^{\text{accT}},\nonumber \\
P_{H}^{+\textrm{total}} & = & P_{H}^{+\text{thermal}}+P_{H}^{+\text{shear}}+P_{H}^{+\text{accT}},\nonumber \\
P_{H}^{-\textrm{total}} & = & P_{H}^{-\text{thermal}}+P_{H}^{-\text{shear}}+P_{H}^{-\text{accT}},\label{eq:AllEffect}
\end{eqnarray}
where the upper indies stand for the helicity polarization contributed
by the thermal vorticity, shear viscous tensor and fluid acceleration,
respectively.

Similarly, we also introduce the local spin polarization at different
momentum rapidity range, 
\begin{eqnarray}
P^{i} & (\phi_{p})= & \frac{2}{\mathcal{M}(+\Delta Y, -\Delta Y)}\int_{-\Delta Y}^{+\Delta Y}dY \int_{p_{T}^{min}}^{p_{T}^{max}}dp_{T} p_T [ N \text{ }\mathcal{S}^{i}(\mathbf{p})],\nonumber \\
P^{i+} & (\phi_{p})= & \frac{2}{\mathcal{M}(+\Delta Y, 0)}\int_{0}^{+\Delta Y}dY\int_{p_{T}^{min}}^{p_{T}^{max}}dp_{T} p_T [ N\text{ }\mathcal{S}^{i}(\mathbf{p})],\nonumber \\
P^{i-} & (\phi_{p})= & \frac{2}{\mathcal{M}(0, -\Delta Y)}\int_{-\Delta Y}^{0}dY\int_{p_{T}^{min}}^{p_{T}^{max}}dp_{T} p_T [ N\text{ }\mathcal{S}^{i}(\mathbf{p})].\label{eq:Pz_defination}
\end{eqnarray}
where the $\mathcal{S}^{i}$ is defined in Eq. (\ref{eq:CooperFryeFormula})
and $i=x,y,z$ stands for the polarization along the in-plane, out-of-plane,
and beam directions, respectively. We further decompose the polarization
along the beam direction $P^{z}$, $P^{z+}$, $P^{z-}$ into $3$
term.

\begin{eqnarray}
P^{z} & = & P_{\text{thermal}}^{z}+P_{\text{shear}}^{z}+P_{\text{acc}}^{z},\nonumber \\
P^{z+} & = & P_{\text{thermal}}^{z+}+P_{\text{shear}}^{z+}+P_{\text{acc}}^{z+},\nonumber \\
P^{z-} & = & P_{\text{thermal}}^{z-}+P_{\text{shear}}^{z-}+P_{\text{acc}}^{z-}.
\end{eqnarray}

As proposed in Ref. \citep{Fu:2021pok} and also used in our previous
work \citep{Yi:2021ryh}, we consider two different scenarios, named\textit{
$\Lambda$ }and \textit{$s$ equilibrium scenario\modDY{s}. }In the \textit{$\Lambda$ }equilibrium scenario, we assume
that the $\Lambda$ hyperons are near the local equilibrium after
they are produced at chemical freezeout. We then use the information
in the freeze-out hypersurface to describe the thermodynamic state
of the $\Lambda$ hyperons. In the $s$ equilibrium scenario, according
to the parton models, we assume that the spin polarization of $\Lambda$
is mainly contributed by the $s$ quark, and the spin polarization
of the $s$ quark is close to the spin polarization of $\Lambda$
hyperons. In these two scenarios, the mass of particles are chosen
as $m=m_{\Lambda}=1.116\text{ GeV}$ and $m=m_{s}=0.3\text{ GeV}$
(constituent quark mass) and we use the same information
of the freezeout hyper-surface.

\subsection{Numerical result \label{subsec:Numerical-result}}

We present the numerical results for helicity polarization $P_{H}$
and the polarization along the beam direction $P^{z}$ in Eq. (\ref{eq:PHpl_def})
and Eq. (\ref{eq:Pz_defination}) as functions of azimuthal angle
$\phi_{p}$ and momentum rapidity $Y$ at $\sqrt{s_{NN}}=200\text{ GeV}$
Au-Au collisions in $20-50\%$ centrality. Note that both $P_{H}$ and $P^{z}$ are calculated in laboratory frame in our numerical simulations.

\begin{figure}
\includegraphics[scale=0.35]{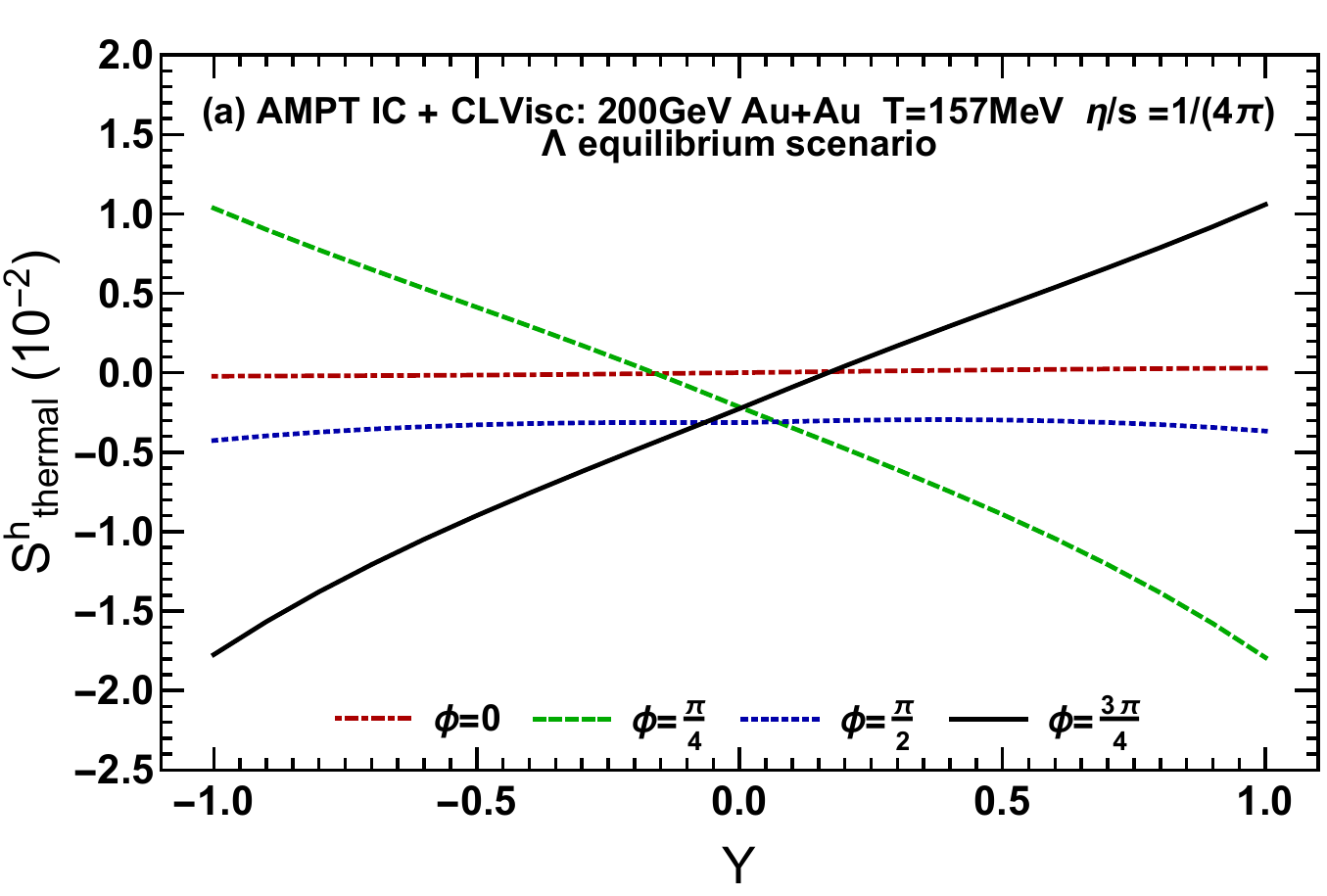}\includegraphics[scale=0.35]{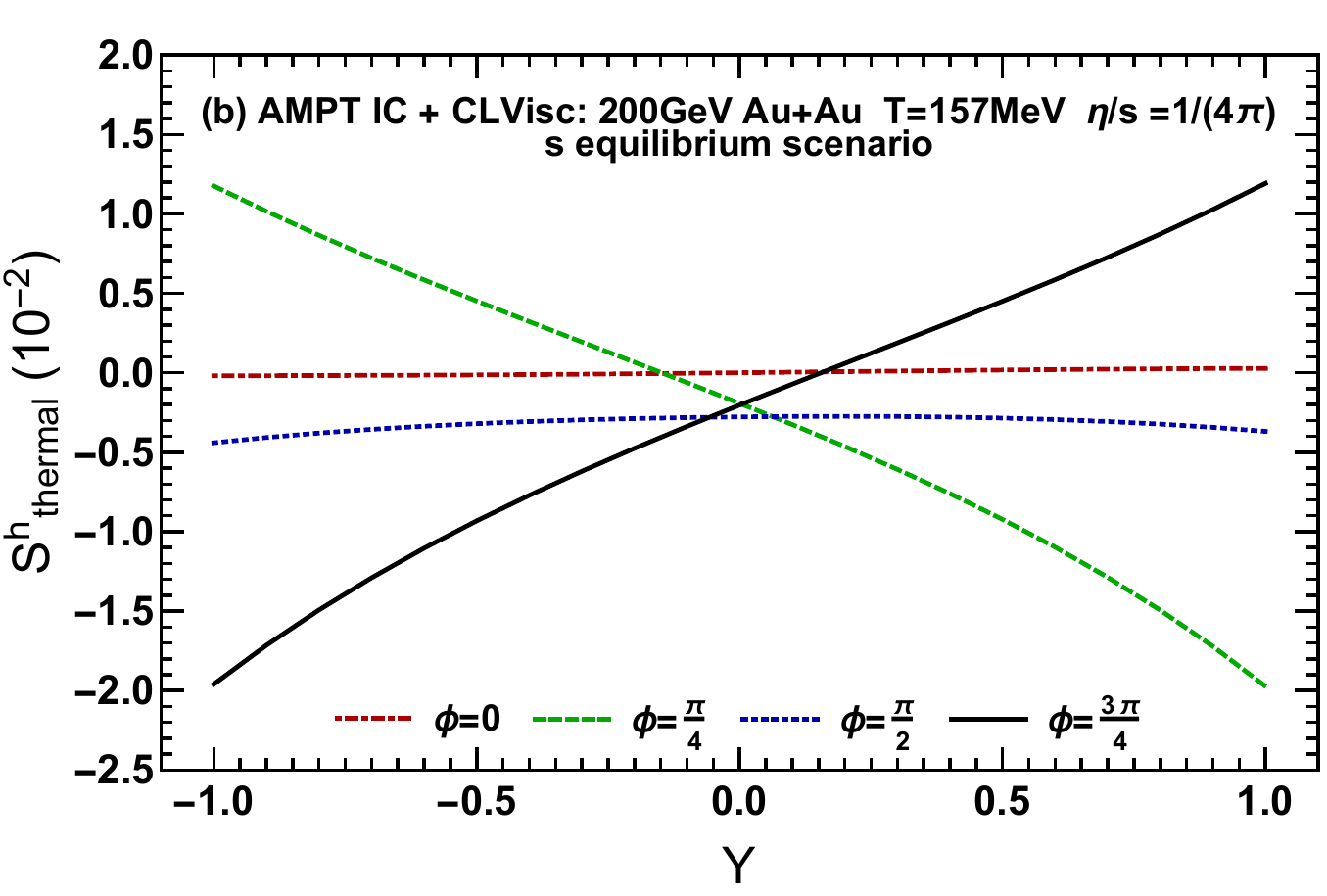}

\includegraphics[scale=0.35]{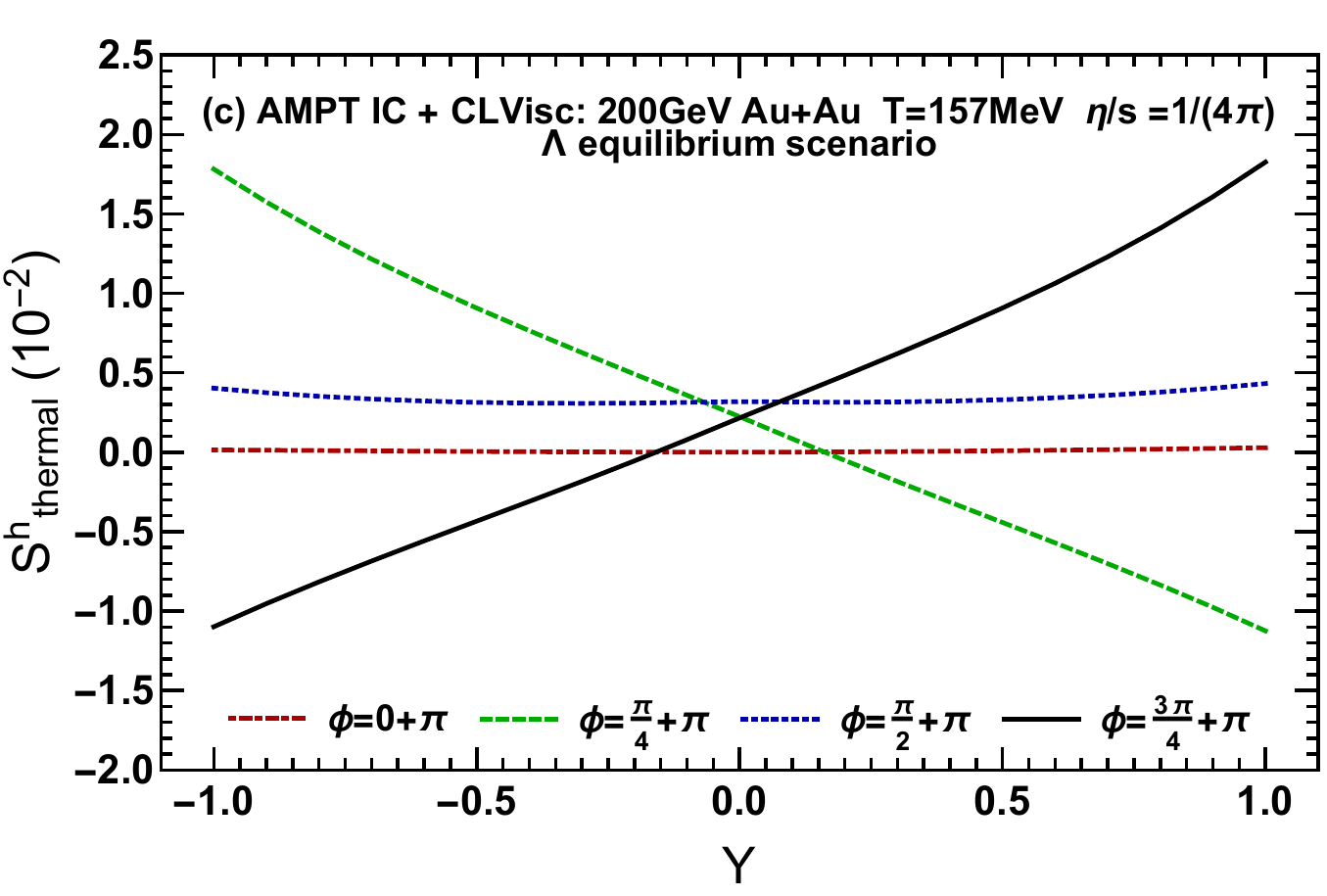}\includegraphics[scale=0.35]{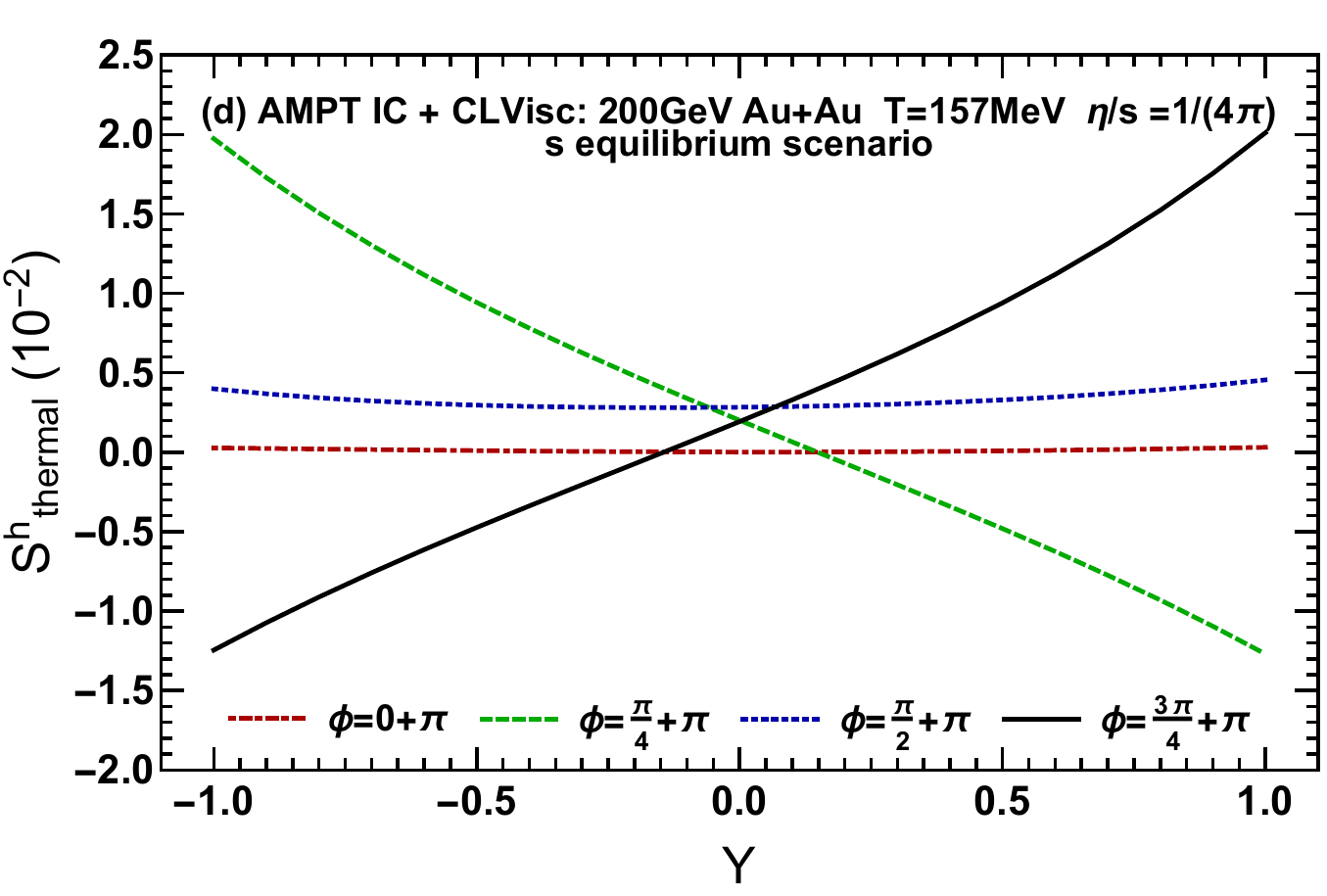}

\caption{The helicity polarization scalar $S_{\text{thermal}}^{h}$ induced by thermal vorticity as a function
of momentum rapidity $Y$ with different azimuthal angle $\phi_{p}$
in $\Lambda$ and $s$ equilibrium scenarios. We use the same setup
as those in Fig. \ref{fig:th_PH_phi}. Colors stand for the different
angle $\phi_{p}$. \label{fig:th_Sh_Y}}
\end{figure}

\begin{figure}
\includegraphics[scale=0.35]{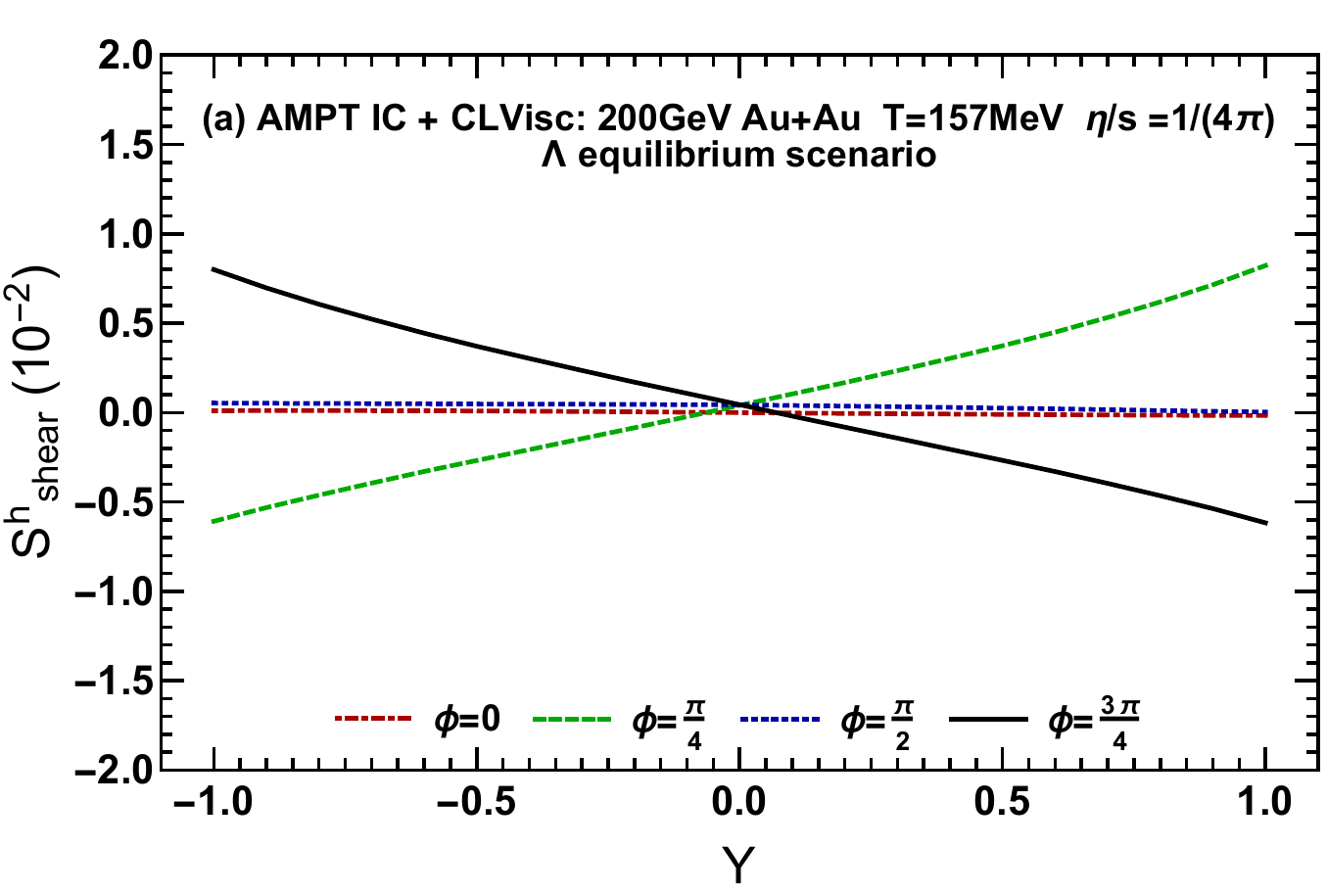}\includegraphics[scale=0.35]{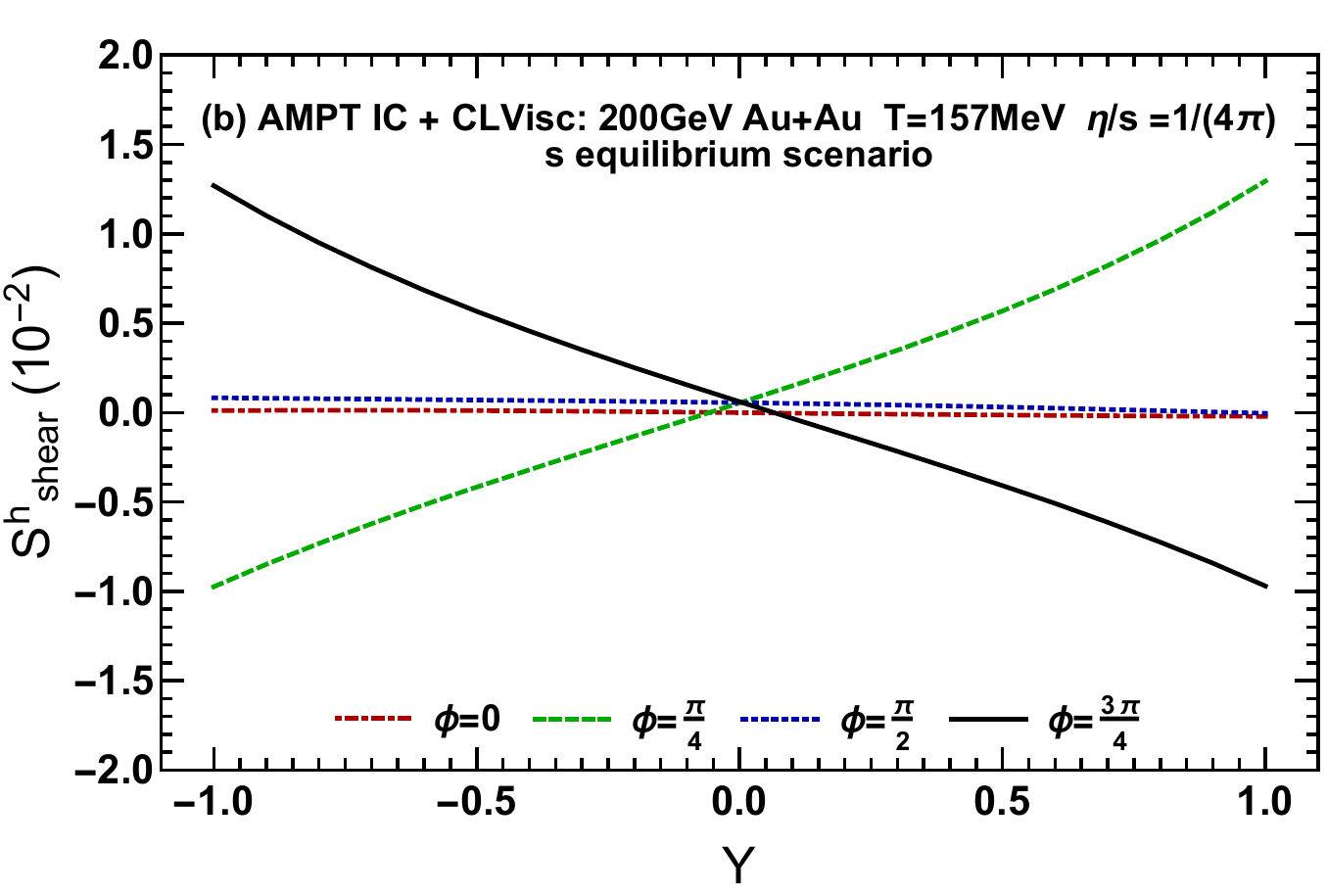}

\includegraphics[scale=0.35]{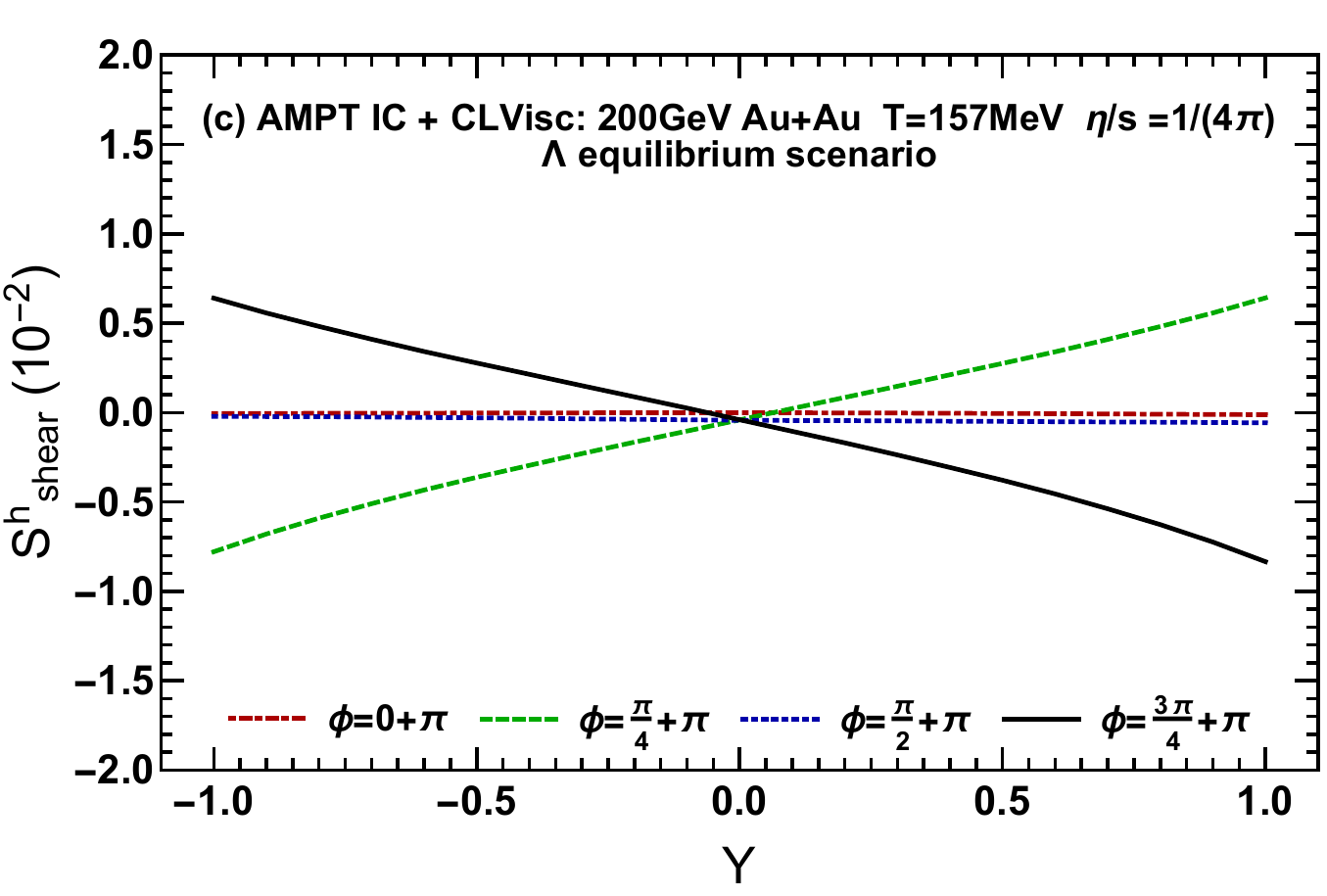}\includegraphics[scale=0.35]{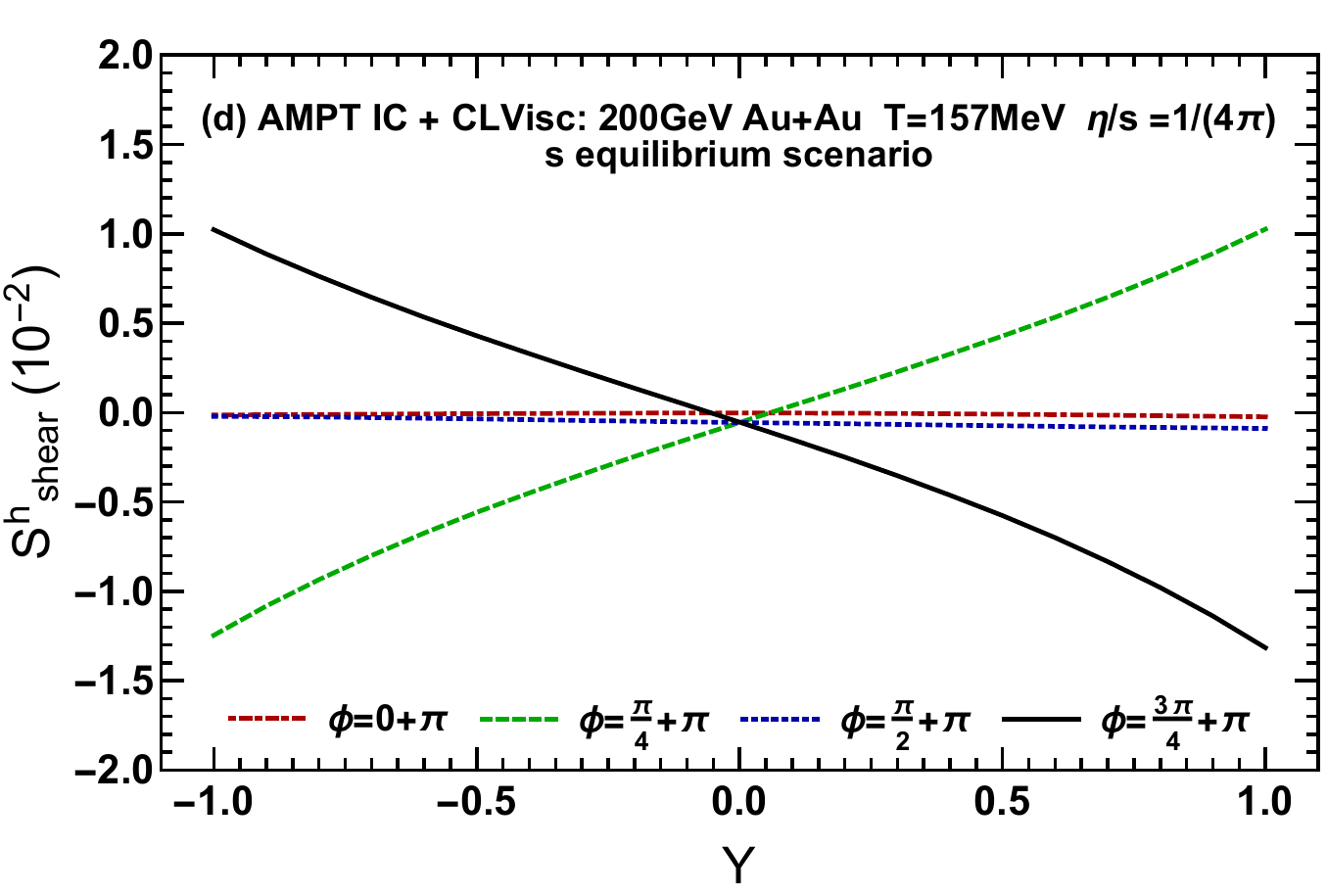}

\caption{The helicity polarization scalar $S_{\text{shear}}^{h}$ induced by
shear viscous tensor as a function of momentum rapidity $Y$ for $\Lambda$
and $s$ equilibrium scenarios. We use the same setup as those in
Fig. \ref{fig:th_PH_phi}. Colors stand for the different angle $\phi_{p}$.
\label{fig:shear_Sh_Y}}
\end{figure}

\begin{figure}
\includegraphics[scale=0.35]{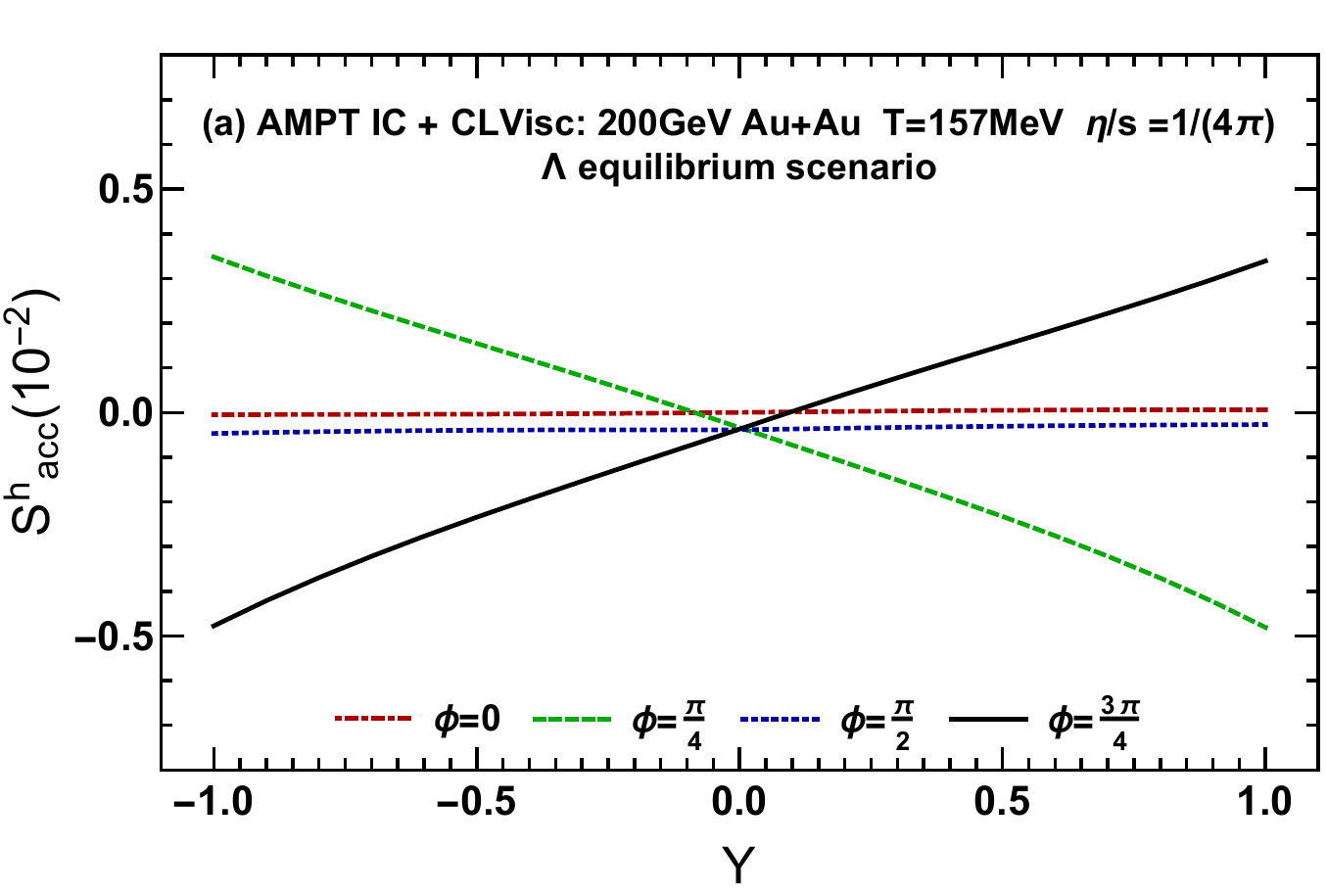}\includegraphics[scale=0.35]{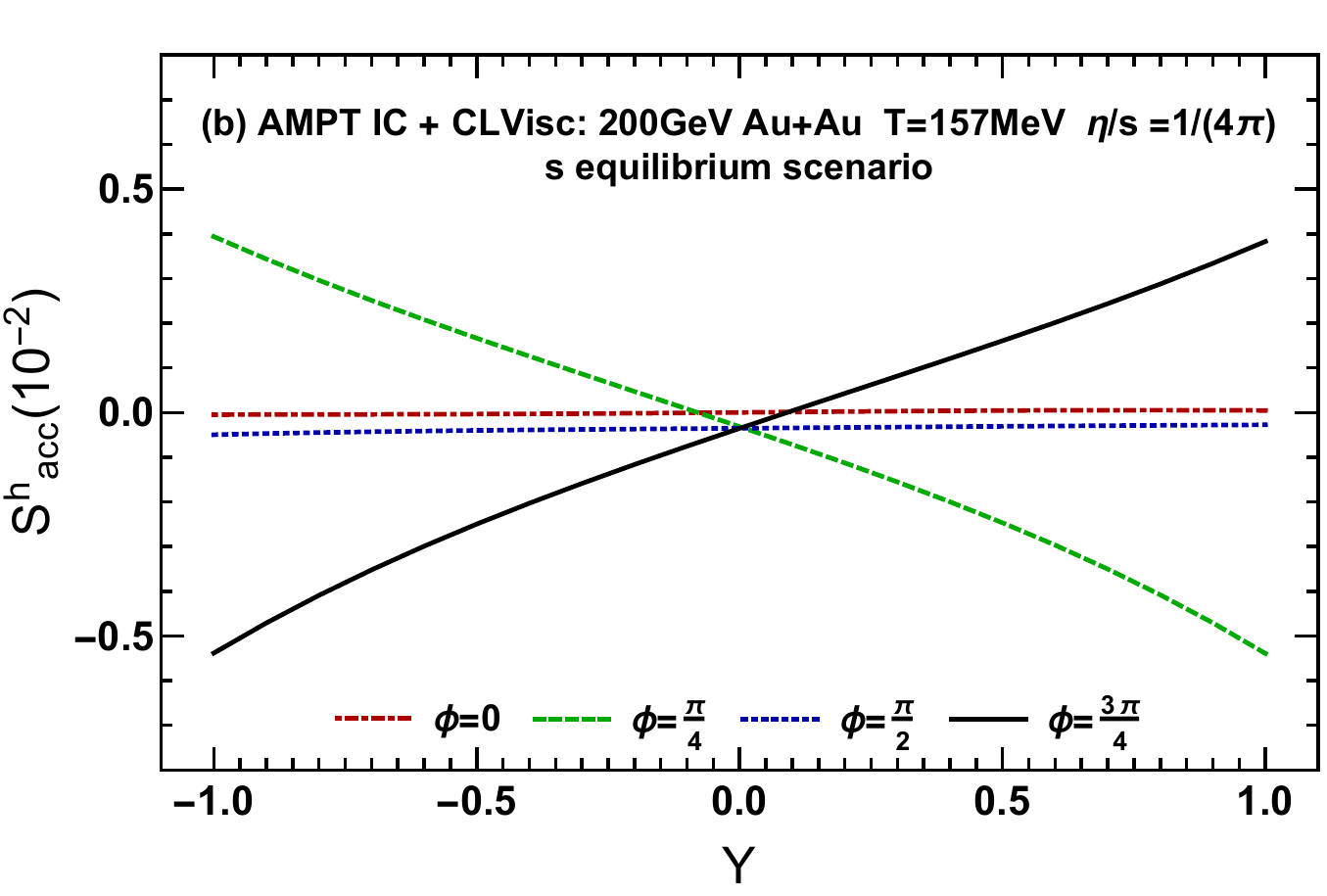}

\includegraphics[scale=0.35]{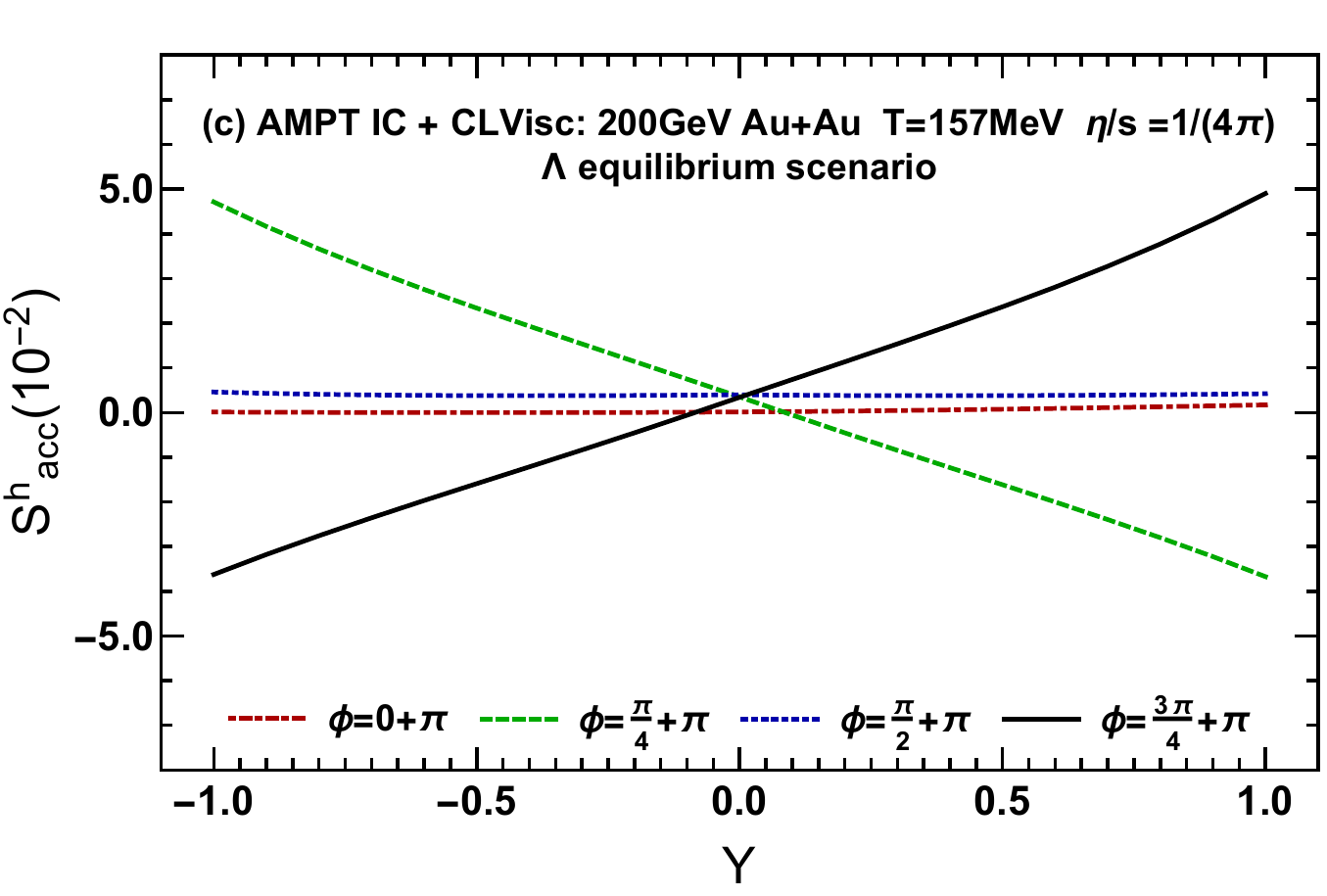}\includegraphics[scale=0.35]{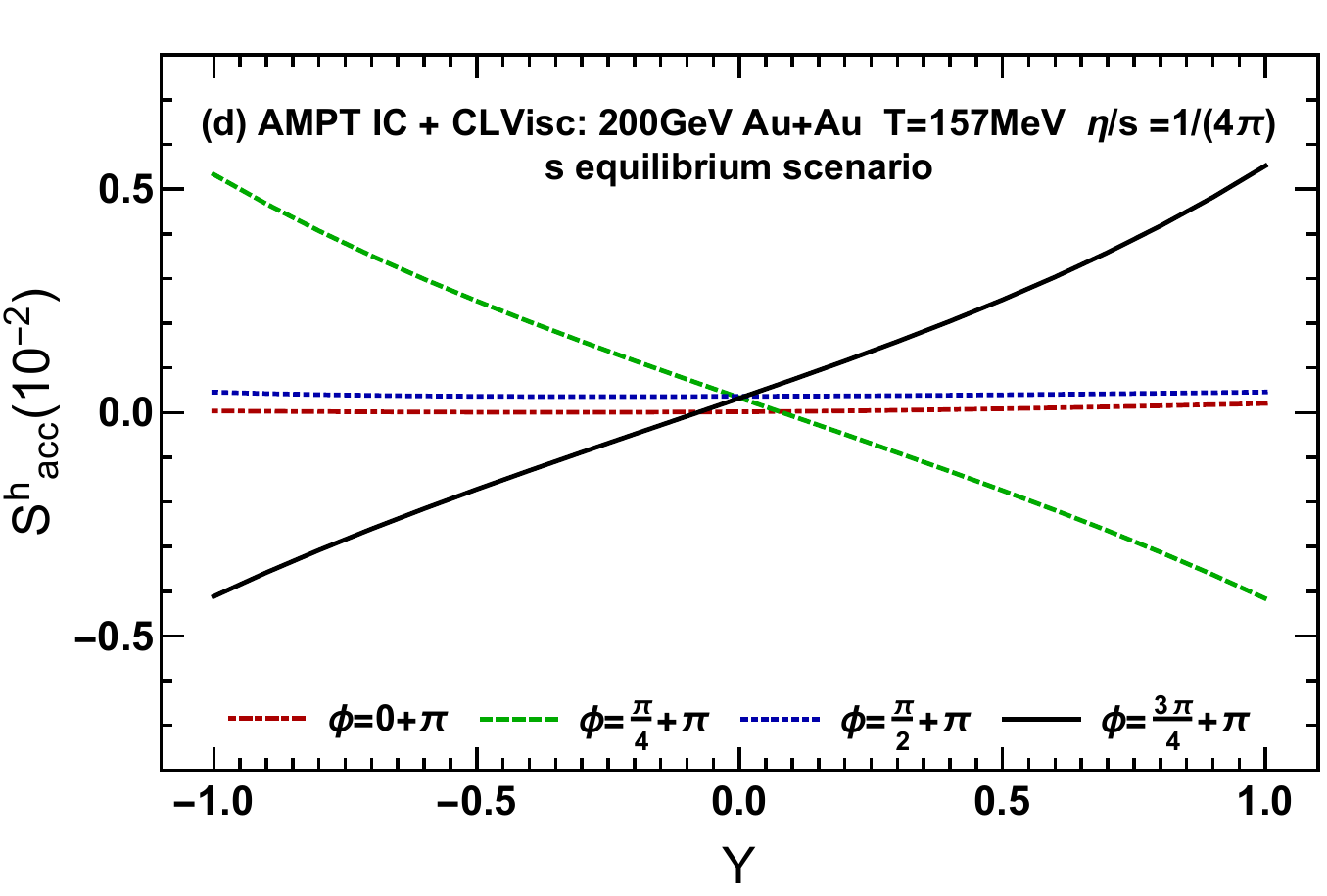}

\caption{The helicity polarization scalar $S_{\text{acc}}^{h}$ induced by
the fluid acceleration as a function of momentum rapidity $Y$ for
$\Lambda$ and $s$ equilibrium scenarios. We use the same setup as
those in Fig. \ref{fig:th_PH_phi}. Colors stand for the different
angle $\phi_{p}$. \label{fig:acc_Sh_Y}}
\end{figure}

\begin{figure}
\includegraphics[scale=0.35]{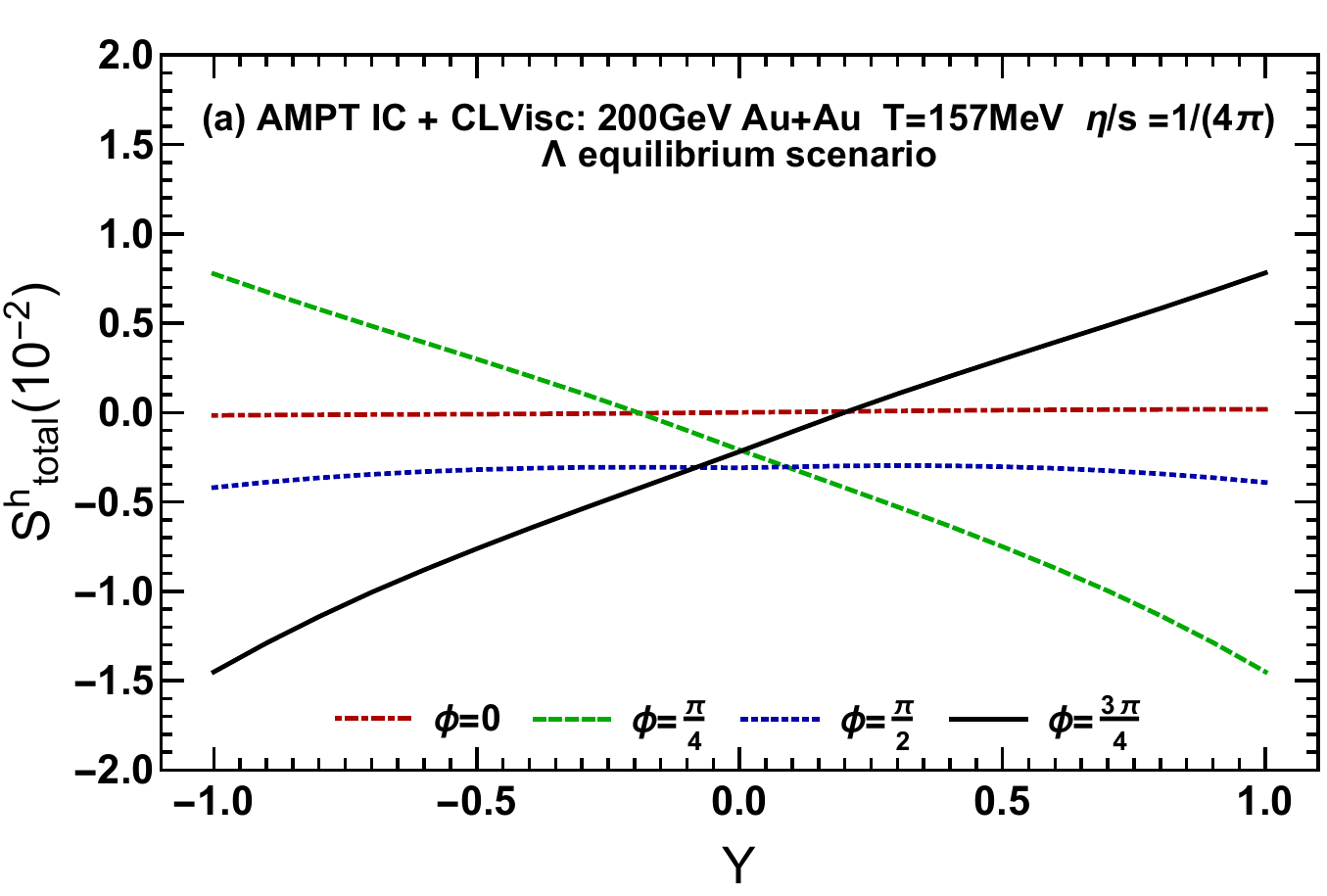}\includegraphics[scale=0.35]{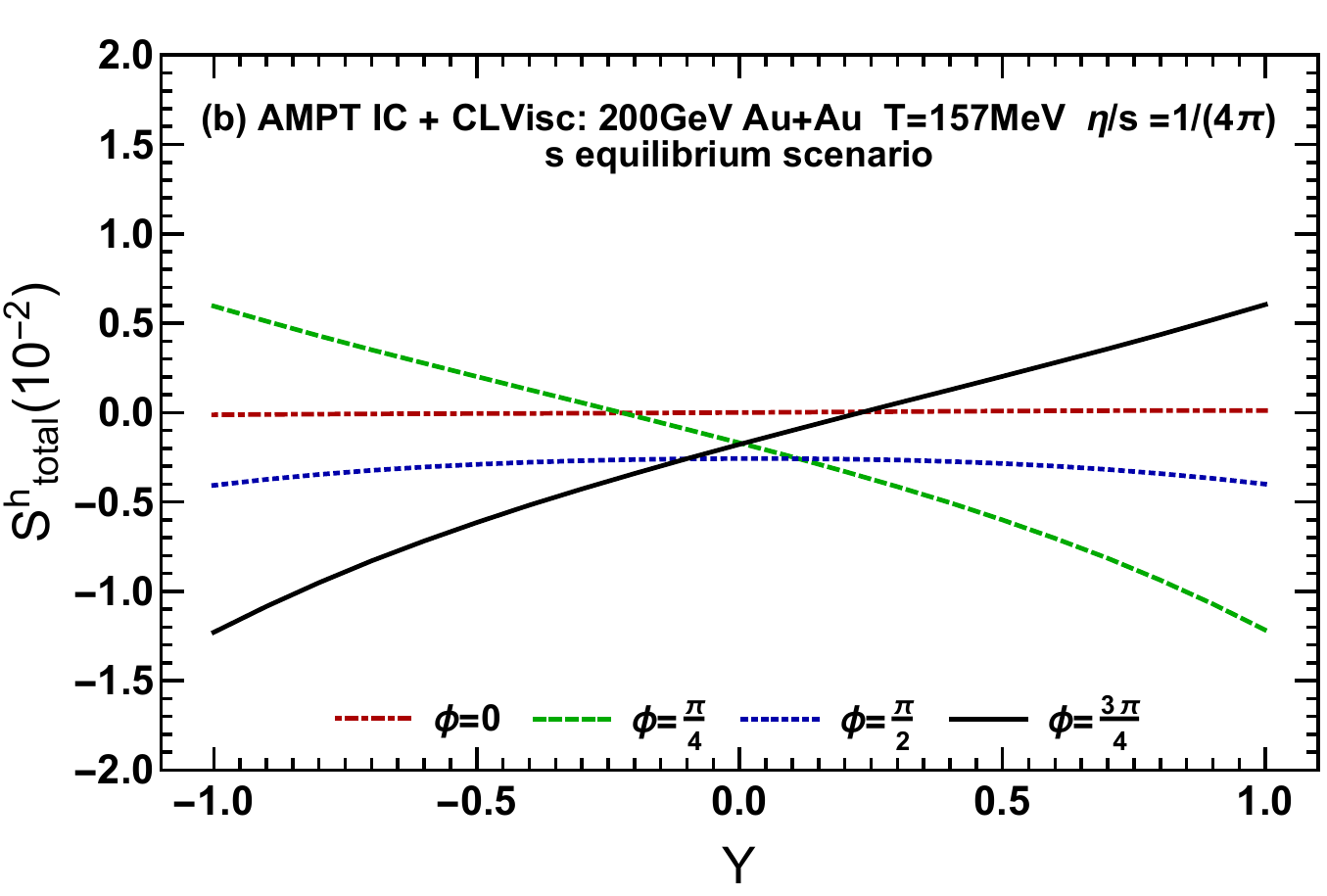}

\includegraphics[scale=0.35]{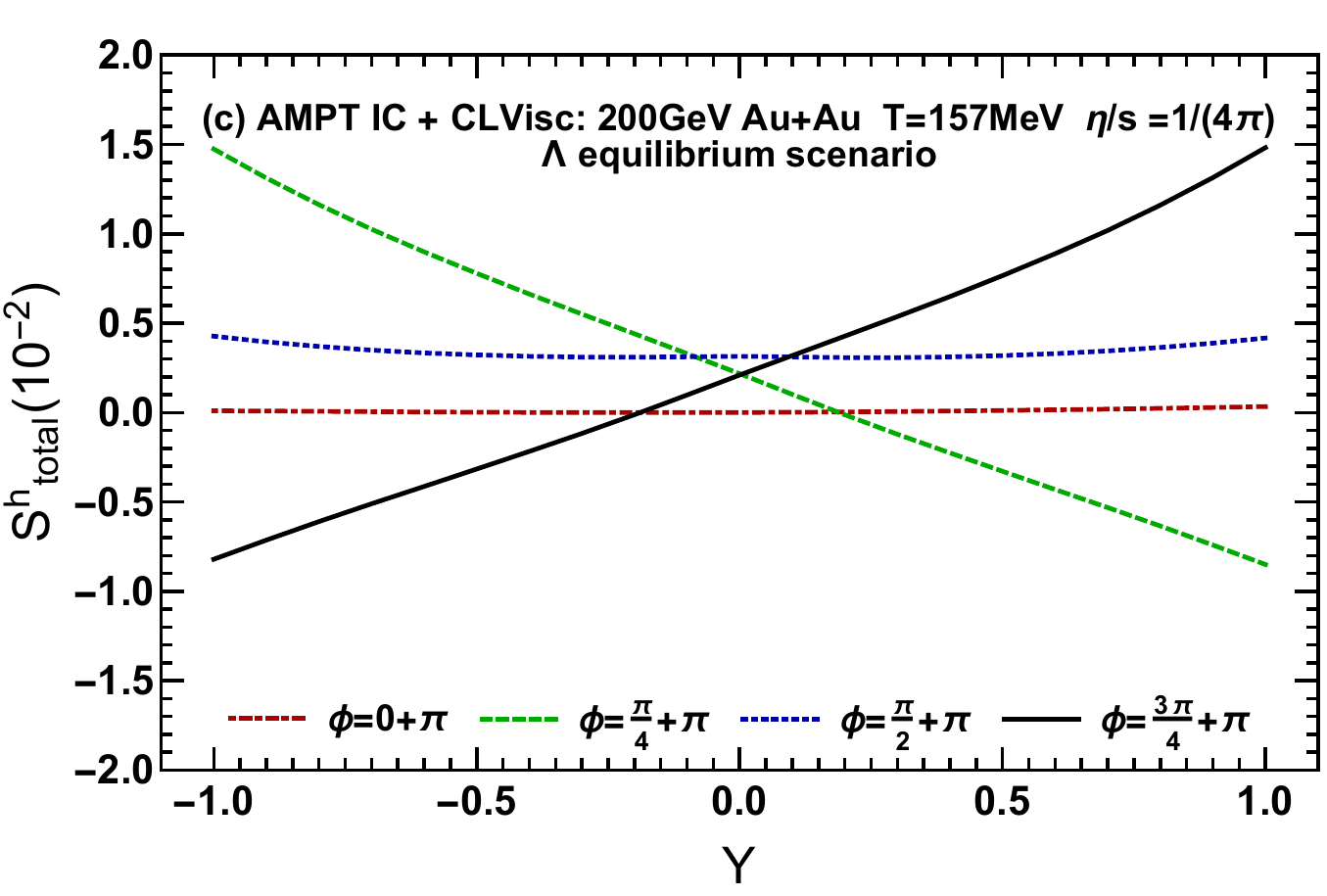}\includegraphics[scale=0.35]{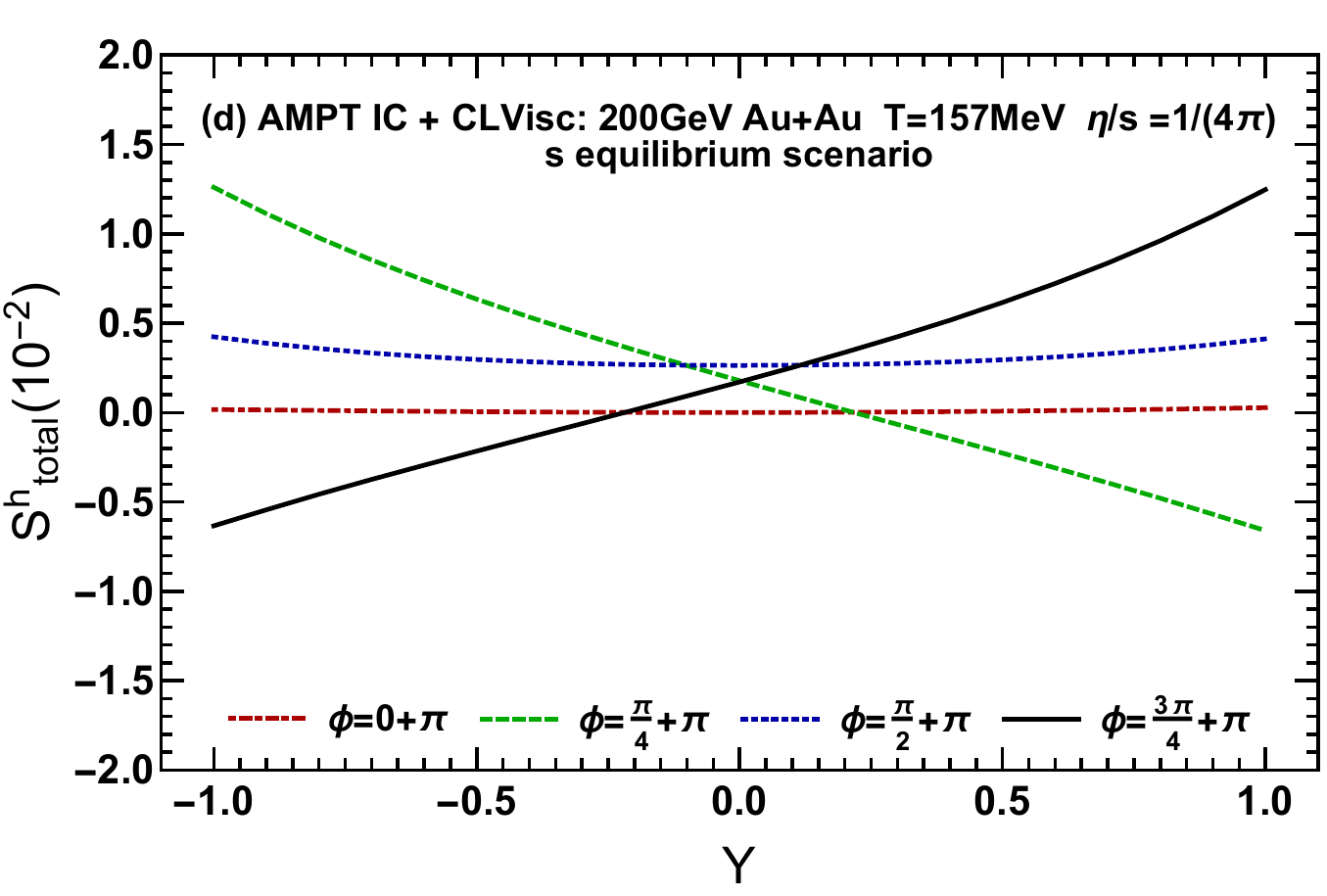}

\caption{The total helicity polarization scalar $S_{\text{total}}^{h}$ as
a function of momentum rapidity $Y$ for $\Lambda$ and $s$ equilibrium
scenario. We use the same setup as
those in Fig. \ref{fig:th_PH_phi}. Colors stand for the different
angle $\phi_{p}$. \label{fig:total_Sh_Y}}
\end{figure}

\subsubsection{Azimuthal angle dependence \label{subsec:Azimuthal-angle-dependence}}


In Fig. \ref{fig:th_PH_phi}, we plot the azimuthal angle dependent
helicity polarization and spin polarization induced by thermal vorticity
with the three different integral areas.
To understand the Fig.  \ref{fig:th_PH_phi}, let us estimate the $P_H^{\textrm{thermal}},P_H^{\pm\textrm{thermal}}$ by using
Eq. (\ref{eq:YSz}). According to the definitions of $P_H^{\pm\textrm{thermal}}, P^z_{\textrm{thermal}}$  in Eqs.(\ref{eq:PHpl_def}, \ref{eq:Pz_defination}), we find that in a Bjorken flow and $Y\simeq 0$ limit,
\begin{equation}
P_H^{\pm, \textrm{thermal}} \simeq \pm (\Delta Y)^{2} P^{z,\pm}_{\textrm{thermal}}.
\end{equation}
For our case, $\Delta Y =1$, we expect that $\textrm{sign}(P_{H}^{\pm\text{thermal}})=\pm\textrm{sign}(P_{\text{thermal}}^{\pm z})$,
$P_{H}^{+\text{thermal}}=-P_{H}^{-\text{thermal}}$,
and $P_{H}^{\text{thermal}}\simeq0$ for arbitrary
$\phi_{p}$. We observe that although the relation for the
signs of $P_{H}^{\pm\text{thermal}}$ and $P_{\text{thermal}}^{\pm z}$
are approximately close to our exception, the difference between $P_{H}^{+\text{thermal}}$
and $-P_{H}^{-\text{thermal}}$ is not negligible for both the $\Lambda$
and $s$ equilibrium scenarios. The reason is that we simulate
the dissipative fluids beyond the assumption of ideal Bjorken flows
used in Eq. (\ref{eq:YSz}). Interestingly, we observe the similar
pattern for helicity polarization $P_{H}$ and spin polarization $P^{z}$
induced by shear viscous tensor and fluid acceleration in Figs. \ref{fig:shear_PH_phi},
\ref{fig:acc_PH_phi} and \ref{fig:total_PH_phi}.


Other important observation in Figs.~\ref{fig:th_PH_phi},\ref{fig:shear_PH_phi},\ref{fig:acc_PH_phi},\ref{fig:total_PH_phi} is that the period
of $P_{H}^{\textrm{thermal}},P_{H}^{\textrm{shear}},P_{H}^{\textrm{accT}},P_{H}^{\textrm{total}}$
as a function of $\phi_{p}$ are approximately $2\pi$, which are
different with $P_{\textrm{ }}^{i}$. Our general discussion is as
follows. According to definition of $S^{h}$ in Eq. (\ref{eq:def_Sh}),
the non-vanishing $P_{H}$ mainly comes from the additional contributions
from $\mathcal{S}{}^{x}$ and $\mathcal{S}^{y}$. Although the period
of $\mathcal{S}(\phi_{P})$ is $\pi$ shown in Eq. (\ref{eq:S_space_reversal}),
the period of $\int_{-\Delta Y}^{+\Delta Y}dY(\widehat{p}^{x}\mathcal{S}^{x}+\widehat{p}^{y}\mathcal{S}^{y})$
should be $2\pi$ instead of $\pi$ since $p^{x}\propto\cos\phi_{p}$
and $p^{y}\propto\sin\phi_{p}$.
To clarify this, we integrate over the rapidity and $p_T$ in Eq.~(\ref{eq:YSz}) 
and obtain that $P_H$ in a Bjorken flow,
\begin{equation}
P_{H}\propto(\Delta Y)^{3}\left.\frac{dT}{d\tau}\right|_{\Sigma}\int dp_{T}p_{T}\left.\frac{\partial v_{1}}{\partial Y}\right|_{Y\rightarrow0}\sin\phi_p,
\end{equation}
which also implies that the period of $P_H$ is approximately $2\pi$ instead of $\pi$.
It is consistent with the analysis
in $Y=0$ limit shown in Ref. \citep{Becattini:2020xbh}.

Remarkably, the future measurement of helicity polarization may help
us to distinguish the thermal-vorticity induced spin polarization
with the others in local spin polarization. In Figs. \ref{fig:th_PH_phi},
\ref{fig:shear_PH_phi}, \ref{fig:acc_PH_phi}, \ref{fig:total_PH_phi},
we observe that the shear and fluid acceleration induced helicity
polarization $P_{H}^{\textrm{shear}},P_{H}^{\textrm{accT}}$ are much
smaller than $P_{H}^{\textrm{thermal}}$, i.e. $P_{H}\simeq P_{H}^{\textrm{thermal}}$
in both $\Lambda$ and $s$ quark scenarios of our simulations. Therefore,
it may be possible to fix the value of thermal induced local spin
polarization $P_{\textrm{thermal}}^{x,y,z}$ by matching the results
from numerical simulations for $P_{H}^{\textrm{thermal}}$ with the
data of $P_{H}$ from the future experiments. 

To understand the above observation, let us take a close look to the
expression of $S_{\textrm{shear}}^{h}$, $S_{\textrm{accT}}^{h}$
and $S_{\textrm{thermal}}^{h}$ in Eqs. (\ref{eq:helcity_decomp_01}).
We can further decompose $S_{\textrm{thermal}}^{h}$ as two parts,
\begin{eqnarray}
S_{\textrm{thT}}^{h}(\mathbf{p}) & = & \int d\Sigma^{\sigma}F_{\sigma}\frac{p_{0}}{T^{2}}\widehat{\mathbf{p}}\cdot
({\mathbf{u}}\times\nabla T),\nonumber \\
S_{\textrm{thU}}^{h}(\mathbf{p}) & = & \int d\Sigma^{\sigma}F_{\sigma}\frac{p_{0}}{T}\widehat{\mathbf{p}}\cdot{\bm{\omega}},
\end{eqnarray}
where $\nabla$ represents the spatial component of $\partial_{\mu}$ and $\bm{\omega}=\nabla\times\mathbf{u}$ denotes the fluid
vorticity. Note that, $S_{\textrm{shear}}^{h}$, $S_{\textrm{accT}}^{h}$
and $S_{\textrm{thT}}^{h}$ are proportional to the integration of
fluid velocity $\mathbf{u}$, while $S_{\textrm{thU}}^{h}$ is proportional
to the integration of the space derivative of $\mathbf{u}$ only.
From Figs. \ref{fig:shear_PH_phi},\ref{fig:acc_PH_phi} and \ref{fig:decompose-thermal},
it seems that these effects proportional to fluid velocity $\mathbf{u}$
are approximately symmetric in the rapidity area $[-1,0]$ and $[0,+1]$.
The difference of $S_{\textrm{thU}}^{h}$ in rapidity area $[-1,0]$
and $[0,+1]$ are significant. That incomplete cancellation of $S_{\textrm{thU}}^{h}$
in two rapidity area causes the thermal induced helicity polarization
dominates in $P_{H}$.

It is also clear to see $P_{\text{H}}^{\text{thU}} \gtrsim 2 P_{\text{H}}^{\text{thT}}$
as shown in Fig.~\ref{fig:decompose-thermal} 
(in central rapidity $[-1,1]$). 
We would like to comment that in low-energy collisions the contributions from fluid vorticity are expected to 
be significant enhanced  due to the nuclear-stopping
effect \cite{Jiang:2016woz,Deng:2016gyh}, while the gradient of temperature may reduce. 
In such a case, we expect that in low-energy collisions $P_{\text{H}}^{\text{total}}$
is mostly led by the contribution from fluid vorticity. Accordingly,
by measuring $P_{\text{H}}^{\text{total}}$, one may estimate the
local magnitude of $|\bm{\omega}|$ in low energy collisions. We also test these results
in different sets of parameters and find that the conclusion holds.


At last, we discuss the results in two different scenarios. We observe
that $P_{H}^{+\text{thermal}}$, $P_{H}^{-\text{thermal}}$ and $P_{H}^{\text{thermal}}$
are almost the same in two scenarios shown in Fig. \ref{fig:th_PH_phi}.
Both the expression of $S_{\text{thermal}}^{h}$ or $\mathcal{S}_{\textrm{thermal}}^{\mu}$
in Eqs. (\ref{eq:S_all}, \ref{eq:helcity_decomp_01}) and the previous
numerical simulations in different studies \citep{Fu:2021pok,Yi:2021ryh}
indicate that $\mathcal{S}_{\textrm{thermal}}^{x,y,z}$ in the lab
frame is insensitive to the particles' mass expect for the
over-all factor $m_{\Lambda}$ in the denominator. Our findings
are consistent with the previous results. For the similar reason,
the difference of $P_{H}^{+\text{accT }}$, $P_{H}^{-\text{accT}}$
and $P_{H}^{\text{accT}}$ in two scenarios is also small. However,
the shear induced local spin polarization are very sensitive to the
mass \citep{Fu:2021pok,Yi:2021ryh} due to the extra $(u\cdot p)\sim m$
in the denominator inside the integral shown in Eq.~(\ref{eq:S_all}).
That is why we observe a significant enhancement of $P_{H}^{\textrm{shear}}$
in $s$ quark scenario in Fig. \textcolor{magenta}{\ref{fig:shear_PH_phi}}.

\subsubsection{Momentum rapidity dependence \label{subsec:Momentum-rapidity-dependence}}


In the Figs. \ref{fig:th_Sh_Y}, \ref{fig:shear_Sh_Y}, \ref{fig:acc_Sh_Y},
\ref{fig:total_Sh_Y}, we plot the $S_{\text{thermal}}^{h},S_{\text{shear}}^{h},S_{\textrm{accT}}^{h},S_{\text{total}}^{h}$
as a function of momentum rapidity $Y$ at different angle $\phi_{p}=0,\pi/4,\pi/2,3\pi/4,\pi,5\pi/4,3\pi/2,7\pi/2$
for the $\Lambda$ and the $s$ \textit{\emph{equilibrium scenario}}s
and observe the space reversal symmetry. According to the space reversal
symmetry in Eq.(\ref{eq:thermal_property_02}), it is clearly shown
in these figures that $S^{h}(Y,\phi_{p})=-S^{h}(-Y,\phi_{p}+\pi)$.

Interestingly, we find at $\phi_{p}=\pi/2,3\pi/2$,only $S_{\textrm{thermal}}^{h}(\phi_{p}=\pi/2)$
is nonzero, while $S_{\textrm{shear}}^{h}$ and $S_{\textrm{accT}}^{h}$
are almost vanishing. It is straightforward to explain this
behavior. When $\phi_{p}=\pi/2,3\pi/2$, since $\widehat{p}^{x}=0$
and $\mathcal{S}^{z}\simeq0$ from both numerical simulations \citep{Fu:2021pok,Yi:2021ryh}
and experimental data \citep{Niida:2018hfw,Adam:2019srw}, $S^{h}(\phi_{p}=\pi/2)\simeq\widehat{p}^{y}\mathcal{S}^{y}(\phi_{p}=\pi/2)$
with the help of Eq. (\ref{eq:def_Sh}), while the $\mathcal{S}_{\text{shear}}^{y}$
and $\mathcal{S}_{\text{accT}}^{y}$ are found to be close to zero
at $\phi_{p}=\pi/2$ \citep{Yi:2021ryh}. Eventually, the nonzero
$\mathcal{S}_{\text{thermal}}^{y}$ gives
finite $S_{\textrm{thermal}}^{h}$ at $\phi_{p}=\pi/2$. When $\phi_{p}=0,\pi$,
$\mathcal{S}^{x,z}$ is found to be zero and $\widehat{p}^{y}=0$,
which leads to $S^{h}=0$ from Eq. (\ref{eq:def_Sh}). The
arguments above explain what we observed in these figures.

Next, we discuss the slopes of $S^{h}$ at several special angles.
At $\phi_{p}=0,\pi/2,\pi,3\pi/2$, $\mathcal{S}^{z}$ vanishes, $S^{h}$
should mainly comes from $\mathcal{S}^{x,y}$. Since $\widehat{p}^{x},\widehat{p}^{y}$
are independent on $Y$ and $\mathcal{S}^{y}$ is almost independent
on $Y$ from experimental observations \citep{Adam:2018ivw,STAR:2021beb,Adams:2021yob},
we find slow variation of $S^{h}$ as a function
of $Y$  at $\phi_{p}=0,\pi/2,\pi,3\pi/2$.
Since $\mathcal{S}^{z}$ at other $\phi_{p}$ is nonzero and it contributes
to the final $S^{h}$ through $\widehat{p}^{z}\sim\text{sin }Y$,
$S^{h}$ is very sensitive to
$Y$ when $\phi_{p}\neq0,\pi/2,\pi,3\pi/2$.



\section{Conclusions and discussions \label{sec:Conclusion-and-discussion}}

We have studied the helicity polarization in hydrodynamic approaches.
Following our previous work \citep{Yi:2021ryh}, we decompose the
helicity polarization $S^{h}$ introduced in Refs. \citep{Becattini:2020xbh,Gao:2021rom}
into several components in local equilibrium, such as helicity
polarization induced by thermal vorticity $S_{\textrm{thermal}}^{h}(\mathbf{p})$,
shear viscous tensor $S_{\textrm{shear}}^{h}(\mathbf{p})$, fluid
acceleration $S_{\textrm{accT}}^{h}(\mathbf{p})$, other terms related
to electromagnetic fields $S_{\textrm{EB}}^{h}(\mathbf{p})$, and
the gradient of the ratio to a vector chemical potential and
temperature $S_{\textrm{chemical}}^{h}(\mathbf{p})$. For simplicity,
we neglect possible contributions from an axial chemical
potential. We then obtain the space reversal symmetry of $S^{h}$
in Eq.(\ref{eq:thermal_property_02}) and discuss the property of
$S_{\textrm{thermal}}^{h}(\mathbf{p})$ in the ideal Bjorken flow
shown in Eq. (\ref{eq:YSz}). We then implement the (3+1) dimensional
viscous hydrodynamic package CLVisc with AMPT initial conditions at
$20\%-50\%$ centrality of $\sqrt{s_{NN}}=200\text{ GeV}$ Au-Au collisions
to study helicity polarization. We neglect $S_{\textrm{EB}}^{h}(\mathbf{p})$
and $S_{\textrm{chemical}}^{h}(\mathbf{p})$ in current studies and
analyze the azimuthal angle $\phi_{p}$ and the momentum rapidity
$Y$ dependence of helicity polarization contributed by $S_{\textrm{thermal}}^{h}(\mathbf{p})$,
$S_{\textrm{shear}}^{h}(\mathbf{p})$ and $S_{\textrm{accT}}^{h}(\mathbf{p})$
in \textit{$\Lambda$ }and \textit{s }\textit{\emph{equilibrium scenarios.}}

We find that the hydrodynamic simulations are beyond the theoretical
expectation in Eq. (\ref{eq:YSz}) for the ideal Bjorken flow. Different
from the local spin polarization vectors $\mathcal{S}^{\mu}$,
the helicity polarization $P_{H}$ has a period $2\pi$ instead of
$\pi$. Remarkably, we find the thermal induced helicity polarization
$P_{H}^{\textrm{thermal}}$ dominates  total
$P_{H}$. In particular, the helicity polarization contributed
by fluid vorticity, $S_{\textrm{thU}}^{h}$, is much larger than
the contributions from other components. Similar to local spin polarization,
only shear induced helicity polarization has the significant enhancement
in the $s$ equilibrium scenario. We also observe the strict space
reversal symmetry for $S^{h}$ expected in Eq. (\ref{eq:thermal_property_02}).

As a first attempt, our studies provide the baseline for the future
investigation on the correlation of helicity polarization
induced by the axial chemical potential, which is a possible signal
of local parity violation proposed by Ref. \citep{Becattini_prl2018_bk,Becattini:2020xbh,Gao:2021rom}.
Meanwhile, since we find that the helicity polarization $P_{H}$ mainly
comes from the thermal induced local spin polarization $\mathcal{S}_{\textrm{thermal}}^{\mu}$.
In the future measurements of helicity polarization, one
might match the numerical simulations of $\mathcal{S}_{\textrm{thermal}}^{\mu}$
or $P_{H}^{\textrm{thermal}}$ with the experimental data of $P_{H}$.
It may help us to distinguish the $\mathcal{S}_{\textrm{thermal}}^{\mu}$
from local spin polarization induced by other effects. 

Furthermore, 
$P_{H}$ serves as a more direct signal to characterize locally how
vortical the quark gluon plasma is and we may extract the magnitude
of local fluid vorticity $|\bm{\omega}|$.
In current study, we find that $P_{H}^{{\rm thU}}$ is much larger than other components in the helicity polarization. Because the enhancement
of fluid vorticity in low-energy collisions due to the nuclear-stopping
effect \cite{Jiang:2016woz,Deng:2016gyh}, the $P_{H}^{{\rm thU}}$ is expected to dominate $P_{H}$ in low-energy collisions.
Therefore, it is tentative to further
investigate the helicity polarization in both theory and experiment
to extract possibly strongest local fluid vorticity from the beam
energy scan.

\begin{acknowledgments}
We would like to thank Xiang-yu Wu, Hong-zhong Wu
for helpful discussion. S.P. is supported by National Natural Science
Foundation of China (NSFC) under Grants No. 1207523 and No. 12135011.
J.H. Gao was supported in part by the National Natural Science Foundation of China  under Grant  Nos. 11890710, 11890713 and 12175123, and the Major Program of Natural Science Foundation of Shandong   Province under Grant  No. ZR2020ZD30.
 D.-L. Y. was supported by the Ministry of Science
and Technology, Taiwan under Grant No. MOST 110-2112-M-001-070-MY3. 
\end{acknowledgments}

 \bibliographystyle{h-physrev}
\bibliography{spinhydro}

\begin{thebibliography}{100}

\bibitem{RevModPhys.7.129}
S.~J. Barnett,
\newblock Rev. Mod. Phys. {\bf 7}, 129 (1935).

\bibitem{ZTL_XNW_2005PLB}
Z.-T. Liang and X.-N. Wang,
\newblock Phys. Lett. {\bf B629}, 20 (2005), nucl-th/0411101.

\bibitem{ZTL_XNW_2005PRL}
Z.-T. Liang and X.-N. Wang,
\newblock Phys. Rev. Lett. {\bf 94}, 102301 (2005), nucl-th/0410079,
\newblock [Erratum: Phys. Rev. Lett.96,039901(2006)].

\bibitem{STAR:2017ckg}
STAR, L.~Adamczyk {\em et~al.},
\newblock Nature {\bf 548}, 62 (2017), 1701.06657.

\bibitem{Becattini:2007nd}
F.~Becattini and F.~Piccinini,
\newblock Annals Phys. {\bf 323}, 2452 (2008), 0710.5694.

\bibitem{Becattini:2007sr}
F.~Becattini, F.~Piccinini, and J.~Rizzo,
\newblock Phys. Rev. C {\bf 77}, 024906 (2008), 0711.1253.

\bibitem{Becattini:2013fla}
F.~Becattini, V.~Chandra, L.~Del~Zanna, and E.~Grossi,
\newblock Annals Phys. {\bf 338}, 32 (2013), 1303.3431.

\bibitem{Fang_2016PRC_polar}
R.-H. Fang, L.-G. Pang, Q.~Wang, and X.-N. Wang,
\newblock Phys. Rev. {\bf C94}, 024904 (2016), 1604.04036.

\bibitem{Karpenko_epjc2017_kb}
I.~Karpenko and F.~Becattini,
\newblock The European Physical Journal C {\bf 77}, 213 (2017).

\bibitem{XieYilong_prc2017_xwc}
Y.~Xie, D.~Wang, and L.~P. Csernai,
\newblock Phys. Rev. C {\bf 95}, 031901 (2017).

\bibitem{LiHui_prc2017_lpwx}
H.~Li, L.-G. Pang, Q.~Wang, and X.-L. Xia,
\newblock Phys. Rev. C {\bf 96}, 054908 (2017).

\bibitem{Sun:2017xhx}
Y.~Sun and C.~M. Ko,
\newblock Phys. Rev. {\bf C96}, 024906 (2017), 1706.09467.

\bibitem{Shi_plb2019_sll}
S.~Shi, K.~Li, and J.~Liao,
\newblock Physics Letters B {\bf 788}, 409  (2019).

\bibitem{WeiDexian_prc2019_wdh}
D.-X. Wei, W.-T. Deng, and X.-G. Huang,
\newblock Phys. Rev. C {\bf 99}, 014905 (2019).

\bibitem{Shi:2019wzi}
S.~Shi, H.~Zhang, D.~Hou, and J.~Liao,
\newblock (2019), 1910.14010.

\bibitem{Fu:2020oxj}
B.~Fu, K.~Xu, X.-G. Huang, and H.~Song,
\newblock Phys. Rev. C {\bf 103}, 024903 (2021), 2011.03740.

\bibitem{Ryu:2021lnx}
S.~Ryu, V.~Jupic, and C.~Shen,
\newblock (2021), 2106.08125.

\bibitem{Lei:2021mvp}
A.~Lei, D.~Wang, D.-M. zhou, B.-H. Sa, and L.~P. Csernai,
\newblock Phys. Rev. C {\bf 104}, 054903 (2021), 2110.13485.

\bibitem{Aziz:2021HADES}
F.~K. for HADES~Collaboration,
\newblock in {\em {talks given at Strangeness Quark Matter 2021, Online, May
  17-22,2021 and the 6th International Conference on Chirality, Vorticity and
  Magnetic Field in HIC, Online, Nov. 1-5,2021}}.

\bibitem{STAR:2021beb}
STAR, M.~S. Abdallah {\em et~al.},
\newblock (2021), 2108.00044.

\bibitem{Ivanov_prc2019_its}
Y.~B. Ivanov, V.~D. Toneev, and A.~A. Soldatov,
\newblock Phys. Rev. C {\bf 100}, 014908 (2019).

\bibitem{Deng:2020ygd}
X.-G. Deng, X.-G. Huang, Y.-G. Ma, and S.~Zhang,
\newblock Phys. Rev. C {\bf 101}, 064908 (2020), 2001.01371.

\bibitem{Guo:2021udq}
Y.~Guo, J.~Liao, E.~Wang, H.~Xing, and H.~Zhang,
\newblock Phys. Rev. C {\bf 104}, L041902 (2021), 2105.13481.

\bibitem{Ayala:2021xrn}
A.~Ayala, I.~Dom\'\i{}nguez, I.~Maldonado, and M.~E. Tejeda-Yeomans,
\newblock (2021), 2106.14379.

\bibitem{Deng:2021miw}
X.-G. Deng, X.-G. Huang, and Y.-G. Ma,
\newblock (2021), 2109.09956.

\bibitem{Niida:2018hfw}
STAR, T.~Niida,
\newblock Nucl. Phys. A {\bf 982}, 511 (2019), 1808.10482.

\bibitem{Adam:2019srw}
STAR, J.~Adam {\em et~al.},
\newblock Phys. Rev. Lett. {\bf 123}, 132301 (2019), 1905.11917.

\bibitem{Becattini_prl2018_bk}
F.~Becattini and I.~Karpenko,
\newblock Phys. Rev. Lett. {\bf 120}, 012302 (2018).

\bibitem{Xia:2018tes}
X.-L. Xia, H.~Li, Z.-B. Tang, and Q.~Wang,
\newblock Phys. Rev. C {\bf 98}, 024905 (2018), 1803.00867.

\bibitem{Xia:2019fjf}
X.-L. Xia, H.~Li, X.-G. Huang, and H.~Z. Huang,
\newblock Phys. Rev. C {\bf 100}, 014913 (2019), 1905.03120.

\bibitem{Becattini_epjc2019_bcs}
F.~Becattini, G.~Cao, and E.~Speranza,
\newblock The European Physical Journal C {\bf 79}, 741 (2019).

\bibitem{Li:2021jvn}
H.~Li, X.-L. Xia, X.-G. Huang, and H.~Z. Huang,
\newblock {Global hyperon polarization and effects of decay feeddown},
\newblock in {\em {19th International Conference on Strangeness in Quark
  Matter}}, 2021, 2108.04111.

\bibitem{Liu:2019krs}
S.~Y.~F. Liu, Y.~Sun, and C.~M. Ko,
\newblock (2019), 1910.06774.

\bibitem{Wu:2020yiz}
H.-Z. Wu, L.-G. Pang, X.-G. Huang, and Q.~Wang,
\newblock Nucl. Phys. A {\bf 1005}, 121831 (2021), 2002.03360.

\bibitem{Wu:2019eyi}
H.-Z. Wu, L.-G. Pang, X.-G. Huang, and Q.~Wang,
\newblock Phys. Rev. Research. {\bf 1}, 033058 (2019), 1906.09385.

\bibitem{Voloshin_2018epjWeb}
{Voloshin, Sergei A.},
\newblock EPJ Web Conf. {\bf 171}, 07002 (2018).

\bibitem{Hattori:2019lfp}
K.~Hattori, M.~Hongo, X.-G. Huang, M.~Matsuo, and H.~Taya,
\newblock Phys. Lett. B {\bf 795}, 100 (2019), 1901.06615.

\bibitem{Fukushima:2020qta}
K.~Fukushima and S.~Pu,
\newblock Lect. Notes Phys. {\bf 987}, 381 (2021), 2001.00359.

\bibitem{Fukushima:2020ucl}
K.~Fukushima and S.~Pu,
\newblock Phys. Lett. B {\bf 817}, 136346 (2021), 2010.01608.

\bibitem{Li:2020eon}
S.~Li, M.~A. Stephanov, and H.-U. Yee,
\newblock Phys. Rev. Lett. {\bf 127}, 082302 (2021), 2011.12318.

\bibitem{She:2021lhe}
D.~She, A.~Huang, D.~Hou, and J.~Liao,
\newblock (2021), 2105.04060.

\bibitem{Montenegro:2017lvf}
D.~Montenegro, L.~Tinti, and G.~Torrieri,
\newblock Phys. Rev. D {\bf 96}, 076016 (2017), 1703.03079.

\bibitem{Montenegro:2017rbu}
D.~Montenegro, L.~Tinti, and G.~Torrieri,
\newblock Phys. Rev. D {\bf 96}, 056012 (2017), 1701.08263,
\newblock [Addendum: Phys.Rev.D 96, 079901 (2017)].

\bibitem{Florkowski:2017ruc}
W.~Florkowski, B.~Friman, A.~Jaiswal, and E.~Speranza,
\newblock Phys. Rev. C {\bf 97}, 041901 (2018), 1705.00587.

\bibitem{Florkowski:2018myy}
W.~Florkowski, E.~Speranza, and F.~Becattini,
\newblock Acta Phys. Polon. B {\bf 49}, 1409 (2018), 1803.11098.

\bibitem{Becattini:2018duy}
F.~Becattini, W.~Florkowski, and E.~Speranza,
\newblock Phys. Lett. B {\bf 789}, 419 (2019), 1807.10994.

\bibitem{Florkowski:2018fap}
W.~Florkowski, R.~Ryblewski, and A.~Kumar,
\newblock Prog. Part. Nucl. Phys. {\bf 108}, 103709 (2019), 1811.04409.

\bibitem{Yang:2018lew}
D.-L. Yang,
\newblock Phys. Rev. D {\bf 98}, 076019 (2018), 1807.02395.

\bibitem{Bhadury:2020puc}
S.~Bhadury, W.~Florkowski, A.~Jaiswal, A.~Kumar, and R.~Ryblewski,
\newblock (2020), 2002.03937.

\bibitem{Shi:2020qrx}
S.~Shi, C.~Gale, and S.~Jeon,
\newblock {From Chiral Kinetic Theory To Spin Hydrodynamics},
\newblock in {\em {28th International Conference on Ultrarelativistic
  Nucleus-Nucleus Collisions (Quark Matter 2019) Wuhan, China, November 4-9,
  2019}}, 2020, 2002.01911.

\bibitem{Gallegos:2021bzp}
A.~D. Gallegos, U.~G\"ursoy, and A.~Yarom,
\newblock (2021), 2101.04759.

\bibitem{Hongo:2021ona}
M.~Hongo, X.-G. Huang, M.~Kaminski, M.~Stephanov, and H.-U. Yee,
\newblock JHEP {\bf 11}, 150 (2021), 2107.14231.

\bibitem{Florkowski:2017dyn}
W.~Florkowski, B.~Friman, A.~Jaiswal, R.~Ryblewski, and E.~Speranza,
\newblock Phys. Rev. D {\bf 97}, 116017 (2018), 1712.07676.

\bibitem{Florkowski:2018ahw}
W.~Florkowski, A.~Kumar, and R.~Ryblewski,
\newblock Phys. Rev. C {\bf 98}, 044906 (2018), 1806.02616.

\bibitem{Florkowski:2019qdp}
W.~Florkowski, A.~Kumar, R.~Ryblewski, and R.~Singh,
\newblock Phys. Rev. C {\bf 99}, 044910 (2019), 1901.09655.

\bibitem{Florkowski:2019voj}
W.~Florkowski, A.~Kumar, R.~Ryblewski, and A.~Mazeliauskas,
\newblock Phys. Rev. C {\bf 100}, 054907 (2019), 1904.00002.

\bibitem{Bhadury:2020cop}
S.~Bhadury, W.~Florkowski, A.~Jaiswal, A.~Kumar, and R.~Ryblewski,
\newblock Phys. Rev. D {\bf 103}, 014030 (2021), 2008.10976.

\bibitem{Shi:2020htn}
S.~Shi, C.~Gale, and S.~Jeon,
\newblock Phys. Rev. C {\bf 103}, 044906 (2021), 2008.08618.

\bibitem{Singh:2020rht}
R.~Singh, G.~Sophys, and R.~Ryblewski,
\newblock Phys. Rev. D {\bf 103}, 074024 (2021), 2011.14907.

\bibitem{Wang:2021ngp}
D.-L. Wang, S.~Fang, and S.~Pu,
\newblock (2021), 2107.11726.

\bibitem{Liu:2020ymh}
Y.-C. Liu and X.-G. Huang,
\newblock Nucl. Sci. Tech. {\bf 31}, 56 (2020), 2003.12482.

\bibitem{Peng:2021ago}
H.-H. Peng, J.-J. Zhang, X.-L. Sheng, and Q.~Wang,
\newblock (2021), 2107.00448.

\bibitem{Florkowski:2021wvk}
W.~Florkowski, R.~Ryblewski, R.~Singh, and G.~Sophys,
\newblock (2021), 2112.01856.

\bibitem{Copinger:2022jgg}
P.~Copinger and S.~Pu,
\newblock (2022), 2203.00847.

\bibitem{Wang:2021wqq}
D.-L. Wang, X.-Q. Xie, S.~Fang, and S.~Pu,
\newblock (2021), 2112.15535.

\bibitem{Gao:2019znl}
J.-H. Gao and Z.-T. Liang,
\newblock Phys. Rev. {\bf D100}, 056021 (2019), 1902.06510.

\bibitem{Weickgenannt:2019dks}
N.~Weickgenannt, X.-L. Sheng, E.~Speranza, Q.~Wang, and D.~H. Rischke,
\newblock Phys. Rev. D {\bf 100}, 056018 (2019), 1902.06513.

\bibitem{Weickgenannt:2020aaf}
N.~Weickgenannt, E.~Speranza, X.-l. Sheng, Q.~Wang, and D.~H. Rischke,
\newblock (2020), 2005.01506.

\bibitem{Hattori:2019ahi}
K.~Hattori, Y.~Hidaka, and D.-L. Yang,
\newblock Phys. Rev. {\bf D100}, 096011 (2019), 1903.01653.

\bibitem{Wang:2019moi}
Z.~Wang, X.~Guo, S.~Shi, and P.~Zhuang,
\newblock Phys. Rev. {\bf D100}, 014015 (2019), 1903.03461.

\bibitem{Yang:2020hri}
D.-L. Yang, K.~Hattori, and Y.~Hidaka,
\newblock JHEP {\bf 07}, 070 (2020), 2002.02612.

\bibitem{Weickgenannt:2020sit}
N.~Weickgenannt, X.-L. Sheng, E.~Speranza, Q.~Wang, and D.~H. Rischke,
\newblock {Wigner function and kinetic theory for massive spin-1/2 particles},
\newblock in {\em {28th International Conference on Ultrarelativistic
  Nucleus-Nucleus Collisions (Quark Matter 2019) Wuhan, China, November 4-9,
  2019}}, 2020, 2001.11862.

\bibitem{Li:2019qkf}
S.~Li and H.-U. Yee,
\newblock Phys. Rev. {\bf D100}, 056022 (2019), 1905.10463.

\bibitem{Liu:2020flb}
Y.-C. Liu, K.~Mameda, and X.-G. Huang,
\newblock (2020), 2002.03753.

\bibitem{Weickgenannt:2021cuo}
N.~Weickgenannt, E.~Speranza, X.-l. Sheng, Q.~Wang, and D.~H. Rischke,
\newblock (2021), 2103.04896.

\bibitem{Wang:2021qnt}
Z.~Wang and P.~Zhuang,
\newblock (2021), 2105.00915.

\bibitem{Sheng:2021kfc}
X.-L. Sheng, N.~Weickgenannt, E.~Speranza, D.~H. Rischke, and Q.~Wang,
\newblock (2021), 2103.10636.

\bibitem{Huang:2020wrr}
A.~Huang {\em et~al.},
\newblock Phys. Rev. D {\bf 103}, 056025 (2021), 2007.02858.

\bibitem{Stephanov:2012ki}
M.~A. Stephanov and Y.~Yin,
\newblock Phys. Rev. Lett. {\bf 109}, 162001 (2012), 1207.0747.

\bibitem{Son:2012zy}
D.~T. Son and N.~Yamamoto,
\newblock Phys. Rev. {\bf D87}, 085016 (2013), 1210.8158.

\bibitem{Gao:2012ix}
J.-H. Gao, Z.-T. Liang, S.~Pu, Q.~Wang, and X.-N. Wang,
\newblock Phys. Rev. Lett. {\bf 109}, 232301 (2012), 1203.0725.

\bibitem{Chen:2012ca}
J.-W. Chen, S.~Pu, Q.~Wang, and X.-N. Wang,
\newblock Phys. Rev. Lett. {\bf 110}, 262301 (2013), 1210.8312.

\bibitem{Manuel:2013zaa}
C.~Manuel and J.~M. Torres-Rincon,
\newblock Phys. Rev. {\bf D89}, 096002 (2014), 1312.1158.

\bibitem{Manuel:2014dza}
C.~Manuel and J.~M. Torres-Rincon,
\newblock Phys. Rev. {\bf D90}, 076007 (2014), 1404.6409.

\bibitem{Chen:2014cla}
J.-Y. Chen, D.~T. Son, M.~A. Stephanov, H.-U. Yee, and Y.~Yin,
\newblock Phys. Rev. Lett. {\bf 113}, 182302 (2014), 1404.5963.

\bibitem{Chen:2015gta}
J.-Y. Chen, D.~T. Son, and M.~A. Stephanov,
\newblock Phys. Rev. Lett. {\bf 115}, 021601 (2015), 1502.06966.

\bibitem{Hidaka:2016yjf}
Y.~Hidaka, S.~Pu, and D.-L. Yang,
\newblock Phys. Rev. {\bf D95}, 091901 (2017), 1612.04630.

\bibitem{Hidaka:2017auj}
Y.~Hidaka, S.~Pu, and D.-L. Yang,
\newblock Phys. Rev. {\bf D97}, 016004 (2018), 1710.00278.

\bibitem{Mueller:2017lzw}
N.~Mueller and R.~Venugopalan,
\newblock Phys. Rev. {\bf D97}, 051901 (2018), 1701.03331.

\bibitem{Hidaka:2018ekt}
Y.~Hidaka and D.-L. Yang,
\newblock Phys. Rev. D {\bf 98}, 016012 (2018), 1801.08253.

\bibitem{Hidaka:2018mel}
Y.~Hidaka, S.~Pu, and D.-L. Yang,
\newblock Nucl. Phys. {\bf A982}, 547 (2019), 1807.05018.

\bibitem{Gao:2018wmr}
J.-H. Gao, Z.-T. Liang, Q.~Wang, and X.-N. Wang,
\newblock Phys. Rev. {\bf D98}, 036019 (2018), 1802.06216.

\bibitem{Huang:2018wdl}
A.~Huang, S.~Shi, Y.~Jiang, J.~Liao, and P.~Zhuang,
\newblock Phys. Rev. {\bf D98}, 036010 (2018), 1801.03640.

\bibitem{Liu:2018xip}
Y.-C. Liu, L.-L. Gao, K.~Mameda, and X.-G. Huang,
\newblock Phys. Rev. {\bf D99}, 085014 (2019), 1812.10127.

\bibitem{Lin:2019ytz}
S.~Lin and A.~Shukla,
\newblock JHEP {\bf 06}, 060 (2019), 1901.01528.

\bibitem{Lin:2019fqo}
S.~Lin and L.~Yang,
\newblock Phys. Rev. D {\bf 101}, 034006 (2020), 1909.11514.

\bibitem{Yamamoto:2020zrs}
N.~Yamamoto and D.-L. Yang,
\newblock Astrophys. J. {\bf 895}, 56 (2020), 2002.11348.

\bibitem{Hidaka:2022dmn}
Y.~Hidaka, S.~Pu, Q.~Wang, and D.-L. Yang,
\newblock (2022), 2201.07644.

\bibitem{Zhang:2019xya}
J.-j. Zhang, R.-h. Fang, Q.~Wang, and X.-N. Wang,
\newblock Phys. Rev. C {\bf 100}, 064904 (2019), 1904.09152.

\bibitem{Huang:2020kik}
X.-G. Huang, P.~Mitkin, A.~V. Sadofyev, and E.~Speranza,
\newblock JHEP {\bf 10}, 117 (2020), 2006.03591.

\bibitem{Hattori:2020gqh}
K.~Hattori, Y.~Hidaka, N.~Yamamoto, and D.-L. Yang,
\newblock JHEP {\bf 02}, 001 (2021), 2010.13368.

\bibitem{Lin:2021mvw}
S.~Lin,
\newblock (2021), 2109.00184.

\bibitem{Luo:2021uog}
X.-L. Luo and J.-H. Gao,
\newblock JHEP {\bf 11}, 115 (2021), 2107.11709.

\bibitem{Muller:2021hpe}
B.~M\"uller and D.-L. Yang,
\newblock (2021), 2110.15630.

\bibitem{Yang:2021fea}
D.-L. Yang,
\newblock (2021), 2112.14392.

\bibitem{Liu:2020dxg}
S.~Y.~F. Liu and Y.~Yin,
\newblock Phys. Rev. D {\bf 104}, 054043 (2021), 2006.12421.

\bibitem{Liu:2021uhn}
S.~Y.~F. Liu and Y.~Yin,
\newblock JHEP {\bf 07}, 188 (2021), 2103.09200.

\bibitem{Becattini:2021suc}
F.~Becattini, M.~Buzzegoli, and A.~Palermo,
\newblock Phys. Lett. B {\bf 820}, 136519 (2021), 2103.10917.

\bibitem{Fu:2021pok}
B.~Fu, S.~Y.~F. Liu, L.~Pang, H.~Song, and Y.~Yin,
\newblock (2021), 2103.10403.

\bibitem{Becattini:2021iol}
F.~Becattini, M.~Buzzegoli, A.~Palermo, G.~Inghirami, and I.~Karpenko,
\newblock (2021), 2103.14621.

\bibitem{Yi:2021ryh}
C.~Yi, S.~Pu, and D.-L. Yang,
\newblock Phys. Rev. C {\bf 104}, 064901 (2021), 2106.00238.

\bibitem{Wu:2022mkr}
X.-Y. Wu, C.~Yi, G.-Y. Qin, and S.~Pu,
\newblock (2022), 2204.02218.

\bibitem{Sun:2021nsg}
Y.~Sun, Z.~Zhang, C.~M. Ko, and W.~Zhao,
\newblock (2021), 2112.14410.

\bibitem{Liu:2021nyg}
Y.-C. Liu and X.-G. Huang,
\newblock (2021), 2109.15301.

\bibitem{Florkowski:2021xvy}
W.~Florkowski, A.~Kumar, A.~Mazeliauskas, and R.~Ryblewski,
\newblock (2021), 2112.02799.

\bibitem{Becattini:2020xbh}
F.~Becattini, M.~Buzzegoli, A.~Palermo, and G.~Prokhorov,
\newblock (2020), 2009.13449.

\bibitem{Gao:2021rom}
J.-H. Gao,
\newblock Phys. Rev. D {\bf 104}, 076016 (2021), 2105.08293.

\bibitem{Vilenkin:1980fu}
A.~Vilenkin,
\newblock Phys. Rev. D {\bf 22}, 3080 (1980).

\bibitem{Kharzeev:2007jp}
D.~E. Kharzeev, L.~D. McLerran, and H.~J. Warringa,
\newblock Nucl. Phys. {\bf A803}, 227 (2008), 0711.0950.

\bibitem{Fukushima:2008xe}
K.~Fukushima, D.~E. Kharzeev, and H.~J. Warringa,
\newblock Phys. Rev. {\bf D78}, 074033 (2008), 0808.3382.

\bibitem{Kharzeev:2015znc}
D.~E. Kharzeev, J.~Liao, S.~A. Voloshin, and G.~Wang,
\newblock Prog. Part. Nucl. Phys. {\bf 88}, 1 (2016), 1511.04050.

\bibitem{STAR:2009wot}
STAR, B.~I. Abelev {\em et~al.},
\newblock Phys. Rev. Lett. {\bf 103}, 251601 (2009), 0909.1739.

\bibitem{STAR:2021mii}
STAR, M.~Abdallah {\em et~al.},
\newblock (2021), 2109.00131.

\bibitem{Jacob:1987sj}
M.~Jacob and J.~Rafelski,
\newblock Phys. Lett. B {\bf 190}, 173 (1987).

\bibitem{Ambrus:2019khr}
V.~E. Ambrus and M.~N. Chernodub,
\newblock (2019), 1912.11034.

\bibitem{Ambrus:2020oiw}
V.~E. Ambrus and M.~N. Chernodub,
\newblock (2020), 2010.05831.

\bibitem{Ambrus:2019ayb}
V.~E. Ambrus,
\newblock JHEP {\bf 08}, 016 (2020), 1912.09977.

\bibitem{Fang:2016uds}
R.-h. Fang, J.-y. Pang, Q.~Wang, and X.-n. Wang,
\newblock Phys. Rev. {\bf D95}, 014032 (2017), 1611.04670.

\bibitem{Becattini:2015ska}
F.~Becattini {\em et~al.},
\newblock Eur. Phys. J. C {\bf 75}, 406 (2015), 1501.04468,
\newblock [Erratum: Eur.Phys.J.C 78, 354 (2018)].

\bibitem{Deng_2016PRC}
W.-T. Deng and X.-G. Huang,
\newblock Phys. Rev. {\bf C93}, 064907 (2016), 1603.06117.

\bibitem{MahajanPRL2010}
S.~M. Mahajan and Z.~Yoshida,
\newblock Phys. Rev. Lett. {\bf 105}, 095005 (2010).

\bibitem{Gao:2014coa}
J.-H. Gao, B.~Qi, and S.-Y. Wang,
\newblock Phys. Rev. D {\bf 90}, 083001 (2014), 1406.1944.

\bibitem{Yang:2017asn}
Y.-g. Yang and S.~Pu,
\newblock Acta Phys. Polon. Supp. {\bf 10}, 771 (2017), 1709.08002.

\bibitem{Wang:2020ewj}
J.~Wang and S.~Pu,
\newblock Nucl. Phys. Rev. {\bf 37}, 679 (2020), 2008.07789.

\bibitem{Pang:2012he}
L.~Pang, Q.~Wang, and X.-N. Wang,
\newblock Phys. Rev. C {\bf 86}, 024911 (2012), 1205.5019.

\bibitem{Lin:2004en}
Z.-W. Lin, C.~M. Ko, B.-A. Li, B.~Zhang, and S.~Pal,
\newblock Phys. Rev. C {\bf 72}, 064901 (2005), nucl-th/0411110.

\bibitem{Huovinen:2009yb}
P.~Huovinen and P.~Petreczky,
\newblock Nucl. Phys. A {\bf 837}, 26 (2010), 0912.2541.

\bibitem{Deng:2012pc}
W.-T. Deng and X.-G. Huang,
\newblock Phys. Rev. {\bf C85}, 044907 (2012), 1201.5108.

\bibitem{Roy:2015coa}
V.~Roy and S.~Pu,
\newblock Phys. Rev. {\bf C92}, 064902 (2015), 1508.03761.

\bibitem{Jiang:2016woz}
Y.~Jiang, Z.-W. Lin, and J.~Liao,
\newblock Phys. Rev. {\bf C94}, 044910 (2016), 1602.06580,
\newblock [Erratum: Phys. Rev.C95,no.4,049904(2017)].

\bibitem{Deng:2016gyh}
W.-T. Deng and X.-G. Huang,
\newblock Phys. Rev. C {\bf 93}, 064907 (2016), 1603.06117.

\bibitem{Adam:2018ivw}
STAR, J.~Adam {\em et~al.},
\newblock Phys. Rev. {\bf C98}, 014910 (2018), 1805.04400.

\bibitem{Adams:2021yob}
J.~R. Adams and M.~A. Lisa,
\newblock (2021), 2109.14726.

\end{thebibliography}

\end{document}